**Heterogeneity of tumor biophysical properties and their potential role as prognostic markers.**

Anja Madleine Markl[1-4], Daniel Nieder[2], Diana Isabel Sandoval-Bojorquez[2], Anna Taubenberger [5,6], Jean-François Berret [7], Artur Yakimovich [8,9], Eduardo Sergio Oliveros-Mata[3], Larysa Baraban[2,6], Anna Dubrovska[1,4, 10-12,*]

1 - Helmholtz-Zentrum Dresden-Rossendorf, Institute of Radiooncology-OncoRay, 01328 Dresden, Germany.

2 - Institute of Radiopharmaceutical Cancer Research, Helmholtz-Zentrum Dresden-Rossendorf, 01328, Dresden, Germany.

3 - Institute of Ion Beam Physics and Materials Research, Helmholtz-Zentrum Dresden-Rossendorf, 01328, Dresden, Germany.

4 - OncoRay-National Center for Radiation Research in Oncology, Faculty of Medicine and University Hospital Carl Gustav Carus, Technische Universität Dresden and Helmholtz-Zentrum Dresden-Rossendorf, 01309 Dresden, Germany.

5 - Biotechnology Center (BIOTEC) and Center for Molecular and Cellular Bioengineering (CMCB), Technische Universität Dresden, Dresden, Germany.

6 - Medical Faculty Carl Gustav Carus, Technische Universität Dresden, Dresden, Germany.

7 - Université Paris Cité, CNRS, Matière et systèmes complexes 75013, Paris, France.

8 - Center for Advanced Systems Understanding (CASUS), 02826 Görlitz, Germany.

9 - Helmholtz-Zentrum Dresden-Rossendorf, 01328 Dresden, Germany.

10 - German Cancer Research Center (DKFZ), 69120 Heidelberg, Germany.

11- German Cancer Consortium (DKTK), Partner Site Dresden, 01309 Dresden, Germany.

12- National Center for Tumor Diseases (NCT), Partner Site Dresden, 01307 Dresden, Germany.

*- corresponding author.

**Abstract**

Progress in our knowledge of tumor mechanisms and complexity led to the understanding of the physical parameters of cancer cells and their microenvironment, including the mechanical, thermal, and electrical properties, solid stress, and liquid pressure, as critical regulators of tumor progression and potential prognostic traits associated with clinical outcomes. The biological hallmarks of cancer and physical abnormalities of tumors are mutually reinforced, promoting a vicious cycle of tumor progression. A comprehensive analysis of the biological and physical tumor parameters is critical for developing more robust prognostic and diagnostic markers and improving treatment efficiency. Like the biological tumor traits, physical tumor features are characterized by inter- and intratumoral heterogeneity. The dynamic changes of physical tumor traits during tumor progression and as a result of tumor treatment highlight the necessity of their spatial and temporal analysis in clinical settings. This review focuses on the biological basis of the tumor-specific physical traits, the state-of-the-art methods of their analyses, and the perspective of clinical translation. The importance of tumor physical parameters for disease progression and therapy resistance, as well as current treatment strategies to monitor and target tumor physical traits in clinics, is highlighted.

**Outline:**

## 1. Introduction

Since the first radical mastectomy conducted by William Halsted in 1882 to treat breast cancer (1), cancer treatment and patient survival have been substantially improved (2). However, the battle against cancer is by far not yet won. One of the central challenges for cancer treatment is identifying patients who are more likely to respond favorably to a given anticancer therapy. Clinical and pathological parameters routinely used for diagnosing, predicting outcomes, and treatment selection, often fail to describe tumor heterogeneity comprehensively. A consensus is that the combination of pathological and clinical parameters with biological prognosticators better explains the heterogeneity in the treatment outcomes. Identification of these biological stratifiers, as well as delivering more tailored treatment, is the ultimate goal of personalized and precise medicine. As the indicators of tumor presence and progression, cancer biomarkers are tightly connected to the functional capabilities of tumor cells, which Hanahan and Weinberg describe as cancer hallmarks (3, 4). These hallmarks of tumor functions include enabling chronic proliferation and immortality, circumventing growth suppression and cell death, activating angiogenesis and metastasis, genome instability and epigenetic reprogramming, transitory senescence and immune evasion, tumor-inducing inflammation and tumor-promoting microbiota, cellular phenotypic plasticity and reprogrammed cellular metabolism (3, 4). As our knowledge of tumor mechanisms and complexity progressed, the physical parameters of cancer cells and their microenvironment gained appreciation as critical regulators of tumor progression and therapy resistance and potential prognostic traits associated with clinical outcomes in patients with different types of malignancies.

The physical traits of cancer include the characteristics of tumor cells, such as mechanical, thermal, electrical properties and the physical microenvironment, including solid and liquid pressure, tissue stiffness and architecture (5, 6). During tumor development, the physical abnormalities of tumors emerge as a consequence of biological hallmarks of cancer but also induce more aggressive functional tumor capabilities and result in a vicious cycle driving tumor progression. Thus, a comprehensive analysis of both biological and physical cancer parameters is critical for developing more robust prognostic stratifiers, discovering yet unexplored therapeutic targets and improving treatment efficiency. Similar to the biological hallmarks, physical abnormalities of tumors have a high inter- and intratumoral heterogeneity and can be dynamically changed during the natural tumor history and as a result of treatment. The heterogeneity of biophysical tumor characteristics and their clinical relevance highlight the necessity of their spatial and temporal analysis in cancer patients. This review focuses on the role of physical parameters of tumor cells and microenvironment for tumor development, progression and therapy resistance, and potential treatment strategies to target tumor physical traits. The analytic approaches for the assessment of these parameters and their potential implementation in clinical practice will also be discussed.

## 2. Mechanical properties of cancer cells

### 2.1. Introduction to the topic

The idea that tissue mechanical properties can inform on certain tissue abnormalities, including neoplastic lesions, is as old as the practice of tissue palpation. Thereby, the differential mechanical properties of the probed tissues are revealed qualitatively (7). Quantitative data can be obtained with higher sensitivity and reproducibility using more sophisticated physical tools that allow for *in situ* mapping of the tissue of interest, e.g., elastography techniques (8, 9), or *ex situ*, e.g., by shear rheometry (10) and atomic force microscopy (AFM) (11). Of relevance, these changes in mechanical properties are characteristic of the development and progression of tumors and can thereby reflect the tumor progression state. This is supported by previous reports showing correlations between tissue mechanical properties and breast tumor subtype and stage (12) or studies suggesting the value of tissue stiffness as diagnostic biomarkers e.g., for prostate cancer (13, 14) and liver cancer (15). In many cases, solid tumors are found to be stiffer than normal tissues. However, there are also studies where no significant differences between normal and tumor tissues can be seen, as in the case of cervical cancer (16).

Biological tissues are soft, complex, and heterogeneous materials composed of various cell types and extracellular matrix (ECM) that provide a scaffold to embedded cells. Mechanical changes in tumors span from subcellular to tissue scales and can be heterogeneous at all levels. At the tissue scale, they often arise from alterations in ECM architecture and mechanical characteristics. In some solid tumors, including breast, prostate and pancreatic cancer, a denser and more rigid collagen network is commonly seen and this desmoplastic response can be associated with poorer patient outcomes (17-20). Increased collagen deposition and cross-linking are driven by cancer-associated fibroblasts that are characterized by a more contractile phenotype (21, 22). Besides increased matrix stiffness, interstitial fluid pressure, solid stress, and cellular interactions can contribute to overall altered tissue mechanical properties. Interstitial fluid pressure, driven by blood and lymphatic fluids, differs between normal tissues and tumors (5, 23). Interstitial fluids can also flow due to pressure differences between blood and lymphatic systems, imposing wall shear stresses on surrounding cells and tissues (24, 25). A tissue mechanical parameter that appears relevant with regard to diagnostics is tissue fluidity- also accessible through elastography measurements. Tumor tissues can be more fluid-like or more solid-like, which is not simply related to the tissue's water content but also to collective cell behaviors, e.g., cell-cell and cell-ECM interactions (26). These collective cell behaviors can be physically described as jamming and unjamming transitions that arise also in non-biological systems (27, 28). Last but not least, the tumor constituent cells themselves can exhibit cell-to-cell heterogeneities but also internal heterogeneity in their biochemical composition, consisting of a liquid or gel phase interlaced with fibrillar networks building the cytoskeleton. Depending on the tissue, these properties can largely affect the tissue mechanical response as, for instance, recently shown for breast and cervix (16) and brain tumors (29). Due to their structural complexity, it is expected that cells, both healthy and cancerous, display considerable variability in their viscoelastic properties - a phenomenon amply confirmed by experimental studies (6, 30-32). This variability depends in particular on the cell types, cell state (cell cycle, shape) and the spatial and time scale at which these properties are measured. A consensus has emerged defining three characteristic length scales in cell mechanics (32): the intracellular scale (< 10 µm), the whole-cell scale (~ 10-30 µm), and the tissue scale (> 30 µm) (33).

Here, the focus will be on cell-scale mechanical changes and measurement techniques assessing apparent elastic and viscoelastic changes in cells at the intracellular and whole-cell scales. Specifically, a meta-analysis of previous studies that have embarked on comparative analyses of healthy and cancerous tissue and cells will be provided. Furthermore, the biological bases of viscoelastic alterations, commonly used tools to assess cell mechanical properties at different scales, will be described.

## 2.2.    Essential terminology and definitions

In the field of cell biomechanics, physical quantities like stiffness, elastic modulus, and viscosity are widely used to describe the mechanical response of cells and tissues (34, 35). The stiffness coefficient (K) is defined as the ratio of the applied force to the resulting displacement at the point of application. K depends not only on the material elastic properties but also on the sample shape and dimensions. The elastic modulus, in contrast, is an intrinsic bulk property of the biological material, denoted G for shear and E for extension/compression and does not depend on geometry. In many biomechanical experiments, including AFM, it is often challenging to determine whether the measured value corresponds to the stiffness coefficient or the elastic modulus. The convention adopted by many research groups is to refer to these measurements as the apparent Young's modulus, $E_{App}$ (30, 33-35). Additionally, certain techniques generate quantitative parameters, such as shear wave speed or Brillouin frequency shifts (36, 37), that are indirect stiffness indicators, rather than physical quantities known from continuum mechanics or rheology. The above quantities, K, G, and E can also be measured as a function of frequency, leading to complex variables K*(ω), G*(ω), and E*(ω). The frequency dependence also allows the data to be compared with predictions from the constitutive equations of known viscoelastic models (38). Active or passive microrheology using internalized micron-sized probes, in turn, measures the shear viscosity ($η_0$) and elastic modulus (G) of the cytoplasm (39). When experimental conditions are not at low-frequency and in shear mode, the terms apparent viscosity ($η_{App}$) or modulus ($G_{App}$) are used instead.

## 2.3. A meta-analysis of comparative studies assessing cell mechanical properties

### 2.3.1. Cell cortical stiffness measurements

The first reports that cancer cells are softer than normal cells were published almost two decades ago (30, 33, 34, 40, 41). Since then, multiple studies have been conducted on the mechanical analysis of established cancer cell lines and patient-derived cells. A meta-analysis on a broader collection of studies contrasting cancer and normal cell stiffness is presented in **Figure 1**. It comprises a total of 30 studies across six different cancer types: breast (n = 14), pancreas (n = 3), bladder (n = 8), prostate (n = 3), and ovarian (n = 2) cancer. Although there are many more published studies that report on differences depending on disease state, we focused here on studies reporting comparable parameters such as the apparent Young's modulus $E_{App}$. Among the selected studies, AFM-based methods dominate, since AFM is by far the most commonly employed technique to assess cell mechanical properties. In **Figure 1**, apparent Young's moduli are plotted for the abovementioned tumor types. Most studies focused on the assessment of established cancer cell lines, mostly breast cancer cells, and a high degree of redundancy becomes apparent. In many cases, even the same cell lines were measured by different research groups, suggesting a relatively high reproducibility of results. However, the limited number of studied models also limits the possibility of drawing more general conclusions for other tumor entities. Of note, the prevalence of cancer cell lines compared to patient-derived material appears critical when cell mechanical parameters are to be evaluated with regard to their clinical relevance. The underlying reasons for using cell lines are obvious, since the use of clinically relevant samples is comparatively more challenging since it involves ethical approvals, collaborations with clinicians providing samples, and patient-to-patient heterogeneities, among other challenges. Nevertheless, it will be important to expand studies on patient-derived cells in the future.

In most cases, there is accordance that normal cells have higher apparent Young's moduli compared to cancer cells. In some cases, the more invasive cell lines or metastatic primary cells had even lower stiffness values compared to less invasive cells. There are some exceptions, however, e.g., where increased cellular stiffness values are shown for invasive compared to low invasive prostate cancer cells (42). Of note, $E_{App}$ across studies widely ranges from approximately 100 to up to 30.000 Pa. Studies reporting higher values up to 100.000 Pa were excluded here, as they appear unrealistic for soft biological tissues. Generally, AFM indentation tests with sharp indenters on spread cells can result in higher values than measurements on rounded cells. In some cases, especially when spread cells with flat cellular extensions are tested, the underlying stiff substrate also might have affected the obtained values.

Although the viscoelastic nature of cells is well known (43, 44), most cell mechanics surveys to date have been limited to an apparent elastic response (> 80% of all studies), with only a small fraction addressing cytoplasmic viscosity properties. Here, we extended our literature search to studies that assess cell viscosity, aiming to explore its potential as a biomarker for cancer diagnostics.

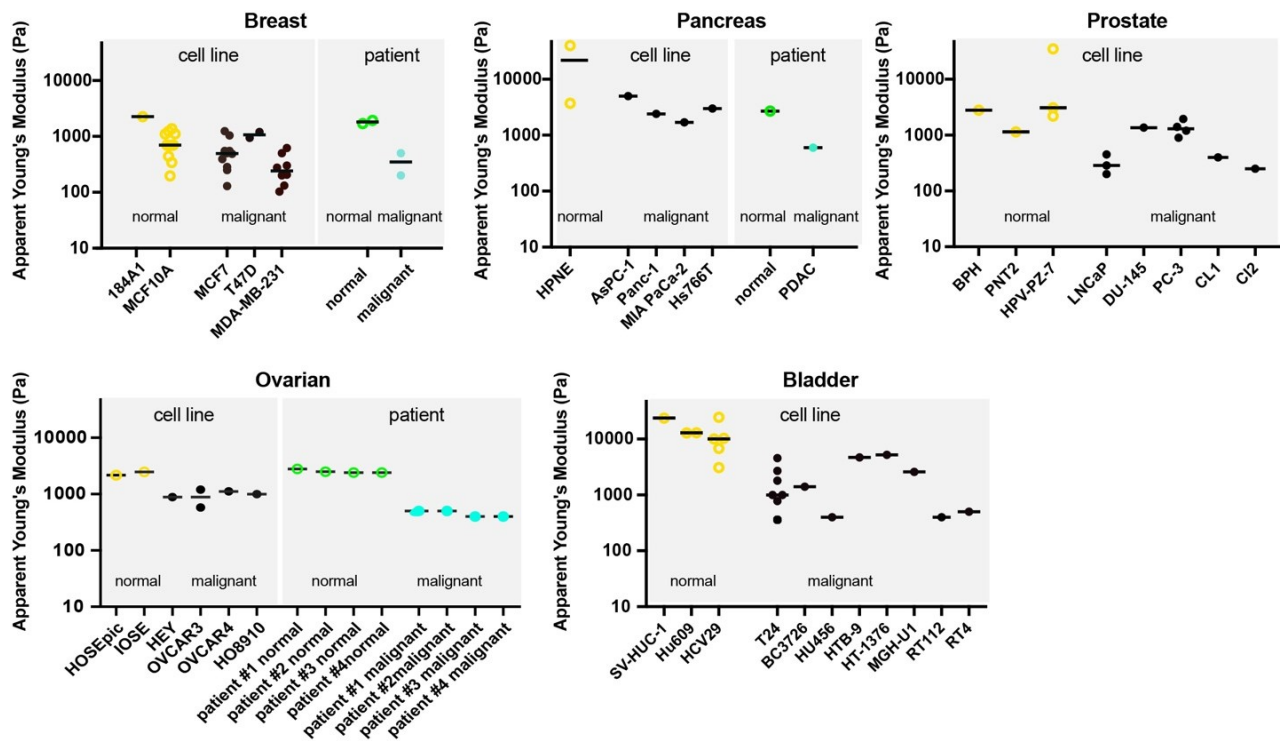

**Figure 1.** Meta-analysis of cell indentation measurements on whole cells conducted by AFM. $E_{app}$ measured for different types of cancer cells are given, with data taken from studies on breast (40, 45-56), pancreas (31, 56, 57), bladder (30, 58-64), prostate (42, 48, 65, 66), and ovarian cancer cells (56, 67, 68). Cell lines and patient-derived cells are displayed separately. Dots represent average or median values taken from respective studies. Where multiple studies report on the same cell lines, multiple dots are shown, and horizontal lines indicate medians.

### 2.3.2. Whole-cell viscosity

To measure the whole-cell apparent viscosity $\eta_{App}$, researchers employ microrheology techniques inspired by stress relaxation or creep tests, commonly used in conventional rheology (38). Our meta-analysis encompasses a broader sample compared to intracellular viscosity, including nine studies across six different cancer types: breast, kidney, prostate, thyroid, ovarian, and liver. Overall, the data reveal a strong preference for AFM-based methods, which account for 80% of the studies (45-47, 50, 69-72). Other techniques are micropipette aspiration (72) and micro-electromechanical systems (MEMS) resonant sensor (47). Similar to the intracellular data, whole-cell viscosity measurements exhibited significant variability, with $\eta_{App}$ ranging from 3 to 500 Pa s (3000 to 500 000 times the viscosity of cell culture medium (73). This variation is likely due to differences in the cell and cancer types studied, and in the case of AFM, also influenced by the specific protocols and tips employed (30, 74). The viscosity ratio $\eta_{Canc}/\eta_{Norm}$ between cancerous and healthy cells was calculated for six cancer types to allow for comparison (**Figure 2 A**). This figure consistently reveals a clear pattern: cancer cells exhibit lower viscosities than healthy cells. Overall, this reduction in viscosity averaged 44%, comparable to the decrease observed at the intracellular level (**Figure 2 B**). For cancers with data on cells of increasing metastatic potential, only ovarian cancer cells (early, intermediate, and late MOSE cells) show a distinction between low and high invasiveness. This finding indicates that cancer cells undergo a fluidization of their flow properties compared to normal cells, a pattern that persists across different scales.

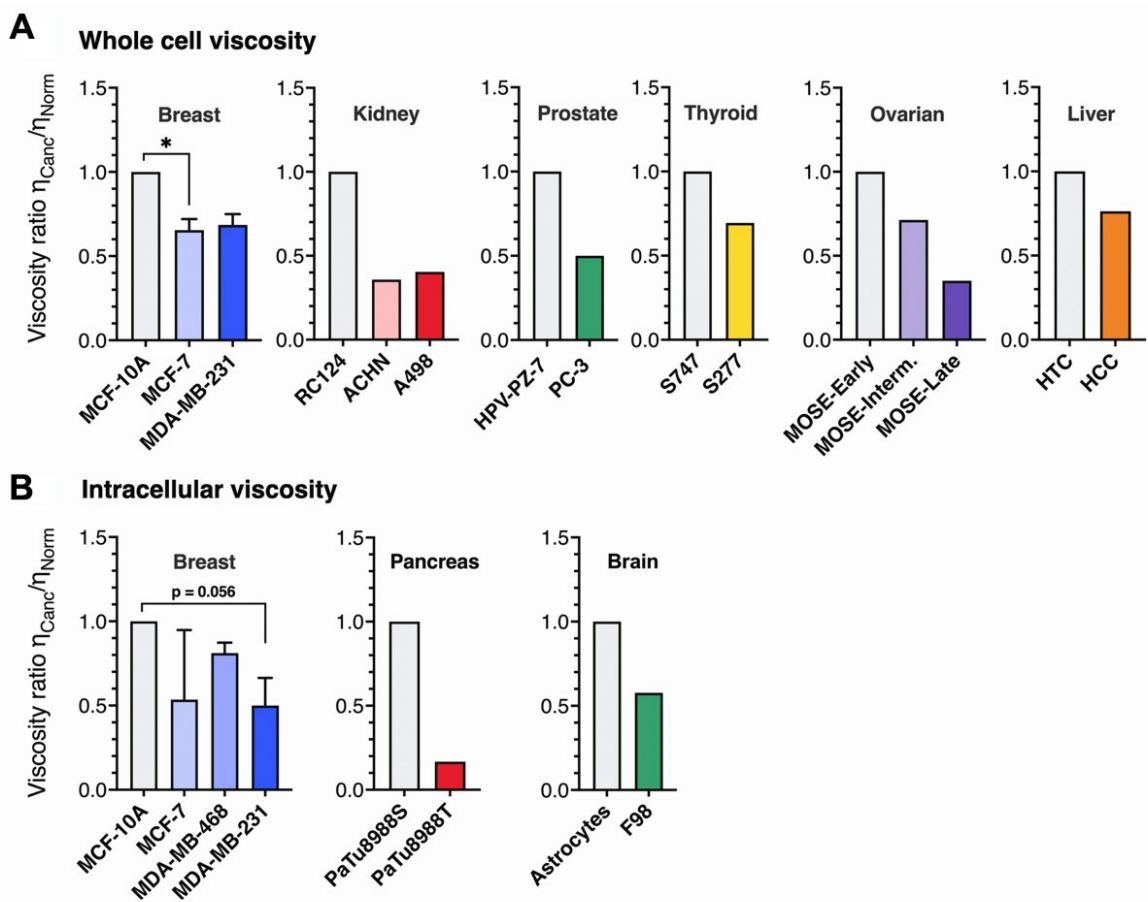

**Figure 2. A.** Analysis of the viscosity ratio between cancerous and healthy cells. Whole-cell viscosity ratio $\eta_{Canc}/\eta_{Norm}$ between cancerous and healthy cells obtained from AFM (45-47, 50, 69-72), micropipette aspiration (72) and micro-electromechanical systems (MEMS) resonant sensor (47) on six different cancers: breast (45-47, 50), kidney (71), prostate (75), thyroid (70), ovarian (69), and liver (72). Error bars in the left panel indicate the standard error of the mean from four different surveys on MCF-10A, MCF-7 and MDA-MB-231 cells; * - p value < 0.05. **B.** Intracellular viscosity ratio $\eta_{Canc}/\eta_{Norm}$ between cancerous and healthy cells obtained for breast (76-81), pancreas (80), and brain (82) cancers, where $\eta_{Canc}$ and $\eta_{Norm}$ denote the static or $\eta_{App}$, respectively. The first bar in each histogram is set to 1 by definition. The techniques used for measuring the cytoplasm viscosity were particle-tracking microrheology (78, 80, 81), magnetic rotational spectroscopy (77) and optical tweezers (76, 82). Error bars in the left panel indicate standard errors of means from six different surveys on MCF-10A, MCF-7, MDA-MB-468, and MDA-MB-231 cells.

### 2.3.3. Cytoplasmic viscosity

To measure the cytoplasmic viscosity of healthy and cancerous cells, particle-tracking microrheology, optical tweezers, and magnetic rotational spectroscopy were applied across seven studies covering three cancer types: breast, brain, and pancreas (76-82). In these techniques, micron or submicron-sized particles are dispersed in the cytosol, and their passive or active movements are tracked, allowing local viscosity measurement. Similar to the above-mentioned stiffness measurements, breast cancer cells dominate the data, specifically four cell lines with increasing metastatic potential, MCF-10A, MCF-7, MDA-MB-468, and MDA-MB-231, respectively, accounting for 75% of all intracellular assays. For conventions, the static shear viscosity, such as that obtained in particle-tracking microrheology and magnetic rotational spectroscopy, will be noted $\eta_0$ (38), and $\eta_{App}$, will be used when the experimental conditions are not low-frequency shear.

According to literature data, particle-tracking data (78-81) show viscosity values ranging from 0.3 to 6 Pa s - 300 to 6000 times that of cell culture medium at physiological temperature (73), while active methods report significantly higher viscosities, between 10 and 50 Pa s (76, 77, 82). These data heterogeneities could be attributed to the size of the probes used, from the order of a hundred nanometers for the former and a micron for the latter (77). To account for the differences in absolute values across techniques and to allow for comparison, we expressed existing intracellular results as the ratio: $\eta_{Canc}/\eta_{Norm}$, where $\eta_{Canc}$ and $\eta_{Norm}$ represent the viscosities of cancerous and normal cells, respectively. **Figure 2 A** displays the $\eta_{Canc}/\eta_{Norm}$ for breast, pancreas, and brain cancer types, with the first bar in each histogram set to 1 by definition. The figure highlights a consistent decrease in $\eta_{Canc}/\eta_{Norm}$ across all cancer types, corresponding to an average 51% reduction in viscosity. This suggests a general trend toward increased cytoplasmic fluidization in cancerous cells.

For breast cells, Young's moduli of MCF-7 and MDA-MB-231 tumorigenic lines were lower than those of healthy cells, but the difference was insufficient to differentiate between malignant and non-malignant cells clearly. Magnetic rotational spectroscopy did, however, reveal a significant difference, with MCF-7 cells showing five times higher intracellular viscosity than MDA-MB-231 (77). This underscores cytoplasmic viscosity sensitivity to metastatic potential, suggesting that it could serve as a mechanical biomarker for cancer cells with high metastatic potential. The preceding analysis shows that findings on cancer cell viscosity have been insufficiently documented and that further research is required, especially by expanding the scope to include more diverse cell lines and cancer types. In the following Sections, we will discuss the techniques utilized to measure the viscoelastic cell properties, which are summarized in Table 1.

**Table 1.** The comparative analysis of the methodologies to assess viscoelastic cell properties.

| Techniques | Whole-cell / Intracellular | Contact / Non-contact | Active / Passive | Shear/Compression-Elongation | What is measured? | Frequency range (Hz) | Time scale (s) | Key strengths | Weak points |
|---|---|---|---|---|---|---|---|---|---|
| **Atomic Force Microscopy (AFM)** | Whole cell | Contact | Active | Compression | Apparent Young's modulus, shear storage and loss modulus, apparent viscosity | 0.1 - 200 | 0.001 - 10 | High spatial resolution; Versatility; Minimal sample preparation; Force control | Low throughput Contact-based Dependence on probe geometry Frequency range limitations Interpretation complexity |
| **Micropipette aspiration** | Whole cell | Contact | Active | Elongation | Apparent Young's modulus, apparent | 0.01 - 1 | 1 - 100 | Direct measurement; Simple | Low throughput Limited spatial |
| | | | | | viscosity | | | setup; Physiological relevance; Wide applicability | resolution Geometric constraints Frequency range User-dependent variability |
| **Microfluidic techniques** | Whole cell | Contact | Active | Compression / Elongation | Apparent Young modulus | 0.1 - 1000 | 0.001 - 10 | High throughput; Dynamic measurements; Quantitative and diverse parameters; Integration and automation; Customizability | Complex fabrication Indirect measurement Limited frequency range Standardization challenges Potential physiological artifacts |
| **Optical stretcher** | Whole cell | Non-contact | Active | Elongation | Apparent Young modulus | 0.1 - 10 | 0.1 - 10 | Non-invasive; Precise force application; Single-cell resolution; High-frequency probing; Versatility | Low throughput Limited deformation range High technical complexity Heat generation Context-specific limitations |
| **Brillouin microscopy** | Intracellular | Non-contact | Passive | Longitudinal compression | Longitudinal modulus | $10^9 - 10^{10}$ | $10^{-10} - 10^{-9}$ | Non-invasive; High spatial resolution; Label-free 3D imaging capability; Simultaneous measurement | Low sensitivity to bulk properties Limited correlation with standard metrics Weak signal in soft tissues High technical complexity Limited throughput |
| **Particle-tracking microrheology** | Intracellular | Contact | Passive | Shear | Static shear viscosity | 0.1 - 10 | 0.1 - 10 | Non-invasive; High spatial resolution; Sensitive to | Limited to low-stress conditions Probe localization |
| | | | | | | | | heterogeneity; Minimal external manipulation | bias Low throughput Interpretation challenges Dependence on probe size and type |
| **Optical tweezers** | Intracellular | Contact | Active | Shear | Elastic modulus, Apparent viscosity | 0.1 - 1000 | 0.001 - 10 | High Precision; Non-invasive; Single-Cell Resolution; Force Measurement | Limited to small forces Potential photodamage Limited throughput Requires optical transparency? |
| **Magnetic rotational spectroscopy** | Intracellular | Contact | Active | Shear | Static shear viscosity, shear elastic modulus | 0.0001 - 100 | 0.01 - 10000 | Quantitative cytoplasm rheology; Non-invasive; Very broad dynamic Range; High Sensitivity; No Need for Contact | Magnetic particle requirement Low throughput |

## 2.4. Methodology

### 2.4.1. Atomic Force Microscopy (AFM)

Of all the mechanobiology techniques available today, AFM is by far the most widely used (30, 34, 41). AFM is an ultrasensitive technique to measure forces down to the piconewton range (83, 84). It relies on a force sensor, the AFM cantilever that can be equipped with a nano- or macro-sized indenter (e.g., a pyramidal tip or also larger e.g., a colloidal probe in the range of 5-10 µm diameter). To probe cell mechanical properties, the cantilever is brought into contact with a cell (either round or in adherent state) with a pre-set force or indentation depth, and from the resultant force-distance curve the force-indentation relation can be derived. By fitting the force indentation curve to a model of choice (most commonly a Hertz model for simple indentation tests (85, 86), an apparent elastic modulus $E_{App}$ can be quantified, which represents the most commonly reported parameter for cell and tissue mechanical measurements. Alternative models include, for instance, the Oliver Pharr model (11), and the liquid droplet model (87). In a configuration where the stress/strain response is measured over time, AFM can also be employed to determine the apparent viscosity of single cells (45, 46, 50, 70, 75). Such measurements include stress relaxation (88), creep compliance, or oscillatory probing protocols (46, 89, 90). In such cases, a viscoelastic mechanical response model, such as the standard linear liquid or solid model can be employed to derive viscoelastic cell parameters from force versus time traces, such as apparent modulus, relaxed modulus, and an apparent viscosity (33), or cortical tension, cortical stiffness and phase shift (89). Alternatively, information on the viscoelastic properties can be obtained from the indentation part of the curve using viscoelastic relaxation models (91, 92). The strength of AFM lies in its ability to integrate high-resolution topographical imaging with force mapping, an essential configuration for biomechanical studies. AFM can also be combined with optical imaging or spectrometric techniques, e.g., advanced optical techniques such as confocal microscopy (93), total internal reflection fluorescence (TIRF) microscopy (93), calcium imaging (42), or Raman microscopy (94) to obtain additional information on cell state. While AFM is versatile with regards to cell shapes (both rounded and adherent), it is typically restricted to probing the cell surface, and furthermore relatively low in throughput (typically 100 cells/hour).

### 2.4.2. Micropipette aspiration

This technique applies negative pressure to draw a part of a single cell into a narrow glass pipet (95). The living cell, suspended in solution, is first immobilized at the pipet tip before suction is applied. Using optical microscopy, the portion of the cell entering the pipet for a given pressure is monitored over time. The kinetic time course of cell deformation resembles creep experiments in rheology, where stress is applied and deformation is recorded over time (38). Using viscoelastic models to fit the length time series, the cell viscoelastic properties, including its apparent viscosity $\eta_{App}$, can be calculated (72).

### 2.4.3. Microfluidic techniques

To overcome the limitation of low throughput inherent to some aforementioned assays, microfluidic techniques can be attractive, e.g., for cell deformability cytometry (96-99), microconstriction arrays (100), or shear flow deformation cytometry (101) studies. These techniques have in common that suspended cells are passing through a microfluidic channel where they deform under fluid shear stress (99, 101) or have to squeeze under pressure through narrow channel constrictions (100). Resultant cell deformations or passage time are quantifiable parameters that can be assessed in timelapse videomicrocopy recordings. When the applied shear stresses are known, the viscoelastic properties of the cells can be inferred. High rates of deformation in the millisecond range (as in deformability cytometry) typically result in increased elastic moduli values compared to deformations in the second range (102). Advantageously, these techniques can also be combined with fluorescent detection, which allows the mapping of mechanical parameters to specific markers (98). Recent real-time deformation cytometry (RT-DC) based studies have revealed mechanical changes of transformed mammary cells (103, 104), mechanical effects of different drugs on leukemic cells (100), and native and malignant cells in pleural fluids (99). In addition, differences between normal and cancer cells were seen in cells derived from solid tumor tissues through digestion or mechanical dispersion (105).

### 2.4.4. Optical stretcher

Using the optical stretcher, cells are mechanically deformed in a contact-less manner using a dual-beam laser trap (106). In a configuration combined with a microfluidic pump system, cells enter the optical flow chamber one by one, where they are trapped and stretched along the axis of the opposing laser beams (107). The resultant cell elongation under stress is monitored over time and information about the cells' viscoelastic properties of the cell can be quantitatively derived (108). Using the optical stretcher, the mechanical properties of the breast (16, 106), cervix (16), and brain cancer cells (26) were measured and compared to normal cells. Automated setups also provide the opportunity to assess the mechanical properties of hundreds or thousands of cells and to distinguish subpopulations to reveal cell mechanical heterogeneity (26, 29).

### 2.4.5.  Brillouin microscopy

While most of the above-mentioned techniques require physical contact between the probe and the material to be probed, optical elastography probes have the advantage that they can interrogate cells in a rather non-invasive manner. Brillouin scattering, discovered nearly a century ago, has recently been adapted for biomechanical measurements of biological specimens by Brillouin microscopy (36, 109-112). Brillouin microscopy measures the inelastic Brillouin scattering that arises from the interaction of photons with acoustic vibrations in a material. In most biomechanical studies, the Brillouin frequency shift, which is related to the longitudinal modulus, is used as a quantitative parameter to describe the elastic properties of biological specimens. Despite its non-invasiveness, an advantage of Brillouin microscopy is its high spatial resolution (in the subcellular range, depending on the optical setup), allowing for in situ 3D mapping of the mechanical properties of biological specimens. This technique can be employed on single cells but advantageously also on transparent tissues or multicellular structures such as organoids (36, 111, 113, 114). Calculation of a longitudinal modulus from the Brillouin frequency shift requires measurements of the refractive index, e.g., by optical diffraction tomography (115). A limitation of the technique is that it is not possible to directly convert the longitudinal modulus at high frequency (GHz range) to shear moduli obtained by common probing techniques, e.g., AFM (110). Nevertheless, previous studies have revealed correlations between results obtained by AFM and Brillouin, e.g., when osmotically perturbing cells and cell aggregates, and when applying drugs that interfere with major cellular structural elements, such as F-actin (113, 116). Differences between normal and cancer cells were also reported (36).

### 2.4.6. Particle-tracking microrheology

Particle-tracking microrheology has been a well-established method for several decades (39, 95) to measure the static viscosity $\eta_0$ in confined environments, particularly in living cells (117). Particle-tracking microrheology employs high-speed optical microscopy to track the movement of particles in the 100 nm range embedded in the cytoplasm. To enhance particle detection and prevent their internalization into endosomes, which can lead to directed movement of the probes, fluorescent particles are used and introduced via ballistic injection (80, 118). The particle trajectories are analyzed in terms of the mean-squared displacement, and interpreted thanks to the Stokes-Einstein equation that relate the diffusion coefficient to $\eta_0$ (119). The technique is non-invasive, requires minimal material (< 1pL), and can be applied in both passive and active matter.

### 2.4.7. Optical tweezers

More recently, advanced methods such as optical tweezers (76, 120) have expanded the range of intracellular rheological techniques for actively probing cytoplasmic viscous properties. This technique uses a fixed optical trap and microscope to study intracellular mechanics (120). An infrared laser traps a micron-sized bead within the cytoplasm, allowing steady or oscillating manipulation. For viscosity measurement, the bead is displaced by moving the microscope stage, and the optical tweezers apply a spring-like force, causing the bead to relax back toward the trap center. This time-dependent relaxation is analyzed using the Standard Linear Liquid model (38) and allows simultaneous determination of the elastic modulus $G_0$ and apparent viscosity $\eta_{App}$ of the intracellular medium. Experiments on breast (76) and brain (82) cancer cells reveal mitigated viscosity and elasticity properties compared to healthy ones.

### 2.4.8. Magnetic rotational spectroscopy (MRS)

This technique leverages the hydrodynamic instability found for anisotropic magnetic objects in a rotating magnetic field (121). Berret *et al.* adapted MRS to intracellular environments using micron-sized magnetic wires that are spontaneously internalized in the cytosol and non-toxic to cells (122, 123). As the angular frequency $\omega$ increases, a transition from synchronous to asynchronous regime occurs at the critical frequency $\omega_C$, which varies inversely with the static viscosity $\eta_0$. In viscoelastic media, the high-frequency wire oscillation amplitude varies inversely with the elastic modulus ($G_0$). MRS hence enables the simultaneous measurement of local viscosity and elasticity, making it a powerful tool for probing local rheological properties of fluids or solids (124, 125), including those of living cells.

### 2.5.  Biological basis of cell mechanical alterations in cancer cells

The measured cellular response to external forces is dominated by the cytoskeleton, which is made up of dynamic and entangled polymer networks of F-actin, intermediate filaments and microtubules (126, 127), and internal structures such as the nucleus (128), and other organelles (129). Cancer progression is associated with remodeling of the actin cytoskeleton (89, 104, 130) and the microtubule network (51). Also, changes in vimentin levels have been linked to cancer progression and even proposed as predictors of patient outcomes in lung carcinomas (131). Recent studies have reported qualitative changes in cytoskeletal structures in normal and cancer cells that are associated with the measured changes in cell viscoelasticity (69, 132). Recently, induction of oncogene expression and epithelial-to-mesenchymal transition (EMT) has been demonstrated to directly affect cancer cell cytoskeleton and the mechanical properties of cancer cells (89, 103, 104).

To which extent the different cytoskeletal structural elements contribute to the cellular response to force during the measurement is also dependent on geometric features of the measurement technique (e.g., indenter size and shape), the probed cell's shape, the time scale of the measurement technique (milliseconds to seconds), the cellular region being probed (nucleus versus lamellipodium, cytoplasmic or nuclear internal structures (52) and cell cycle stage (89, 103, 133). For instance, AFM experiments are typically dominated by the actin cortex, which is located beneath the cell membrane (134). Experiments perturbing the actin cytoskeleton (e.g., by cytochalasin D, latrunculin B) can have drastic effects when indenting the cell surface using an AFM cantilever (52) or when deforming suspended cells (97), albeit latter depends on the timescale: while cells deforming under moderate flow rates (97) are sensitive to actin-depolymerizing drugs, assays employing high rates were shown to be insensitive to these changes (99, 102). Similarly, interfering with actomyosin contractility (e.g., by blebbistatin) typically softens cells, at least when they are adherent to a surface where they form stress fibrils (89, 135). However, targeting myosin activity in suspended cells can even have the opposite, stiffening effect (136).

Also, the location where the cell is probed can matter, particularly in adherent cells. Previous studies have, for instance, revealed different mechanical properties when probing nuclear and perinuclear regions (52). For thin regions probed by AFM, thin layer corrections should be taken into account (88, 137). Deeper indentations by an AFM tip can also reveal contributions of the intermediate filaments (138) or the nuclear lamina (along with F-actin depolymerization) (139). Manipulations of microtubules can have opposing effects, where nocodazole can soften cells (76), have no effect, or even stiffen cells (140). The response could be cell type dependent but also be related to crosstalk between microtubules and F-actin (126). In some cases, e.g., for intermediate filaments, selectively targeting cytoskeleton components using drugs is rather challenging; instead, genetic modifications, e.g., gene knockdowns, can be employed (138). Similarly, F-actin modulators can be targeted genetically to reveal their effects on cell mechanics, such as Rho GTPases Rho and Rac (89) or Ena/VASP (104). In the latter two examples, a molecular basis for the cell mechanical changes coming with oncogenic transformation was explored.

Adherent cells can display highly varying stiffness values; these are further influenced by the substrate on which the cell is sitting on, with concomitant cell shape changes (50, 70). To control cell shape during cell mechanical testing, probed cells can be trapped within microwells (50). In addition, 3D matrix stiffness (141, 142) and compressive stresses arising from growth under confinement affect the mechanical properties of cells (135). Since the mechanical properties of cells are less accessible within their 3D context, appropriate methods to assess them have to be chosen, e.g., passive or active microrheology or Brillouin microscopy (114, 141, 142). Moreover, the presence of cell-cell contacts when cells are forming multicellular cell clusters can affect cell mechanical properties (16, 135, 143). Resuspending adherently growing cells for mechanical probing has large effects on the cortical cytoskeleton (144). On the other hand, it can be argued that the suspended state represents a "ground state" that is not affected by unnatural cell adaptions to stiff substates commonly used for 2D cell culture. Over the past years, also interesting links between cell mechanical properties and metabolism have become evident (reviewed in (145). Since metabolic and mechanical changes are both hallmarks of cancer, more studies have to be conducted to reveal the underlying mechanistic links.

In sum, due to the substantial impact of above-discussed factors, i.e., cell state (cell cycle, metabolism, shape), context (2D/3D, matrix stiffness), cell preparation protocols (detachment of adherent cells, time in suspension) and measurement conditions (medium, temperature, deformation rates, force regime, cellular regions) on the mechanical phenotype of cells, the use of cancer cell mechanical markers in diagnostics requires reproducible and stringent protocols and thorough documentation of all experimental parameters.

### 2.6. Cancer cell responses to altered mechanical cues

As outlined in Section 2.1, the mechanical aberrations arising in tumors have multiple contributing factors, from increased ECM stiffness to cellular alterations that are likely to be sensed by neighboring cells. A vast number of studies have focused on stiffness sensing by cancer cells, which can also be studied experimentally through bioengineered 2D and 3D models. These studies have also been subject to recent reviews on the topic (34, 146-148). In the upcoming section, the focus will be on a less discussed subject, the influence of biofluid viscosity and shear stress on cancer cells.

The previous sections indicate that cancer cells exhibit softer mechanical properties compared to healthy cells. Notably, all the data in **Figure 1** and **Figure 2** were collected under static conditions, meaning no external flow was applied to the cellular environment, and conventional culture media were used. However, under physiological conditions, soft tissues—including tumors—are immersed in interstitial fluids, which can exert forces and stresses on surrounding cells. Interstitial fluids can locally have a viscosity 5 to 50 times that of cell culture medium due to the dissolution of extracellular macromolecules (24, 149), making them even more viscous than lymph or blood (3-8 mPa s) (150, 151). Interstitial fluids can also flow, imposing wall shear stresses on surrounding cells and tissues (24, 25). Although these effects have been known for years (24, 152), they remain relatively understudied. Only recently, significant progress has been made in understanding how increased fluid viscosity influences cancer cell behavior.

### 2.6.1. Interstitial fluid viscosity

To increase the viscosity of the extracellular medium in 2D cultures, researchers used biopolymers like alginates, polysaccharides, mucins, and polyethylene glycols, adjusting their molecular weight and concentration (149, 153-156). These experiments aimed to replicate the properties of interstitial fluid in the tumor environment, achieving viscosities as high as 2 Pa s (2000 times that of cell culture medium) (73). Gonzalez-Molina *et al.* used viscosity-enhancing polymers to simulate extracellular fluids, conducting wound healing and cell spreading assays on liver cancer cells (153). Surprisingly, despite the increased hydraulic resistance, cell motility was significantly enhanced on 2D substrates for both healthy and cancerous cells. Structural changes at the cell level, including cytoskeleton rearrangement, cytoplasmic expansion, and nuclear flattening were also observed (153). Maity *et al.* later confirmed these findings, showing a 2.5-fold increase in migration speed compared to conventional cultures (154, 157). More recently, Bera *et al.* conducted an in-depth study, suggesting that the increased cell motility in high-viscosity environments is driven by a mechanotransduction pathway involving cytoskeleton-ion channel interactions, calcium influx, and enhanced cell contractility (149). Importantly, Bera *et al.* demonstrated that this increased motility occurs *in vivo* using zebrafish, chick embryos and animal models, and affects metastasis. These findings suggest that higher local extracellular fluid viscosity could be a factor in cancer cell migration within the tumor microenvironment. Given the current knowledge of interstitial fluid viscosity, technologies for measuring their flow and stress *in vivo* for prognostic purposes are still unavailable to patients. However, developing such methods could offer valuable insights into key aspects of metastatic biology (34). This contrasts with blood, which also plays a crucial role in the metastasis spread and has been extensively studied for its hemorheological changes related to cancer (25, 158). Well-established research shows that rheological alterations occur in advanced cancers and are often linked to disease stage and prognosis. For various cancers, including breast, lung, ovarian, and cervical cancers (158), it has been consistently observed that whole blood viscosity is higher in cancer patients compared to healthy controls (150). This increase is attributed to elevated plasma viscosity and red blood cell aggregation, both of which are associated with metastasis development. A recent study on whole blood viscosity in patients with hepatocellular carcinoma found that both systolic and diastolic whole blood viscosities increased by 16% and 25%, respectively, compared to healthy individuals (150). These increases are significant enough to improve clinical diagnosis for a range of cancers.

### 2.6.2. Interstitial fluid shear stress

Interstitial fluid flow in tumors is driven by pressure differences between blood and lymphatic systems. The flows of interstitial fluids and blood are governed by distinct physical models based on the medium they traverse. Interstitial fluid follows Darcy law for heterogeneous media, characterized by low velocities (1 $\mu$m s$^{-1}$) and shear stresses in the range 0.01-0.1 Pa. Blood, in contrast, follows a pulsatile Poiseuille flow with velocities 10 to $10^5$ times higher than interstitial flow, with shear stresses range from 0.1 to 10 Pa (25). As a result, the behavior of cancer cells within a tumor, or circulating in the bloodstream varies significantly (24, 34). Recent studies have shown that cancer cells can respond to a wide range of interstitial stresses (159-163), with experiments conducted using *in vitro* microfluidic devices designed to replicate the tumor microenvironment. Most commonly, single microfluidic channels are used, where cells line the channel walls and a controlled flow is applied. Relevant work includes that of Calibasi Kocal *et al.*, who developed a microfluidic platform capable of generating low shear stress around $10^{-5}$ Pa on esophageal cancer cells, showing that laminar flow induces epithelial/mesenchymal hybrid transition and increases mechanotransduction protein expression, in contrast to static cultures (159). Other recent reports explored how shear stress in the range of 0.005 Pa to 0.5 Pa influences ovarian, breast, and prostate cancer cell behavior through the activation of mechanosensitive sodium and calcium ion channels, enhancing motility and tumor growth, or allowing tumor cells to evade high shear regions during intravasation (162, 163). The impact of fluid shear stress on cervical cancer cells also revealed that moderate shear stress boosts cell proliferation and resistance to chemotherapy, while high shear stress suppresses growth, indicating that fluid shear stress significantly influences cancer cell behavior and drug resistance during metastasis (160).

### 3. Electrical properties of cancer cells

The investigation of electrical properties of cancer cells is a rapidly evolving area with promising implications for cancer diagnosis, prognosis, and monitoring of cancer treatment (164-166). Compared to normal cells, cancer cells exhibit distinct morphological and functional features (167-170) due to the accumulation of charged metabolites and ions, ion channel activity, and alterations in membrane composition and cell size (171-174). These biomolecular variations can serve as versatile markers for cell proliferation, apoptosis, adhesion, cell cycle determination, and other cellular processes but also substantially affect the electrical properties of cells.

One of the hallmarks of tumor metabolism, known as the Warburg effect, has a high influence on the pH values of tumor tissues and cell cytosol due to the activation of the ion channel proteins and accumulation of the extracellular lactate. For instance, the pH changes and activity of ion channels lead to the depolarizing of the cell membrane potential (175, 176). Thus, deregulation of the ion channels and transporters results in uncontrolled tumor cell growth and spread (177).

As it is difficult to attribute the changes on a molecular level with the resulting electrical signal, the electric properties of normal and tumor cells are often correlated to key biological functions or cell phenotypes. In this way, the influence of the cytoskeleton on the resistance of the inner part of the cell (178) has been investigated. Also, the observation throughout the cell cycle revealed differences in phases, which have been attributed to the increase of membrane surface area and the rearrangements of large structures inside the cell (179). Interestingly, there are assumptions that morphological changes can be associated with the electrical properties of the cells (180). In particular, enhanced plasma membrane damage has been associated with increased motility and invasiveness of tumor cells (181), and the electrical measurement of the damaged membranes could be used as a marker of tumor aggressiveness. Furthermore, morphological features such as cell size, and cell functional properties, e.g., necrosis and apoptosis, can also be detected at low frequencies (182).

Understanding the 'electrical signatures' of cancer provides valuable insights into cellular physiology and transformation, tumor heterogeneity, and proliferative behavior (165, 183). Tracking electrical parameters, such as conductivity and permittivity, enables diagnosis, monitoring, and treatment surveillance by identifying different cell types and distinguishing between healthy cells and those exhibiting abnormal behavior. Therefore, it is crucial to explore how the changes in electrical properties are related to cancer types and cancer development.

### 3.1. Biological basis of the tumor-specific cell electrical properties

The electrical properties of cells reflect their physiology and pathology, influenced by both their intracellular and extracellular environments (183). When applied to cancer, tumor-specific cell electrical profiles emerge, differing significantly from those of healthy cells. Particular interest represents changes in the mechanisms regulating pH, ion levels, and membrane potential, leading to the alteration of various cellular processes and abnormal behavior (**Figure 3**).

#### 3.1.1. pH

Dysregulation of pH balance, intracellularly and in the tumor microenvironment, contributes to uncontrolled proliferation and immune evasion (183, 184). Variations in pH values become fundamental to the survival of cancer cells. Note, that the intracellular pH in healthy cells (6.99-7.05) is kept lower than the extracellular pH (7.35-7.45), while cancer cells display a "proton gradient reversal" with extracellular values (6.2-6.8) more acidic than the intracellular pH (7.2-7.8) (185). Unlike normal cells, tumor cells thrive in environments with high lactate and hydrogen ions ($H^+$) production, conditions that would typically induce cell death (186). However, tumor cells counteract this threat with the enhanced expression of several intracellular pH-regulating systems (187). In contrast to tumors, healthy cells can barely survive in this hostile environment (186). pH changes in cancer cells influence the activity of cellular channels, leading to different ionic concentrations and depolarizing the cell membrane potential (175). Additionally, studies have shown a relationship between intracellular pH changes and the dynamics of microtubules and actin filaments, further highlighting the impact of pH on cancer cell function (188, 189).

### 3.1.2. Role of dysregulation of ion channels

Dysregulation of the functions of ion channels and transporters is a condition that supports the uncontrolled growth and spread of cancer (177). The role of ion channel malfunction in cancer progression and metastasis is currently of great interest, considering that not all tumors share the same pattern of ion channel expression (183). Overall, changes in calcium, sodium, and potassium promote a cascade of events that result in uncontrolled proliferation, abnormal cell physiology, and ultimately contribute to the formation of aggressive, metastatic tumors (183).

Precise concentration and localization of calcium ($Ca^{2+}$) ions govern cell proliferation and apoptosis (190). Cancer cells capitalize and maintain calcium levels that avoid death. Calcium levels in tumors also favor cell proliferation and angiogenesis, which enhances nutrient distribution and growth (190-192). Cancer and wounded cells are known for excessive intake of water and sodium ($Na^+$) (193, 194). This feature affects the cell towards a bigger size and more spherical shape. This altered shape influences cell signaling contributing to cancer progression and metastasis (194). Cancer cells often exhibit enhanced expression of potassium ($K^+$)-calcium ($Ca^{2+}$) channels (183). Altering the cell's response to hypoxia and modifying cell adhesion, migration, and apoptosis, tumor progression is favored (195-197). The aforementioned ionic changes also influence the membrane potential. Cancer cells have a more depolarized resting membrane potential ($V_m$) compared to healthy cells, contributing to an increased proliferation and migration capacity of tumor cells (198).

### 3.1.3. Membrane potential

The membrane potential, determined by the ion concentrations, is highly affected by ion channels' permeability, expression, and activity. Cancerous cells show a tendency to have a more depolarized $V_m$ compared to healthy cells. This depolarization is associated with poorer patient prognosis. Depolarized cells increase their proliferative behavior in cancer, influencing the cell cycle progression (198, 199). Hyperpolarization is required to initiate apoptosis, and the depolarized state disrupts this process increasing the survival rate of cancer cells (198, 200). Furthermore, the depolarized phenotype of cancer cells is associated with stem-cell-like behavior characterized by self-renewal, inhibition of cell differentiation, and migration. All these factors contribute to increased tumor aggression and metastasis (198).

## 3.2. Impedance of cancer cells

The transformed state of cancer cells is also reflected in a change in membrane capacitance and conductivity (**Figure 3 A**). Cancer cells frequently display an abundance of sialic acid-rich glycoproteins, resulting in a more negative surface charge (194, 201, 202). This acts as an electrical shield that protects cancer cells from negatively charged immune effector cells (194). The changes in composition, shape, and permeability anticipate a differential charge storage capacity with respect to healthy cells (203). This relative difference is a more reliable marker for malignancy than the absolute capacitance of the cells (203). In support of this conclusion, studies have linked variations in membrane capacitance to patient survival rates in head and neck cancer (204). Beyond capacitance, it is observed that conductivity also has increased values in cancer development and malignancy; this suggests that impedimetric measurements can be used to assess the prognosis of cancer based on impedimetric biomarkers (165, 205).

Impedance (Z) is measured by applying an alternating current (AC) potential enabling to distinguish cancer and non-cancerous cells in a real-time and label-free manner. To understand the concept, a rather simplified explanation will be used; for a more comprehensive understanding, please consult Lazanas et al. (2023) (206).

Impedance is a complex property of materials and can be separated into a real (resistance – $R$) and imaginary (reactance – $X$) part; with its complex character and frequency dependence, it provides the possibility to derive multiple other cell-specific parameters that can be used to gain important information about cell processes and properties.

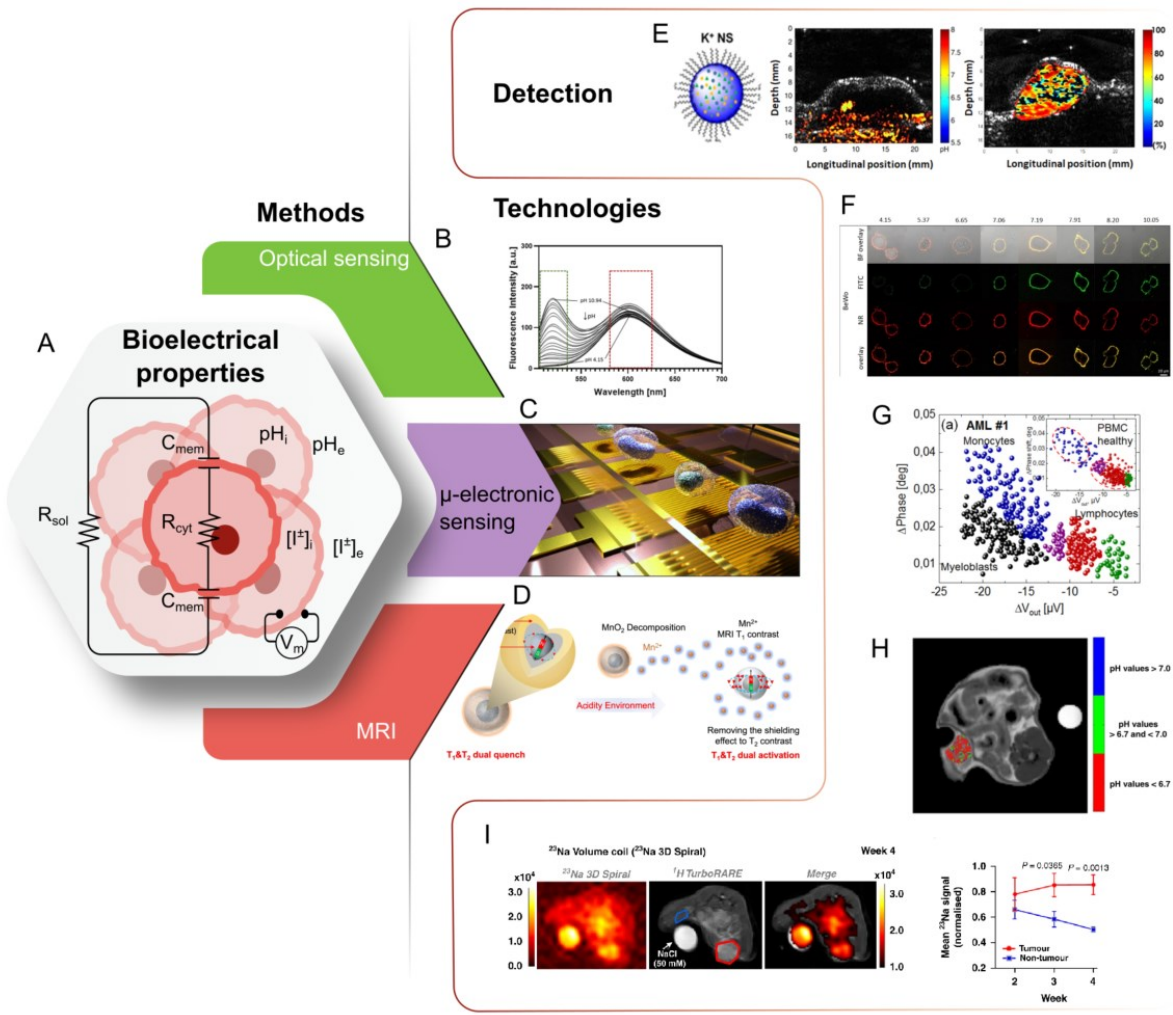

**Figure 3.** Bioelectrical properties, as prognostic biomarkers for cancer. **A.** Schematics of cancer cells and tissue impedance, reflected as a change in membrane capacitance $C_m$ and conductivity; pH, ionic concentration [$I^+$], and membrane potential ($V_m$). These biomarkers have been studied using MRI, optical, and microelectronic probing. The increasing interest in the prognosis potential of biophysical markers has driven the development of technologies for the detection of properties at cellular, tissue, in vitro, and in vivo levels. Examples include **B.** ratiometric fluorescent dyes (207), **C.** micro/nanoelectronic detection systems (164), and **D.** dual positive and negative contrast agents for MRI imaging (208). Selected examples of these advancements have enabled the detection of **E.** increased ionic potassium [K+] levels within tumorous tissue (209, 210), **F.** nanosensors comprising biocompatible polystyrene nanoparticles

loaded with a pH-inert reference dye (Nile red), and surface-functionalized with a pH-responsive fluorescein dye, enabling the visualization of BeWo cells (human choriocarcinoma) (207). Direct probing approaches have also facilitated the development of **G.** portable microfluidic cytometers designed to detect and quantify myeloblasts in peripheral blood samples (164). MRI-based methods have further advanced diagnostic capabilities, including **H.** imaging of pH in tumorous models (211) and **I.** mapping of sodium ion [$^{23}$Na] distribution to investigate metastatic potential, enhance diagnostic accuracy, and monitor treatment responses (212).

In common impedance measurements, the magnitude ($|Z|$) and the phase angle between voltage and current ($\theta$) are determined, which can be used to derive the capacitance ($C$), the ability of a system to store electrical charge, conductivity ($\sigma$), the ability of a cell to conduct electric current, and permittivity ($\varepsilon$) the ability to polarize in response to an electric field (**Supplementary Table 1**). All these properties have a high potential to be used to improve the understanding of cellular processes. The simplified models are used for the measurement evaluation, and the equivalent circuit model (**Figure 4 A**) is one of the most common. The cell compartments are simplified as elements in an electrical circuit, with properties that have minimal impact on the overall signal being neglected. This model consists of two main components: the cell membrane, which primarily contributes capacitance due to its bilayer structure and is represented as a capacitor ($C_{mem}$), and the cytosolic side, which exhibits resistive behavior and is modeled as a resistor ($R_{cyt}$). Both elements are arranged in parallel and combined with components that represent the surrounding environment, such as the ECM and medium (213).

The frequency plays a key role in understanding various properties, as it primarily influences the interaction of current (214). Parts of the cell affected by current can be divided into three frequency ranges: low frequencies, where the cell acts as an insulator with no current transmission but allows cell size determination; mid frequencies, where partial interaction with the cytosolic side occurs, reflecting membrane capacitance; and high frequencies, which penetrate the cell and interact most with the cytoplasm (**Figure 4 B**).

### 3.3.  Methodology

#### 3.3.1.  Optical sensing techniques

Fluorescence microscopy and spectroscopy are well-developed and state-of-the-art techniques for imaging the inner cell processes at high resolution. These techniques, relying on light absorption and emission at a larger wavelength, are carefully adapted to interact with specific cell components (215). By analyzing light intensity, wavelength, and distribution, it is possible to extract information about concentration, binding state, and environmental conditions experienced by the probe in the tumor or cancer cell **(Figure 3 B)** (216, 217).

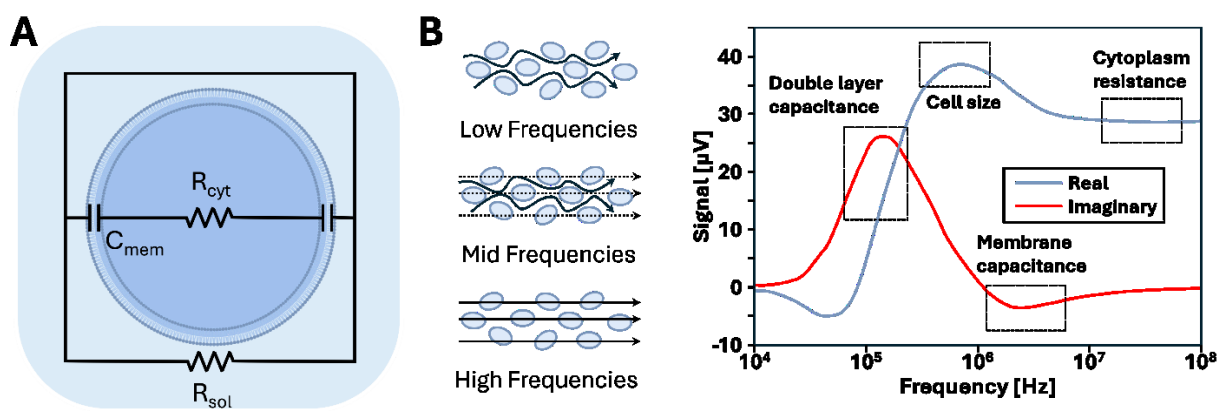

**Figure 4. A**. Equivalent Circuit Model for Estimating Cellular Properties, used to simulate and analyze the electrical properties of a biological cell. The circuit comprises key elements such as resistors and capacitors, representing different components of the cell (e.g., intracellular ($R_{cyt}$) and extracellular resistance ($R_{sol}$), and membrane capacitance ($C_{mem}$)). This model serves as a basis for estimating parameters like membrane integrity, intracellular conductivity, and other cellular properties based on electrical behavior. **B**. Frequency Dependence of Current Interaction and Impedance Correlation with Cellular Properties, shows how the electrical current interacts across different frequencies, highlighting the frequency-dependent behavior of the system. At lower frequencies, the current interaction predominantly reflects properties such as cell size, while higher frequencies penetrate the cell, providing insight into intracellular conductivity. The graph visualizes the relationship between frequency and the impedance response, where the real part of impedance corresponds to resistive components, and the imaginary part of impedance reflects capacitive behavior. This frequency-dependent analysis enables the separation and estimation of specific cellular properties, with each frequency range correlating to distinct physical characteristics of the cell (adapted from (214, 218))

Fluorescent pH indicators, such as SNARF-1 and the genetically encoded pH sensor SypHer2, have been used to study the reversed pH gradient in cancer and assess the efficacy of pH-regulating therapies (186, 219). By calculating the fluorescence ratio at two different wavelengths of the ratiometric dye SNARF-1, the extracellular pH of a tumor was measured in vivo (186). In addition, genetically encoded pH nanosensors can be designed to trace single extra- to intracellular gradients and monitor intracellular pH response to different stimuli, such as treatment with pharmaceuticals. One recent example represents the nanosensor consisting of biocompatible polystyrene nanoparticles loaded with the pH-inert reference dye Nile red, and is surface-functionalized with a pH-responsive fluorescein dye. The nanosensor is equipped with a targeting moiety and can adhere to cell membranes, allowing direct measurement of extracellular pH at the cell surface (**Figure 3 F**) (207).

Ion dynamics can also be studied through fluorescence methods. Ion concentration transients and waves have been detected through calcium-related markers probing the long-distance communication between cancer cells in human colon and prostate cancer cells (217). Further efforts to quantitatively evaluate the tumor microenvironment concentration of potassium ions in vivo have been realized through the combination of fluorescent nanosensors and photoacoustic emission (**Figure 3 E**) (209, 210). This analysis revealed significantly higher levels of potassium ions (29 mM) in the microenvironment compared to the surrounding muscle tissue (19 mM).

By tracking the expression of $K^+$ channels tagged with green fluorescent protein (GFP) probes, researchers have studied the effects of manipulating the membrane potential in cancer cells. It was found that hyperpolarization of the membrane potential through $K^+$ channels overexpression increases breast cancer cell migration, invasion, tumor growth, and metastasis (196). The increased polarization creates conditions to upregulate cadherin-11, leading to metastatic behavior (196).

### 3.3.2. Magnetic resonance imaging

The electrical properties of cancer cells have been studied through the magnetic resonance imaging (MRI) technique that utilizes the interaction in high-intensity magnetic fields of magnetic moments of atoms and radiofrequency pulses (220). It can utilize contrast agents with optimized nuclei relaxation times that enhance the contrast between specific ions, molecules, and targets (**Figure 3 D**) (208, 221).

While traditional imaging often overlooks ions, researchers have recognized that altered ion levels within tumors hold valuable diagnostic and prognostic information. Previous investigations have hinted at a prevalence of higher concentrations of sodium ions $Na^+$ in cancer cells compared to the surrounding healthy tissue (193). After analyzing orthotopic breast cancer regions with MRI combined with diffusion-weighted imaging (DWI), it was possible to confirm this observation and identify that the excess of $Na^+$ is concentrated intracellularly in cancer cells (**Figure 3 I**) (212). By employing both techniques, enhanced levels of sodium ions were linked to increased cellularity in breast cancer models that serve as a promising biophysical marker in early tumor development (212).

Advanced MRI-based chemical exchange saturation transfer (CEST) has been shown as a promising tool for the diagnosis of metastases potential in primary tumors based on spatial acidosis studies (**Figure 3 H**) (211). A pH-responsive dual-mode MRI contrast agent was also demonstrated to be effective in detecting metastasis in liver tumors with sizes of 0.5 mm when using the $T_1$-$T_2$ dual-modal MRI contrast agent for cancer imaging (208).

### 3.3.3. Microelectronic probe

While previous methods can infer cellular electrical properties, using microelectronic techniques provides a direct characterization of these parameters (**Figure 3 C**). Direct electrical measurements offer the advantage of real-time quantifiable dynamics of localized events (222). Based on the fact that each biological system has distinct dielectric properties, it is possible to use them as relevant quantitative biophysical markers in cancer prognosis (223).

Microelectrodes have been engineered in recessed-tip, and double-barreled configurations to facilitate negligible perturbations during sensing with small tip sizes and simultaneous measurements of the membrane potential and pH, respectively (224, 225). Due to the compatibility with microfabrication techniques, it has been shown that it is possible to fabricate arrays of sensors for high throughput tracking of the heterogeneity of pH changes in cancer cells (226). The activity of voltage-gated ion channels has also been related to metastatic potential in prostate cancer cells through the patch clamp technique (227). The direct influence of bias voltage and sensing through integrated sources makes possible the control and regulation of specific ion channels (228).

### 3.3.3.1. Electrical impedance spectroscopy (EIS)

The use of electrical impedance spectroscopy (EIS) for analyzing biological samples has a high potential due to its sensitivity and specificity. A variety of electrode designs have been developed to optimize the precision of measurements (229). Frequency sweeping enables the detection of changes in response across a broad range of AC frequencies, providing a comprehensive view of how the system behaves at different frequencies. Depending on the necessary dynamics that must be detected, the range of frequencies can be chosen, and the number of frequencies has to be adjusted accordingly in a relatively quick manner. A clear limitation is the time resolution and capability of complex measurements of the used device.

EIS has already shown its versatile character in cancer research, as already mentioned (229). Multiple studies have shown that EIS can differentiate between cancerous and benign cells (167, 230, 231). When treated with anti-cytoskeletal drugs, breast cancer cell lines with increasing malignancy displayed distinct resistive behavior, highlighting the impact of microtubules and actin on the cell impedance (232). Cell behaviors, such as migration and invasion, can be monitored, with migration detected by an increase in impedance as cells move onto the electrodes (233, 234), and signal changes across multiple frequencies can be used to quantify invasion. As cancer cells degrade the EMC, their higher impedance compared to the ECM was interpreted as evidence of invasion (235).

### 3.3.3.2. Electrical Impedance Cytometry (EIC)

The ability to polarize and measure the conductivity of cells and their components can be assessed through electrical impedance cytometry. This technique, based on the excitation of an electric field (in the kHz to MHz range) in the area between micro- and nanoelectrodes, can differentiate the capacitive and conductive properties of healthy and cancer cells (223, 236). This approach has proved the successful distinction between healthy tissue and breast carcinoma tissue as a prospective early diagnostic tool (165). This technique can be extended to in vivo tissue, using an array of 90 electrodes to map the dielectric properties directly in patients; researchers performed a direct tomography extracting the cues of abnormal tissue (237).

EIC offers a complementary approach to EIS for analyzing the electrical properties of cancer cells. Unlike impedance spectroscopy, which typically measures the impedance of monolayers or tissues over a range of frequencies, impedance cytometry focuses on individual cells and single frequencies in real-time. This technique utilizes microelectrode arrays to measure changes in electrical impedance as cells pass through a microfluidic channel (238). The advances in device integration have allowed the development of electrical analogous to fluorescence cytometry assays. Impedance cytometry uses capacitive-like electrodes integrated into microfluidic channels that analyze the heterogeneity of single cells through their impedance signatures (238). This label-free method has been demonstrated in a pilot study to identify cancer cells in the peripheral blood of patients with acute myeloid leukemia (**Figure 3 G**) (164). Another pilot study demonstrated that the differentiation of cancer cells in human peripheral blood in patients with acute myeloid leukemia was possible (164). It was also demonstrated that bladder cancer cells encapsulated in microdroplets generated by a microfluidic device were successfully discriminated (239).

Another property that is often determined in EIC is opacity, which represents the transmission of electromagnetic fields (electrical penetration) and can be used to study the cell membrane and size, by plotting the absolute values of the impedance at low and high frequencies against each other (66, 240). These techniques offer a high potential for creating high-throughput miniaturized devices that could be used in point-of-care settings. EIC is particularly advantageous for drug and marker screening. Machine learning is often used to automate the EIC data analysis process and avoid human bias (238, 241).

### 3.3.3.3. Manipulation of cells with dielectrophoretic (DEP)

Dielectrophoresis (DEP) is a technique to manipulate cells based on their dielectric properties when exposed to a non-uniform electric field. Single cells can be affected by negative DEP which deflects them from areas with a high electric field, while positive DEP attracts them towards high-field regions (242).

This method enables the separation and analysis of cancerous cells from healthy ones due to their distinct electrical characteristics (243). DEP can be utilized for the isolation of circulating tumor cells (CTCs) from blood (244), offering a label-free technique for liquid biopsy analysis.

## 4. Thermal properties of cancer cells

### 4.1. Theoretical background and determination of the basic parameters

The thermal properties of healthy tissue cells and tumor cells can differ. These differences arise due to the distinct metabolic activities, structural composition, and microenvironment of the two cell types (245). Tumor cells often have higher metabolic rates than healthy cells (4, 246). This increased metabolic activity can lead to excessive heat production in tumor tissues (247). This phenomenon is partially due to the Warburg effect, where cancer cells preferentially use glycolysis over oxidative phosphorylation, even in the presence of oxygen (247). In addition to the increased metabolic activity, the rapid proliferation contributes to the elevated heat generation, the so-called thermogenesis (248). Tumor tissues may have different compositions compared to healthy tissues, including variations in cell density, ECM, and vascularization (249). These differences can affect the thermal conductivity κ of the tissue (250):

$$\kappa = \frac{Q * L}{A * \Delta T} \qquad \text{(Eq. 1)},$$

where κ - thermal conductivity of the tissue $[\frac{W}{m\,K}]$; Q - rate of energy transfer (or heat flow) through the tissue [W]; A - area of the surface through which heat is transferred [m$^2$]; ΔT difference in temperature across the tissue [K]; L - thickness of the tissue sample or the distance over which the temperature difference is measured [m].

To access the structural differences in cells, including variations in cell density and the organization of cells and extracellular matrix, one can define the thermal diffusivity α (250):

$$\alpha = \frac{\kappa}{\rho * c_p} \qquad \text{(Eq. 2)},$$

where α - thermal diffusivity of the bulk cell culture $[\frac{m^2}{s}]$; κ - thermal conductivity of the bulk cell culture $[\frac{W}{m\,K}]$; ρ - density of the tissue or cell culture $[\frac{kg}{m^3}]$; $c_p$ - specific heat capacity of the tissue $[\frac{J}{kg*K}]$.

Tumor tissues might have a lower thermal diffusivity compared to healthy tissues because of their denser and unorganized structure. The specific heat capacity of a tissue depends on its biochemical composition, including the amounts of water, lipids, and proteins (251). Changes in the biochemical composition of tumor cells, such as altered levels of these components, can affect their specific heat capacity. Consequently, thermal diffusivity measurements can provide insights into the structural and compositional differences between healthy and cancerous tissues. Tumor cells may have altered biochemical compositions, affecting their specific heat capacity (251).

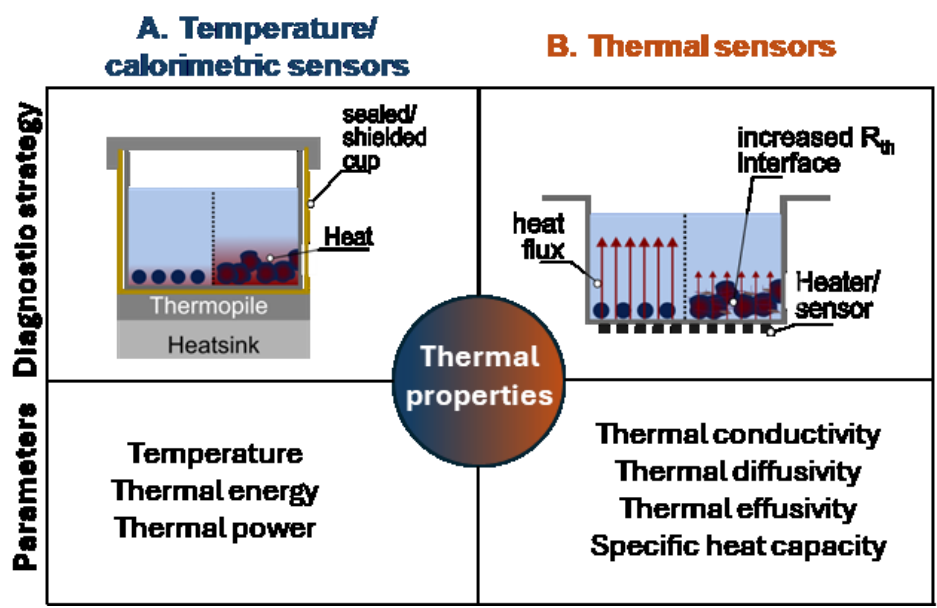

**Figure 5.** Schematic illustration of different sensing strategies utilizing the distinct physiological and metabolic differences between healthy cells and cancer cells. **A**. Temperature or calorimetric sensors detect the thermal heat of biophysical processes, biochemical processes or biological events. **B**. Thermal

sensors monitor the change in physical properties that arises due to biophysical processes, biochemical processes or biological events.

There are two main groups of thermal sensors to detect these physiological differences, temperature/ calorimetric sensors, which measure heat generation or heat transfer, see **Figure 5 A**), thermal sensors, which measure physical properties such as thermal diffusivity, thermal effusivity or thermal conductivity, see **Figure 5 B**). Indirect thermal sensors that derive signals such as force pressure, flow rate, or acceleration do not yet play a major role in cancer research.

While the potential of these thermal properties as prognostic biomarkers is extensive, research in this field has not garnered significant attention. However, some promising techniques have been explored, particularly those utilizing calorimetry and heat transfer methods. It is important to note that the application of thermal sensors has so far been mainly limited to in-vitro studies under controlled laboratory conditions and has not yet been widely implemented in in-vivo or clinical settings. The following sections discuss these methodologies and introduce a recent strategy with potential applications in diagnostic settings.

## 4.2. Methodology

### 4.2.1. Calorimetry

Calorimetry is a label-free technique used to measure the heat generated or absorbed by biological samples, providing insights into their metabolic activity (252). The heat generated can reflect physiological processes in normal cell cultures but can also be employed to detect cellular responses to various stimuli, such as drugs, environmental changes, or stressors, offering a versatile tool for studying cellular behavior in different conditions. Growth of hepatocarcinoma tumor microtissues (based on HepG2 cell line) releasing a maximum thermal power was monitored, which corresponds to a diameter increase of the microtissues from *ca.* 100 to 428 μm (252). Isothermal microcalorimetry, in particular, was also used to investigate the influence of enriched media compared to normal medium on neuroblastoma cells (253). The spiking of fructose, glucose or fructose/glucose to the medium led to increases in the metabolic activity of the cells (253). These studies also show that tested glioblastoma cells (Kelly and SH-EP Tet-21/N) preferred fructose metabolism over glucose metabolism, a beneficial adaption of these cancer cells under low oxygen and low nutrient conditions (253). While differential scanning calorimetry (DSC) is known as a sensitive technique in the field of phase transition studies, during the last decades, it has also been recognized as a tool for cancer diagnosis and monitoring (254-256). DSC has been tested as a tool for lung cancer patient diagnosis and prediction in clinical settings (257). Different lung cancer subtypes and stages were successfully distinguished from control patients (257). With further optimization DSC could provide an accurate, non-invasive, and radiation-free strategy for state-of-the-art low-dose CT scans (257).

### 4.2.2. Heat transfer method (HTM)

HTM is a thermal transducer platform that monitors the thermal interface resistance $R_{th}$ between the solid chip and the supernatant liquid (258). When combined with surface-imprinted polymers (SIPs), HTM has been used to distinguish between human breast cancer cell lines MCF-7 and leukemia Jurkat cells (259), and to detect differences in their glycosylation patterns using modified Chinese hamster ovarian cells (260). The model allowed excessive expression of the transmembrane protein mucin-1 (MUC1) and control of its glycosylation by varying culturing protocols (260). The selective differentiation between cells expressing the MUC1 protein and cells that do not express MUC1 and MUC1 glycosylated and MUC1 non-glycosylation indicate significant differences in the imprints (260). HTM has also been effective in monitoring the quality of cell cultures over time, as demonstrated by comparisons between adherent breast cancer cell culture (ZR-75-1a) and suspension culture (ZR-75-1s) (261).

### 4.2.3. Modified transient plane source (mTPS) sensor

Another noteworthy technique is the mTPS sensor systems, which can combine aspects of calorimetry with the monitoring of changes in thermal properties at the chip interface. Although not yet applied to cancer detection, mTPS has been used for real-time quantification of yeast cell numbers and the metabolic activity of cultures (262).

The label-free and indirect sensing strategy is suited for bi-modal sensing, revealing both thermal and electrical information about the sample (263, 264). The integration of mTPS into microplates for high-throughput screening or microfluidic chips has also been explored (265), and therefore holds great potential for future life science applications.

## 5. Tumor heterogeneity

### 5.1. Intertumoral and intratumoral heterogeneity and cancer stem cells (CSCs)

Tumors are complex living systems where tumor cells and their microenvironment are regulated in a bilateral feedback loop manner. As each patient is unique, so are their tumors, which possess intertumoral (between tumors) heterogeneity (266, 267). Molecular fingerprinting of individual tumors by protein analysis, mRNA expression, or DNA sequencing serves as a basis for the development of cancer biomarkers - measurable indicators of tumor presence and progression (268). In addition to the difference between the patients, each individual tumor possesses a certain level of intratumoral (within a particular tumor) heterogeneity attributed to the tumor cell phenotypes as well as genetic and epigenetic traits (269). Intratumoral genetic diversity is a result of genomic instability, which is an increased frequency of mutations during cell division. Genomic instability is attributed to many types of tumors (270). In tumor bulk, most cancer cells have limited proliferative potential. Thus, not all acquired mutations are passed across multiple cell generations and fuel tumor evolution. In contrast, only a subpopulation of tumor cells called CSCs possesses unlimited self-renewal potential and the ability to differentiate into other tumor and non-tumor cell subsets (271). Due to their unlimited self-renewal potential and ability to recapitulate all tumor cell populations, these cells maintain tumor growth and, therefore, are considered a unit of tumor evolution. A general consensus is that tumor cells with CSC properties not only maintain the primary tumor growth, but evolve, disseminate, and give rise to tumor metastasis (272). The populations of CSCs can be identified and isolated using different plasma membrane and intracellular protein molecules serving as CSC markers, such as CD44, CD133, integrin $\alpha2\beta1$, and others (273-275). CSC-related biological markers are the focus of many clinical trials as promising prognostic indicators and therapeutic targets (276-279). Furthermore, CSCs can be differentiated from non-CSCs by physical properties such as their size, viscoelasticity, and electrical properties (165, 280-283).

CSCs produce cell clones that expand during tumor development and possess different genetic, epigenetic, and functional features (271, 284). Tumor heterogeneity depends on the number of CSCs that contribute to tumor growth. At the same time, genomic instability, the number of acquired genomic alterations, and the heterogeneity of CSCs are increasing during tumor progression. The evolutionary dynamics and spatial tumor heterogeneity can be analyzed by different approaches, such as longitudinal sampling, spatial tumor biopsies, and multi-omics profiling, including whole genome sequencing (WGS), whole exome sequencing (WES), RNA sequencing (RNAseq), assay for transposase-accessible chromatin sequencing (ATAC-seq), single-cell RNA sequencing (scRNAseq) and single-cell mass spectrometry, spatial proteomics and transcriptomics, etc (285, 286). In addition to their distinct molecular and functional characteristics, different clones composing a given tumor possess specific functional features and physical parameters such as cell size, viscoelasticity, and cytoplasmic viscosity, and have different impacts on the microenvironmental components, including hypoxia, acidity, extracellular fluid viscosity and matrix stiffness (11, 287).

## 5.2. An interplay between biological and mechanical heterogeneity and tumor stemness

### 5.2.1. The mechanotransduction signaling

Tumor growth is accompanied by biomechanical changes in the tumor microenvironment, including an increase in matrix stiffness generated by ECM components, and the accumulation of solid stresses from compression or tension during tumor growth (288). Analysis of the genomic mutations in cancer of different types suggested that tumor tissue stiffness correlates with the scale of genomic variations and intratumoral heterogeneity (289). This increased genomic variation can be associated with increased DNA damage in response to the cell and nuclear squeezing upon cell proliferation in a stiff microenvironment or migration through tiny pores of the collagen-enriched ECM or basement membranes (289, 290). The stiffness of tumor tissues depends on the deposition of the ECM components such as collagen, laminin, elastin, fibronectin, and glycosaminoglycans (GAG) by cancer-associated fibroblasts (CAFs) and tumor cells and ECM remodeling by enzymes such as matrix metalloproteinases (MMP), procollagen-lysine, 2-oxoglutarate 5-dioxygenase (PLOD) and lysyl oxidase (LOX) secreted by tumor cells and tumor-associated stroma cells (22, 291). Mechanical stress induces actin stress fiber reorganization in the tumor cells, consequently affecting cell elasticity (292, 293). Mechanical stress also induces large-scale chromatin remodeling and changes in gene expression profiles (294). Mechanical stimuli can activate cell membrane mechanosensors such as transient receptor potential (TRP) channels and Piezo ion channels (295), integrin receptors (296), G-protein coupled receptors (GPCR) (297) and cadherins (296, 298), and corresponding cytoplasmic mechanotransducers including RhoA-mediated actin dynamics (299), focal adhesion kinase (FAK) complexes, transcription factors (e.g., yes-associated protein (YAP)/WW domain-containing transcription regulator 1 (TAZ), mechanistic target of rapamycin complex 1 (mTORC1) (300), phosphoinositide 3-kinase (PI3K)/Akt (301), and β-catenin (298), which induce other downstream signaling mechanisms in tumor and stroma cells (302, 303). The mechanotransduction signaling is activated in most types of cancer. It is involved in the regulation of different aspects of cancer progression, including tumor growth, therapy response, immune evasion, metastasis, and stemness (304).

### 5.2.2. Mechanical properties of CSCs

Indeed, for many cancer types, CSCs are shown to be the dynamic and plastic cell populations. The epigenetic changes, newly acquired mutations, and microenvironmental factors such as hypoxia, nutrient availability, physical parameters including stiffness, shear stress, and microenvironment architecture, as well as different types of treatments, can induce a bi-directional transition between CSC and non-CSC states. CSC surface markers such as CD44, CD133, integrin α2β1, and integrin β4 directly interact with collagen and GAG hyaluronan (HA) and induce intracellular signaling critical for CSC maintenance such as FAK, Akt/mTOR, MAPKs and β-catenin (305-308). Indeed, increasing stiffness leads to the enrichment of CSC populations in breast (281), liver (309), and lung tumors (282), as reviewed recently (310). High stiffness triggers TAZ-mediated mechanotransduction (281) and activates the YAP/β-catenin-dependent transcriptional program and expression of the reprogramming factors Nanog, Oct4, and Sox2 (281, 282). Integrin-mediated activation of FAK and downstream pathways, including AKT/mTOR signaling, contribute to CSC regulation through matrix stiffness (311). Mechanotransduction mediated by Piezo ion channels is critical for the dissemination of cancer cells (162, 312-315), and the high expression of the Piezo mechanoreceptors was correlated with worse clinical outcomes and metastases in many cancer types (313, 315-319). Piezo proteins, the mechanosensitive ion channel, are sensing cell deformation and transforming mechanical stimuli such as pressure or shear stress into biochemical processes. In particular, activation of Piezo channels leads to the influx of extracellular ions, mainly $Ca^{2+}$, triggers the RhoA pathway regulating actin cytoskeleton (299, 320), and activates different intracellular signaling mechanisms, including integrin/FAK, Akt/mTOR, and MAPKs (321-323). Piezo proteins regulate the maintenance of CSC populations. Genetic deletion of Piezo1 inhibits CSCs in glioblastoma and colon cancer preclinical models (317, 324). The expression levels of Piezo1 or Piezo2 correlate with CSC-related protein markers and transcriptional signatures in patients with colorectal and gastric cancer, respectively (318, 324). TRP family of mechanoreceptors also plays a critical role in the regulation of CSCs, as recently reviewed (325).

Given the importance of mechanotransduction for CSC maintenance, it is unsurprising that CSCs can be identified and isolated based on their intrinsic mechanical properties: several studies demonstrated that CSCs is a mechanically soft population compared to non-CSC counterparts. The soft tumor populations were isolated from breast cancer cell lines by using microfluidic devices and characterized using *in vivo* models. These experiments demonstrated that soft tumor cells are more metastatic and tumorigenic *in vivo* and possess CSC properties (326). The previous studies reached a consensus conclusion that metastatic tumor cells are much softer than non-metastatic cells (11, 49, 67, 106), and as discussed in Section 2. A softness of tumor cells was associated with an oxygen deficiency called hypoxia, another feature of aggressive, metastatic, and therapy-resistant tumors (11, 327). In each individual tumor, different spatially defined areas may be perfused differently by functional blood vessels (328). Therefore, hypoxia has substantial inter- and intratumoral heterogeneity (329). The hypoxic microenvironment promotes CSCs and metastatic properties, and one of the key mechanisms of this regulation is the activation of the transcriptional program driven by hypoxia-inducible factors (HIF). Both hypoxia and ECM stiffness induce CSCs, tumor growth and metastases through the mechanotransduction pathways such as β1-integrin/integrin-linked kinase (ILK)/PI3K/Akt mechanism (330). In a feedback loop of this interplay, Piezo1-mediated mechanotransduction is shown to induce HIF1α expression (313).

### 5.2.3. Dynamic changes of the CSC mechanical properties in the metastatic cascade

A subset of CSCs determines metastatic growth through sequential steps, including tumor cell intravasation into the bloodstream, dissemination of the CTCs to distant organs, extravasation at the distant site, and initiating metastatic tumor development (246, 331). At the initial stage of the metastatic progression, immobile cancer epithelial cells acquire mesenchymal phenotype and the ability to migrate in a process called epithelial-mesenchymal transition (EMT) (332, 333). The EMT process is associated with generating a range of intermediate cell phenotypes, including both epithelial and mesenchymal features. EMT is positively associated with other tumor hallmarks, such as genomic instability and hypoxia. This developmental mechanism is associated with an epigenetic reprogramming of the bulk tumor cells into CSCs (333, 334) and contributes to the heterogeneity of CTCs (335). Therefore, EMT is considered to be a key driver of intratumoral heterogeneity (336). The ECM stiffness activates the EMT program via Piezo1/2-mediated $Ca^{2+}$ influx and mechanosensitive ephrin type A receptor 2 (Epha2)/Lyn protein complex followed by the activation of the TGF-β and Akt signaling pathways and transcriptional program driven by Twist1 and Snail (214, 316, 337-340). Consequently, the activation of the EMT program softens the cytoplasm (341, 342), induces actin reorganization, and increases actomyosin traction forces (342), playing a key role in cell migration (343), and, therefore, facilitates tumor cell invasiveness. At the intravasation stage of metastatic progression, the highly migratory and invasive tumor cells leave the primary site and invade through the basement membranes to the blood of lymphatic vessels. The stiffness of the ECM is a limitation encountered by invading cells, and only a subpopulation of tumor cells resistant to deformation and able to move through stiff microenvironment and dense basement membrane might survive and propagate metastases. As discussed in Section 2, highly metastatic cells have lower elasticity and viscosity compared to less metastatic tumor cells. The EMT program induced by stiff microenvironments makes prostate tumor cells more migratory than in soft substrate, enabling them to escape. The physical cell compression upon squeezing through narrow gaps in the wall of blood vessels is associated with nuclear envelop rupture (NER), temporary disruption of a lipid bilayer surrounding cell nuclei (290). NER is associated with DNA damage and is potentially leading to the genomic heterogeneity of the metastatic cells (344). The recent studies described the adaptive longitudinal dynamics of single tumor cell viscoelasticity under different pressures in the microvessels and softening of tumor cell nuclei during transendothelial migration (345, 346). The cells with softer nuclei have advantages in cell migration but a higher possibility of DNA damage and genome instability (347). In addition to its effect on DNA integrity, the stiff microenvironment also induces mechanical adaptations of tumor cells through epigenetic reprogramming. A study of tumor and non-tumor epithelial cell lines revealed that their exposure to a stiff matrix induces migration driven by the activation of the YAP transcriptional program. The cells retained enhanced actomyosin expression and nuclear YAP translocation induced by a stiff matrix even after exposure to a soft secondary matrix, suggesting that matrix stiffness induces mechanical memory through epigenetic changes (348, 349). The high viscosity of the tumor interstitial fluid is associated with the leakage of the lymphatic vessels and increased degradation of ECM, leading to the accumulation of macromolecules (149). This elevated interstitial viscosity induces TRPV4 activation and RhoA-mediated cell contractility. Breast tumor cells exposed to elevated viscosity exhibited increased migration and extravasation potential and high lung colonization in murine xenograft models. Tumor cells previously exposed to elevated viscosity retained high migratory potential after switching the medium viscosity to baseline level. This viscous cell memory was mediated by TRPV4 and Hippo signaling (149). Thus, mechanical memory in the form of epigenetic adaptations induced by the primary tumor microenvironment can influence tumor cell survival during the entire metastatic journey and colonization of distant organs (149, 349).

Tumor cells can spread from the primary site as single cells or clusters and can be detected in the bloodstream in the form of single or clustered CTCs (328). A minimally invasive method of the CTC enumeration in peripheral blood samples called liquid biopsy can be used as a prognostic marker in patients with breast, prostate, and colorectal cancer (350). It is based on immunomagnetic capture of cells positive for the epithelial cell adhesion molecule (EpCAM) in the blood samples. However, some CTCs are EpCAM negative as a result of their phenotypic plasticity, such as EMT (351). This plasticity is induced by different microenvironmental stimuli, including mechanical stresses. The CTCs in the bloodstream are exposed to hydrodynamic shear stress (HSS), which can affect the metastasis-initiating properties of these cell populations. Indeed, HSS has been shown to trigger EMT phenotypes (159, 352), and induce CSC properties through the inhibition of the extracellular signal-related kinase (ERK)/glycogen synthase kinase (GSK)3β signaling (353). Shear stress induces a number of signaling pathways, including PI3K/Akt, MMPs, and YAP/TAZ (354-356), and promotes tumor-forming and migration potential (353, 355, 357). Therefore, CTC exposure to the HSS in the bloodstream might induce their reprogramming toward more aggressive metastasis-inducing populations. In contrast to the highly invasive soft tumor cells, CTCs are mechanically robust (299), allowing them to withstand HSS. Some CTCs are protected from HSS-mediated damage by inducing RhoA signaling, increased formation of cortical F-actin, and activation of myosin II (299). Despite the adaptive mechanisms, metastatic colonization is a rare event, and less than 0.01% of tumor cells entering the bloodstream give rise to macroscopic metastases (358). Different microenvironmental factors, including HSS and cell compression upon squeezing through the wall of blood vessels and small-sized capillaries, might induce NER and cell death. On the other hand, mechanical deformation selects the tumor subpopulation with resilience to cell death induced by mechanical deformation. This cell population possesses highly activated DNA damage response, upregulated proliferation, and resistance against chemotherapeutic drugs (359). Thus, the biomechanical characteristics of tumor cells are dynamically changing through epigenetic reprogramming or selection processes depending on the applied physical forces. This biomechanical plasticity provides tumor cells with high adaptability during tumor growth and metastatic dissemination in the dynamic and heterogeneous microenvironment (34, 360) (**Figure 6**).

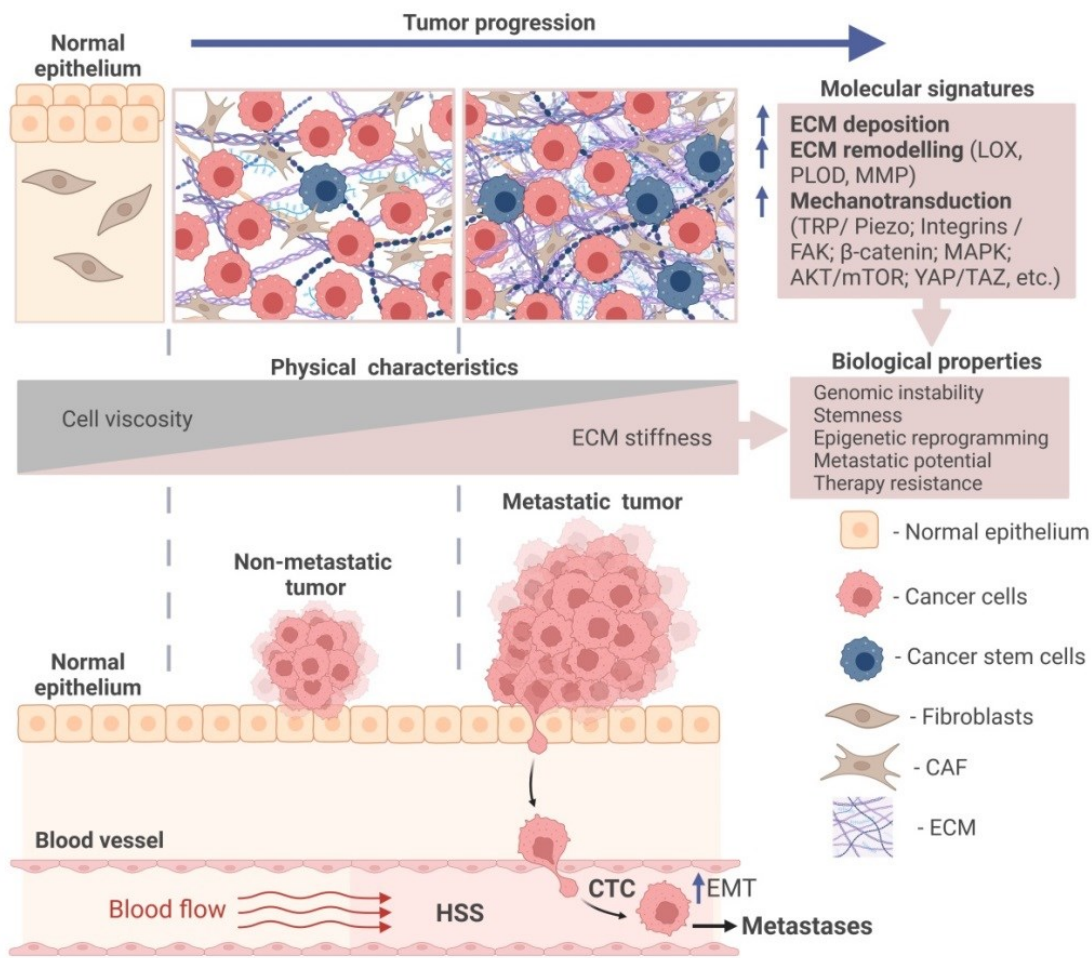

**Figure 6.** Tumor cell adaptability to the mechanical stresses. Dynamic changes in the tumor cell properties provide high adaptability to mechanical stresses during tumor growth and metastatic dissemination. CAF: cancer-associated fibroblasts; CTC: circulating tumor cell; ECM: extracellular matrix; EMT: epithelial-mesenchymal transition; FAK: focal adhesion kinase; HSS: hydrodynamic shear stress; MAPK: mitogen-activated protein kinase; MMP: matrix metalloproteinases; mTORC1: mechanistic target of rapamycin complex 1; LOX: lysyl oxidase; PLOD: procollagen-lysine, 2-oxoglutarate 5-dioxygenase; TAZ: WW domain-containing transcription regulator 1; TRP: transient receptor potential channels; YAP: yes-associated protein 1. Created with BioRender.

### 5.2.4. Electrical features of CSCs

Electric fields regulate the key biological functions of cells, including proliferation, stemness, and differentiation (361, 362). In addition to their distinct mechanical properties, CSCs have unique electrical characteristics. As discussed in Section 3, cancer cells possess unique bioelectric properties, making them distinct from their normal counterparts. The bioelectric properties of tumors and CSCs are defined by the above-mentioned $Ca2^+$ ion channels TRP, Piezo as well as $K^+$, $Na^+$ and $Cl^+$ channels, the intracellular and extracellular pH, and mitochondrial characteristics including pH, membrane potential, and ion concentrations (183, 196, 325, 363, 364). Consequently, the seminal studies demonstrated the unique bioelectrical features of CSC cells.

A depolarization of the cell membrane was suggested as a marker of CSC populations (198), whereas induction of cell membrane depolarization by a suitable electrical environment was associated with CSC differentiation in glioblastoma (GBS) and was proposed as one of the therapeutic approaches to eliminate CSC populations (365). The electrostatic potential difference (EPD) measured on cryosections of different tumor tissues correlates with tumor spread and CSC abundance, suggesting the prognostic potential of EPD measurement (366). Furthermore, galvanotaxis study revealed that CSCs migrate to the regions with negative electrostatic potential (366).

A single-cell impedance measurement revealed that CSC and non-CSC populations from liver and oral squamous carcinoma cell lines can be differentiated within the impedance magnitude ranging between 2 and 20 kHz (283). The stem cell phenotype of these cells was verified by the analysis of CSC biomarkers. Another study measured the bioelectrical impedance of non-cancerous breast epithelial cells and breast cancer cells with different metastatic potential in the range of 200 kHz-5 MHz. This study revealed a high correlation of the conductivity (σ) and permittivity (ε) to the tumor cell characteristics, including metastatic potential and migration capacities, and expression levels of different markers indicative of cell proliferation, microenvironment hypoxia and acidification, and glucose metabolism. Highly metastatic breast cancer cells have shown higher relative ε than non-metastatic tumor cells, and this difference is even more prominent compared to normal epithelial cells. These distinct electric properties can be attributed to the activation of the ion channel proteins such as $Na^+/H^+$ exchanger 1 (NHE1), which transports a proton to the extracellular space, high lactate secretion and, correspondingly, increased extracellular acidification (165). Hence, this methodology can be used to monitor CTCs in the bloodstream as a prognostic marker of metastatic tumors. The importance of electrical signaling for tumor progression was demonstrated by finding that metastatic potential can be inhibited by non-contact induced electric fields (iEFs), suggesting a new treatment strategy to prevent tumor dissemination (367).

## 6. Application of biophysical properties in clinical trials

### 6.1. Imaging and targeting tumor stiffness

The complex interplay between the biological and physical properties of tumor cells and the microenvironment drives tumor progression. Physical forces play a critical role in cancer growth but also can be used for its treatment. Radiation therapy in the form of electromagnetic waves has been one of the mainstays of cancer treatment for almost 130 years since German scientist Wilhelm Conrad Röntgen described the discovery of X-rays in 1895 (368). X-rays are electromagnetic waves inducing tissue ionization and leading to a chain of biochemical reactions, starting from molecular damage to the death of tumor cells. Radiation therapy affects cells directly through DNA damage and indirectly through radiolysis of water and the production of chemically reactive oxygen species (ROS). Today, radiotherapy is applied for about 50–60% of all cancer patients during their treatment (369, 370). X-rays are also broadly used for non-invasive imaging techniques such as computed tomography (CT) (371).

Although the translation of biophysics into clinics is relatively new, the increased stiffness of solid tumors has been known since ancient times. The mechanobiology-based method of tumor palpation is also a part of clinical practice for the initial tumor diagnosis. Imaging techniques such as MRI and ultrasonography (US), which enable the evaluation of tissue stiffness, are critical non-invasive diagnostic and prognostic tools to predict clinical outcomes (372-375). To improve the specificity of MRI, advanced MR imaging techniques, including magnetic resonance elastography (MRE), have been clinically tested to detect tumor tissues in patients with liver (376), brain (374, 377), breast (378) and prostate cancer (379). The endoscopy-guided US, also called echoendoscopy, is routinely used in clinical practice for the visualization of the gastrointestinal (GI) tract and nearby tissues and is a standard imaging tool for tumor diagnostics and treatment application (380). Both MRE and US were proved to be reliable techniques for differentiating benign and malignant lesions with high sensitivity and specificity (376, 379-384). The presence of dense, collagen-rich ECM around a tumor, called desmoplasia, is a typical pathological feature of different types of tumors, including breast cancer (385). Breast tissue density measured with X-ray imaging (mammography) has been strongly associated with increased risk for breast cancer development (385, 386). Collagen deposition rises in response to cancer therapy, serving as a marker of treatment response and being associated with improved outcomes in patients with pancreatic ductal adenocarcinoma (PDAC) (387, 388). The positron emission tomography (PET) [68]Ga probe based on the collagen-binding peptide ([68]Ga-CBP8) was successfully used in a clinical setting for specific targeting and non-invasive quantifying the dynamic changes of type I collagen in tumor ECM (389) (clinical trial number NCT04485286). This type of ECM-specific imaging is a promising tool for non-invasive monitoring treatment response in patients with PDAC and potentially other types of tumors. A high expression of integrins in many cancer types is employed to develop PET tracers based on the integrin-binding peptides. These PET tracers were clinically tested for breast, pancreatic, lung, colon, prostate, head and neck, and other tumors positive for αVβ3 integrin (e.g., clinical trials NCT05013086, NCT05976620, NCT02747290), αvβ6 integrin (e.g., clinical trials NCT05835570, NCT04285996, NCT03164486) (390), and αvβ8 integrin (391). The binding sites for the above-mentioned integrins possess very similar surface-interface properties, potentially challenging the binding specificity of the PET tracers (392).

Given the critical role of stiffness in tumor initiation, progression, and therapy response, targeting ECM components and mechanosignaling is a promising anti-cancer strategy. One of the approaches to decrease of ECM stiffness is the inhibition of lysyl oxidase (LOX), the enzymes mediating crosslinking and stabilization of ECM. LOX are copper amine oxidases inducing covalent crosslinking collagen and elastin proteins by oxidizing their lysine residues (393). LOX proteins are deregulated in several types of cancer, and their expression levels correlate with tumor metastases and therapy resistance (394, 395). The LOXL2-targeted humanized antibodies Simtuzumab (GS-6624) were tested in combination with chemotherapeutic drug gemcitabine for the treatment of metastatic pancreatic adenocarcinoma in phase II clinical trial (trial number NCT01472198). Although this combination treatment was tolerable, it failed to improve clinical outcomes in these patients (396). The later-developed chemical drugs were reported to provide more efficient inhibition of LOX activity than antibodies (397, 398). The chemical inhibitors of LOXL2, PXS-5338, PXS 5382A, and PAT-1251 entered early-phase clinical trials for analysis of their safety and tolerability in healthy subjects and in patients with myelofibrosis (clinical trial numbers NCT02852551, NCT04676529, NCT04183517, and NCT04676529). The administering of PXS-5338 results in efficient inhibition of LOXL2 activity in blood plasma (397). The recent preclinical study described selective and potent pan-lysyl oxidase inhibitor PXS-5505. Treatment with this drug decreased tumor growth and metastases, decreased chemotherapy-induced collagen crosslinking and stiffness of the tumor microenvironment in human patient-derived xenografts (PDX) and genetically engineered mouse models (399). This inhibitor was reported to provide long-lasting, potent inhibition of LOX and to be safe in a phase 1 clinical trial in healthy human subjects (400). Future clinical trials have to evaluate the clinical efficacy of this treatment in patients with malignant diseases.

Another promising approach is targeting mechanotransduction pathways, including inhibition of TRP ion channels and integrins. The expression of the TRPV6 calcium channel is increased in response to the mechanical tension and is associated with the development of different types of cancer (401, 402). The chemical inhibitor of TRV6, SOR-C13, was safe and tolerated, and demonstrated antitumor activity in a phase I clinical study in patients with advanced solid tumors (clinical trial number NCT00839631 and NCT01578564) (403). Another receptor of the TRP family, the TRPV4 calcium channel, can be activated by diverse stimuli, including altered cell volume, shear stress, and temperature changes (404-406). The TRPV4-induced signaling mechanisms are critical for tumor proliferation and metastases (407-409). The TRPV4 inhibitor GSK2798745 was well-tolerated in phase 1 clinical study involving healthy volunteers and stable heart failure patients (clinical trial number

NCT02119260)(410). The cold- and voltage-induced calcium channel TRPM8 is deregulated in tumors. It plays divergent roles in the regulation of cancer cell proliferation and invasion depending on the tumor entity (411). The results from a phase 1 study of D-3263 HCl, a TRPM8 agonist, in patients with advanced solid tumors have shown evidence of on-target activation and disease stabilization in patients with advanced prostate cancer (clinical trial number NCT00839631) (412).

Integrins play a critical role in mechanotransduction by linking ECM and cell cytoskeleton, mediating the signals from other mechanoreceptors, and converting mechanical stimuli into intracellular signaling regulating tumor cell survival, migration, and treatment response (413-415). Since their discovery around four decades ago, different members of the integrin family have been investigated as biomarkers and pharmacological targets. To date, seven drugs targeting integrins have been approved for clinical use by the FDA for the treatment of cardiovascular diseases, multiple sclerosis, plaque psoriasis, inflammatory bowel disease, and dry eye disease (416). Targeting integrins in cancer has not yet been translated efficiently into clinics. The phase 3 multicentre trial testing the selective αvβ3 and αvβ5 integrin inhibitor cilengitide in combination with temozolomide and radiotherapy in patients with glioblastoma did not reveal additional toxic effects from cilengitide but also did demonstrate clinical benefit from combination of cilengitide with temozolomide chemoradiotherapy (clinical trial number NCT00689221) (417). Another study of the monoclonal antibody PF-04605412 directed against the α5β1 integrin in patients with advanced or metastatic solid tumors was terminated due to an unfavorable safety profile (clinical trial number NCT00915278). The phase II study tested the antitumor activity of the monoclonal humanized antibody MEDI-522a (etaracizumab) targeting αVβ3 integrin with or without dacarbazine chemotherapy in patients with metastatic melanoma (clinical trial number NCT00066196); however the results have not been reported yet. Additional integrin-targeting therapies such as α5β1-specific inhibiting antibody volociximab (418), anti-αv antibody abituzumab (419) and intetumumab (420) and several other antibody-based integrin-targeted therapies are being tested in early-phase clinical trials. However, most of the completed studies in patients with cancer did not demonstrate significant clinical efficacy that could be partially explained by the complexity of integrin biology, the multifaceted role of integrins in cancer and non-tumor cells such as immune cells, and the lack of biomarkers to predict patients response to the anti-integrin treatment (416, 421, 422).

### 6.2. Electricity-mediated cancer therapy

As discussed previously in Section 4, tumor and normal cells possess different electric properties. Transdermal exposure of tumors to the 100-400 kHz AC electric fields is a non-invasive treatment called tumor treating fields (TTFields) applied by the transducer arrays placed on the patient's scalp close to the tumor. The anti-cancer effect of TTFields depends on different factors, including treatment durations, used frequency, direction of the fields, and their intensities (V/cm) (423). TTFields therapy induced a broad spectrum of biological processes in tumor cells, including inhibition of cell proliferation and migration, increasing membrane permeability, inducing replication stress and autophagy, and activation of the immune response (424, 425). Since the treatment with TTFields targets highly proliferative cells, tumors can be more sensitive to this therapy than adjacent normal tissues (426-429). The anti-proliferative effect of TTFields was attributed to its impact on the proteins with large dipole moments like tubulin dimers and microtubules. TTFields induces conformational rearrangement and depolymerization of tubulin microtubules critical in cell division (426, 428, 430). TTFields also affect septins, the proteins with high dipole moment required for positioning the mitotic spindle and cytokinetic progression leading to the aberrant cytokinesis, cell cycle block, and cell death (426, 431). The clinical benefits of anti-cancer therapy with TTFileds were mostly extensively analyzed in patients with glioblastoma (GBM) (432). The electric field therapy was well tolerated in patients with GBM. The TTFields therapy in combination with temozolomide chemotherapy was more clinically efficient than therapy with temozolomide alone, which was evidenced by significant improvement in overall survival and progression-free survival (clinical trial number NCT00916409) (433, 434). The efficacy of the TTFields in clinical trials led to its FDA approval for patients with GBM in 2015. Several other clinical trials for TTFields therapy in combination with chemotherapy have shown its clinical benefit without an increase in toxicity for patients with mesothelioma (clinical trial number NCT02397928) (435), advanced non-small cell lung cancer (NSCLC) (clinical trial number NCT00749346) (436), advanced pancreatic cancer (clinical trial number NCT01971281) (437) and some other tumor entities (424). These clinical data suggest that TTFields therapy has great potential to improve the outcomes of patients diagnosed with different types of malignancies. A broad variety of underlying molecular mechanisms enable its combination with different types of anti-cancer treatment (424, 425, 432).

The second clinically approved electricity-assisted therapy is irreversible electroporation (IRE). An external IRE increases cell membrane permeability. The irreversible membrane structural defects induced by IRE make cells more permeable to chemotherapeutic drugs, induce tumor cell death, and activate anti-tumor immune reponses (438, 439). IRE is performed under the US or computer tomography guidance using the NanoKnife IRE generator system  (439, 440). Promising therapeutic effects and safety profile resulted in the clinical approval of The NanoKnife IRE system for ablation of soft tissue tumors. The PRESERVE study for the use of IRE treatment in patients with prostate cancer (clinical trial number NCT04972097) confirmed favorable treatment safety. The final results of the study are required to evaluate the treatment effectiveness (441). NanoKnife IRE treatment is currently analyzed alone or in combination with conventional therapies in several additional clinical trials for different types of malignancies, e.g., pancreatic cancer (clinical trial numbers, e.g., NCT02791503, NCT03105921, NCT03180437), colorectal liver metastases (clinical trial number NCT02082782), breast cancer (clinical trial number NCT02340858), stomach tumors (clinical trial number NCT02430636), bladder cancer (clinical trial number NCT02430623) and others. The published results of these studies confirm the safety and the evidence of anti-tumor efficacy of the NanoKnife IRE therapy (442-445).

### 6.3.    Thermotherapy

Thermoregulation is an evolutionarily developed adaptive mechanism enabling mammals to survive environmental alterations. Consequently, the biochemical reactions in the cells of mammalian organisms are highly temperature-sensitive (446, 447). A tissue heating of just a few degrees induces protein denaturation and aggregation, inhibits protein synthesis and DNA repair, and leads to cell cycle arrest and cell death (446, 448). Cancer cells are more sensitive to temperature deviations than normal cells. Depending on the applied temperature, local heating affects tumor access to a blood supply, and, therefore, availability of oxygen and nutrients; downregulates DNA repair and replication enzymes and induces DNA damage; impacts the tumor microenvironment, and activates anti-tumor immune response (449). Thermotherapy, or thermal ablation, is an exposure of tumor tissue to temperature conditions different from its physiological diapason between 36 and 37.5 °C. Thermotherapy is classified as hyperthermia (applying increased temperature of > 40 $^{o}$C) or hypothermia (applying decreased temperatures of < - 40 $^{o}$C) (450, 451). The anti-cancer effect of hypothermia has been shown for preclinical models of tumors with mutated p53 tumor suppressor (452), but in contrast to the hyperthermia, it was not developed into conclusive clinical trials for cancer treatment. Tumor heating can be achieved by different techniques, including electromagnetic fields with different frequencies, wavelengths, and tissue penetration (including radiofrequency waves of 0.3–30 MHz or microwave systems powered by 433 MHz, 915 MHz or 2450 MHz generators) (453, 454); US heating using acoustic waves with a frequency on 0.5-10 mHz, infrared heating with frequency >300 GHz, laser light or perfusional hyperthermia or local heating techniques through physical contact with heating courses (453, 455). Thermotherapy is a non-invasive and safe therapy with clinically proven anti-cancer activity when used in combination with other treatments. Due to its effect on the blood flow, intratumor hypoxia, and tumor microenvironment hyperthermia is a potent sensitizer of tumor tissues to other anti-cancer treatments such as radiotherapy, chemotherapy, and immune therapy (450, 456, 457). Thermal therapy techniques have been clinically approved for the management of several malignant diseases. A nanoparticle-mediated hyperthermia treatment using superparamagnetic iron oxide nanoparticles (SPIONs) subjected to alternating magnetic fields (AMF), and has been approved by the European Medicine Agency (EMA) for glioblastoma treatment (458). Several clinical trials have evaluated the efficacy of hyperthermia alone or in combination with chemotherapy and radiotherapy (459-461). Most of the studies reported evidence of the clinical benefit from hyperthermia treatment and similar rates of toxicity for the control arm (conventional therapy) and experimental arm (thermotherapy and conventional therapy), and many other trials are currently in progress for different tumor types, including head and neck (clinical trials number NCT00848042), prostate cancer (clinical trials number NCT02680535), and locally advanced cancers (clinical trials number NCT05099809) (459, 462-464). The clinical studies suggest that thermotherapy treatment holds great promise for improving cancer treatment in combination with conventional therapies. Nevertheless, although hyperthermia has been known from ancient times, the mechanisms of its action are still not wholly understood, and much work is required to understand the intracellular and tissue processes mediating the effects of hyperthermia and its possible synergism with other treatment modalities and define the optimal treatment protocols (453, 457).

### 7. Predicting cancer outcomes with biophysical tumor properties and artificial intelligence

Employing machine learning (ML) algorithms to predict clinical outcomes has a longstanding history; some examples include (465, 466). However, conventional ML algorithms like support vector machines require an additional step of manual feature engineering, i.e. qualitative or quantitative description of the diagnostic data. This limits both the scalability and applicability of the conventional ML algorithms to unstructured molecular or imaging data. In contrast, the current generation of AI algorithms is leveraging an approach called representation learning, allowing them to learn features directly from unstructured data algorithmically. One of the starkest examples of representation learning algorithms instrumental in predicting cancer outcomes is the application of convolutional neural networks to biomedical imaging of cancer (reviewed in (467)), such as in digital pathology (468) and MRI (469).

Crucially, management of cancers like lung, brain, breast, and prostate, where traditional radiographic assessment may be limited, has been shown to improve by applying representation learning algorithms (469). Specifically, since tasks like the demarcation of tumor volume and the extraction of characteristic cancer phenotypes can be addressed more consistently, integrating AI-based workflows contributes to the improvement of risk prediction and, ultimately, outcomes. Furthermore, in the case of digital pathology, representation-learning algorithms allow for rapid analysis of large whole-slide images by detecting tumor regions (image segmentation) and categorizing the tumor microenvironments by identifying their respective cellular composition (instance segmentation) (468, 470, 471). Notably, these approaches can lead to new ways of characterizing tumor microenvironments in pathology images through determining factors like tumor purity. In a recent study, Gong et al. have shown that low tumor purity is associated with unfavorable outcomes and immune evasion phenotypes in gastric cancer, highlighting the importance of microenvironment characterization (472).

Beyond imaging, recent advances in representation learning include unstructured multi-omics data, which allows for the leveraging of genetic and molecular profiles, as well as clinical data as predictive hallmarks (473). For example, AI algorithms used in natural language processing have been successfully employed to facilitate the annotation of multi-cancer genomic datasets (474). Finally, representation learning techniques allow for joint training of multi-modal models on diverse data with cancer prognosis as the objective (475).

Furthermore, representation learning has been applied for the analysis of complex data in the field of translational biophysics, for example, for the identification of tumor cells, classification of their phenotype, and therapy response based on their electrical properties measured with EIC (164, 476-478), for correlation of tumor stiffness and cancer treatment efficacy (479, 480) or prediction of patients' risk of metastases based mechanobiological assays (481).

## 8. Challenges and perspectives

The physical and biological properties of tumor cells are tightly interconnected and mutually dependent. Increasing ECM stiffness and hypoxia are the fundamental hallmarks of tumor progression and poor therapeutic response. Together with other associated tumor features such as metabolic programming and lactic acidosis, genomic instability, and upregulation of the oncogenic pathways, including ion channels, they not only fuel tumor aggressiveness and therapy resistance but also alternate the physical parameters of tumor cells and microenvironment, including electrical, mechanical, and thermal properties. Across different solid tumor entities, there is wide accordance that tumors show impedimetric parameters different from normal tissues and characteristic mechanical changes at the intracellular, cellular, ECM, and interstitial fluid levels that bear great potential for new biomarker developments. Recent preclinical developments demonstrated that multiple physical and mechanical properties might serve as promising diagnostic, prognostic, and predictive biomarkers. Still, many questions remain about how to facilitate their clinical translation. A current limitation is the comparatively low number of studies that exploit reasonable numbers of patient-derived materials for single-cell and intracellular experiments. In order to develop clinically applicable routines, a higher level of standardization of measurement protocols, larger patient cohorts, longitudinal sample analysis, and higher throughput of measurements are needed. In addition to probing cancer cells themselves, more future studies may also be expanded on other cell types of the tumor microenvironment, e.g., on stromal cells (482, 483) and immune cells. Furthermore, measurements of cells from solid tumors typically require digestion and dispersion of tissues to get access to single cells (49). On the other hand, testing of patient-derived cancer cells from pleural fluids or blood from liquid biopsy samples does not require enzymatic digestion and can better reflect the native mechanical properties of analyzed cell populations.

The development of improved preclinical models that can recapitulate tumor physical parameters and biological clues is one of the main challenges for translational research. The earlier studies focused on the tumor biophysical properties by employing 2D cell culture. However, these 2D models did not recapitulate the complexity of the tumor interaction with the microenvironment. This concern is partially addressed using 3D in vitro culture, where tumor cells alone or together with stroma cells are embedded in the synthetic and natural biomatrix such as hydrogels or animal-derived matrigel. Although these models better resemble tumor architecture, they fail to fully recapitulate the physical forces exerted on cells in tissues. The recently developed technology, such as organ-on-a-chip, combines tumor cell culture along with ECM components with fluid flow. These models are used to recapitulate tumor mechanobiology better and investigate the role of mechanical forces in the regulation of tumor cell survival, invasion, and therapy resistance (34, 484).

Although the thermal properties of cells exhibit significant changes between healthy and cancerous states, highlighting their potential as prognostic biomarkers, the thermal methods have not yet garnered the same level of attention as electrical or optical techniques in biomedical research and diagnostics. This disparity can be attributed to several challenges, particularly in the experimental setup. Achieving high sensitivity in thermal measurements requires effective thermodynamic shielding from environmental influences, a challenge seen in methods such as calorimetry and HTM method. Despite these limitations, the simplicity of the read-out principles and the potential for indirect and label-free sensing in thermal techniques offer promising avenues for future exploration. With advancements in experimental design and improved sensitivity, thermal methods could emerge as a valuable tool in cellular analysis and cancer prognosis. Further research and innovation are needed to overcome current barriers and fully realize their potential.

Functional and phenotypical diversity of cancer cells within individual tumors is one of the major challenges for efficient tumor treatment. Tumors are highly heterogeneous tissues consisting of different types of cells, with different potentials to initiate and maintain tumor growth, withstand therapies, and metastasize. The eradication of all tumor-maintaining CSC cells is critically required for tumor treatment. Despite stemness being more transient than stable cell populations, CSCs are the critical determinant of tumor growth during cancer progression and its regrowth after treatment (279, 485). Thus, eradication of all CSCs is essential for permanent tumor control (486). However, the biological markers of CSC often lack specificity and are expressed on other types of cells. Due to the lack of precise tools to identify CSCs, the targeting of these populations still remains elusive. The maintenance of CSC in their niches depends on the mechanoenvironment. The intra- and intertumoral mechanical heterogeneity of tumor tissues impacts the distribution and properties of CSCs. The maintenance of CSCs in their niches depends on the microenvironment. The intra- and intertumoral mechanical heterogeneity of tumor tissues impacts the distribution and properties of CSCs. Thus, targeting the mechanical microenvironment can be a promising target for eradicating CSC populations (487). Furthermore, spatial mapping and longitudinal analysis of the mechanical and electrical properties of CSCs can bring an opportunity to develop novel label-free prognostic markers. Analysis of the physical characteristics of CSCs and their correlation with biological, pathological, and clinical parameters can better explain tumor heterogeneity and help in tailoring cancer treatment to individual patients.

Although the biological mechanisms making tumor cells more sensitive to the electricity- and temperature-driven treatment are not fully understood, several clinical studies confirmed the efficacy and safety of hyperthermia and electricity-assisted anti-cancer therapy. Still, much work is required to decipher the comprehensive biological mechanism activated by these treatments in tumors and surrounding tissues and to develop predictive markers for the patients response. To establish a meaningful correlation between complex imaging, molecular and electrical tumor parameters, and therapeutic responses or clinical outcomes, ML models are currently used. The AI and improved computing power are making possible to connect biomarkers with biophysical properties and clinicopathological parameters to develop novel label-free diagnostic and prognostic tools.

## Acknowledgements

J.-F. B. thanks G. Arkowitz, B. Ladoux, R.-M. Mège and M. Reffay for stimulating discussions. This research was supported by the Agence Nationale de la Recherche (ANR) under the contract ANR-24-CE42-6142-03 (VISCOMAG2) (to J.-F. B.) and by the HZDR - Programm für Vernetzungsdoktorand*innen (to L.B. and A. D.).

## Abbreviations

AC – alternating current

AFM - atomic force microscopy

AMF – alternating magnetic fields

ATAC-seq – assay for transposase-accessible chromatin sequencing

CAFs – cancer-associated fibroblasts

CEST – chemical exchange saturation transfer

CSCs – cancer stem cells

CT - computed tomography

CTCs – circulating tumor cells

DEP – Dielectrophoresis

DSC – differential scanning calorimetry

ECM – extracellular matrix

EPD – electrostatic potential difference

EIC – Electrical impedance Cytometry

EIS – Electrical impedance spectroscopy

EMA – European Medicine Agency

EMT – epithelial-mesenchymal transition

EpCAM – epithelial cell adhesion molecule

ERK – extracellular signal-related kinase

FAK – focal adhesion kinase

FDA – Food and Drug Administration

GAG – glycosaminoglycans

GFP – green fluorescent protein

GI – gastrointestinal

GPCR – G-protein coupled receptors

GSK – glycogen synthase kinase

HSS – hydrodynamic shear stress

HTM – heat transfer method

iEFs – induced electric fields

LOX – lysyl oxidase

MEMS – micro-electromechanical systems

ML – machine learning

MMP – matrix metalloproteinases

MRE – magnetic resonance elastography

MRI – magnetic resonance imaging

MRS – magnetic rotational spectroscopy

mTORC1 - mechanistic target of rapamycin complex 1

mTPS - modified Transient plane source

MUC1 – mucin-1

NER – nuclear envelop rupture

PET – positron emission tomography

PI3K/Akt – phosphoinositide 3-kinase

PLOD – procollagen-lysine 2-oxoglutarate 5-dioxygenase

ROS – reactive oxygen species

RT-DC – real-time deformation cytometry

(sc)RNAseq – (single-cell) RNA sequencing

SIPs – surface-imprinted polymers

SPIONs – superparamagnetic iron oxide nanoparticles

TAZ – WW domain-containing transcription regulator 1

TRP – transient receptor potential

TTFields – tumor treating fields

US – ultrasonography

Vm – membrane potential

WES – whole exome sequencing

WGS – whole genome sequencing

YAP – yes-associated protein

$\eta_{App}$ - apparent viscosity

$G_{App}$ - apparent modulus (

$E_{App}$ - apparent Young's modulus

viscosity ratio $\eta_{Canc}/\eta_{Norm}$ represent the viscosities of cancerous and normal cells

Impedance (Z) real (resistance $- R$) and imaginary (reactance $- X$) part

the magnitude ($|Z|$)

phase angle between voltage and current ($\theta$)

capacitance ($C$),

conductivity ($\sigma$)

permittivity ($\varepsilon$)

conductivity κ

thermal diffusivity α

stiffness coefficient K

elastic modulus G/E

**References:**

1.      Halsted CP, Benson JR, Jatoi I. A historical account of breast cancer surgery: beware of local recurrence but be not radical. Future Oncol. 2014;10(9):1649-57.

2.      Siegel RL, Miller KD, Wagle NS, Jemal A. Cancer statistics, 2023. CA Cancer J Clin. 2023;73(1):17-48.

3.      Hanahan D, Weinberg RA. Hallmarks of cancer: the next generation. Cell. 2011;144(5):646-74.

4.      Hanahan D. Hallmarks of Cancer: New Dimensions. Cancer Discov. 2022;12(1):31-46.

5.      Nia HT, Munn LL, Jain RK. Physical traits of cancer. Science. 2020;370(6516).

6.      Wirtz D, Konstantopoulos K, Searson PC. The physics of cancer: the role of physical interactions and mechanical forces in metastasis. Nat Rev Cancer. 2011;11(7):512-22.

7.      Mahoney L, Csima A. Efficiency of palpation in clinical detection of breast cancer. Can Med Assoc J. 1982;127(8):729-30.

8.      Sack I. Magnetic resonance elastography from fundamental soft-tissue mechanics to diagnostic imaging. . Nat Rev Phys. 2023;5(25–42).

9.      Sigrist RMS, Liau J, Kaffas AE, Chammas MC, Willmann JK. Ultrasound Elastography: Review of Techniques and Clinical Applications. Theranostics. 2017;7(5):1303-29.

10.     Mihai LA, Chin L, Janmey PA, Goriely A. A comparison of hyperelastic constitutive models applicable to brain and fat tissues. J R Soc Interface. 2015;12(110):0486.

11.     Plodinec M, Loparic M, Monnier CA, Obermann EC, Zanetti-Dallenbach R, Oertle P, et al. The nanomechanical signature of breast cancer. Nat Nanotechnol. 2012;7(11):757-65.

12.     Chang JM, Park IA, Lee SH, Kim WH, Bae MS, Koo HR, et al. Stiffness of tumours measured by shear-wave elastography correlated with subtypes of breast cancer. Eur Radiol. 2013;23(9):2450-8.

13.     Duan J, Liu Z, Li Z, Wang H, Zhao W. Research on the mean maximum Young's modulus value as a new diagnostic parameter for prostate cancer. Sci Rep. 2024;14(1):16828.

14.     Krouskop TA, Wheeler TM, Kallel F, Garra BS, Hall T. Elastic moduli of breast and prostate tissues under compression. Ultrason Imaging. 1998;20(4):260-74.

15.    Shahryari M, Tzschatzsch H, Guo J, Marticorena Garcia SR, Boning G, Fehrenbach U, et al. Tomoelastography Distinguishes Noninvasively between Benign and Malignant Liver Lesions. Cancer Res. 2019;79(22):5704-10.

16.    Fuhs TW, F.; Fritsch, A.W.; Li, X.; Stange, R.; Pawlizak, S.; Kießling, T.R.; Morawetz, E.; Grosser, S.; Sauer, F.; Lippoldt, J.; Renner, F.; Friebe, S.; Zink, M.; Bendrat, K.; Braun, J.; Oktay, M.H.;  Condeelis, J.; Briest, S.; Wolf, B. ; Horn, L.C.; Höckel, M.; Aktas, B.; Marchetti, M.C.; Manning, M.L.; Niendorf, A.; Bi, D.; Käs, J.A. Rigid tumours contain soft cancer cells. Nature Physics. 2022;18:1510–9.

17.    Acerbi I, Cassereau L, Dean I, Shi Q, Au A, Park C, et al. Human breast cancer invasion and aggression correlates with ECM stiffening and immune cell infiltration. Integr Biol (Camb). 2015;7(10):1120-34.

18.    Levental KR, Yu H, Kass L, Lakins JN, Egeblad M, Erler JT, et al. Matrix crosslinking forces tumor progression by enhancing integrin signaling. Cell. 2009;139(5):891-906.

19.    Ling Y, Li C, Feng K, Palmer S, Appleton PL, Lang S, et al. Second harmonic generation (SHG) imaging of cancer heterogeneity in ultrasound guided biopsies of prostate in men suspected with prostate cancer. J Biophotonics. 2017;10(6-7):911-8.

20.    Tuxhorn JA, Ayala GE, Smith MJ, Smith VC, Dang TD, Rowley DR. Reactive stroma in human prostate cancer: induction of myofibroblast phenotype and extracellular matrix remodeling. Clin Cancer Res. 2002;8(9):2912-23.

21.    Calvo F, Ege N, Grande-Garcia A, Hooper S, Jenkins RP, Chaudhry SI, et al. Mechanotransduction and YAP-dependent matrix remodelling is required for the generation and maintenance of cancer-associated fibroblasts. Nat Cell Biol. 2013;15(6):637-46.

22.    Winkler J, Abisoye-Ogunniyan A, Metcalf KJ, Werb Z. Concepts of extracellular matrix remodelling in tumour progression and metastasis. Nat Commun. 2020;11(1):5120.

23.    Nia HT, Liu H, Seano G, Datta M, Jones D, Rahbari N, et al. Solid stress and elastic energy as measures of tumour mechanopathology. Nat Biomed Eng. 2016;1.

24.    Mitchell MJ, King MR. Computational and experimental models of cancer cell response to fluid shear stress. Front Oncol. 2013;3:44.

25.    Follain G, Herrmann D, Harlepp S, Hyenne V, Osmani N, Warren SC, et al. Fluids and their mechanics in tumour transit: shaping metastasis. Nat Rev Cancer. 2020;20(2):107-24.

26.    Streitberger KJ, Lilaj L, Schrank F, Braun J, Hoffmann KT, Reiss-Zimmermann M, et al. How tissue fluidity influences brain tumor progression. Proc Natl Acad Sci U S A. 2020;117(1):128-34.

27.    Fredberg JJ. On the origins of order. Soft Matter. 2022;18(12):2346-53.

28.    Ilina O, Gritsenko PG, Syga S, Lippoldt J, La Porta CAM, Chepizhko O, et al. Cell-cell adhesion and 3D matrix confinement determine jamming transitions in breast cancer invasion. Nat Cell Biol. 2020;22(9):1103-15.

29.    Sauer F, Fritsch A, Grosser S, Pawlizak S, Kiessling T, Reiss-Zimmermann M, et al. Whole tissue and single cell mechanics are correlated in human brain tumors. Soft Matter. 2021;17(47):10744-52.

30.    Lekka M. Discrimination Between Normal and Cancerous Cells Using AFM. Bionanoscience. 2016;6:65-80.

31.    Massey AE, Doxtater KA, Yallapu MM, Chauhan SC. Biophysical changes caused by altered MUC13 expression in pancreatic cancer cells. Micron. 2020;130:102822.

32.    Ladoux B, Mege RM. Mechanobiology of collective cell behaviours. Nat Rev Mol Cell Biol. 2017;18(12):743-57.

33.    Alibert C, Goud B, Manneville JB. Are cancer cells really softer than normal cells? Biol Cell. 2017;109(5):167-89.

34.    Liang L, Song X, Zhao H, Lim CT. Insights into the mechanobiology of cancer metastasis via microfluidic technologies. APL Bioeng. 2024;8(2):021506.

35.    Guimarães CFG, L.; Marques, A.P.;  Reis R.L. . The stiffness of living tissues and its implications for tissue engineering. Nat Rev Mater. 2020;5:351–70.

36.    Margueritat J, Virgone-Carlotta A, Monnier S, Delanoe-Ayari H, Mertani HC, Berthelot A, et al. High-Frequency Mechanical Properties of Tumors Measured by Brillouin Light Scattering. Phys Rev Lett. 2019;122(1):018101.

37.    Scarcelli G, Polacheck WJ, Nia HT, Patel K, Grodzinsky AJ, Kamm RD, et al. Noncontact three-dimensional mapping of intracellular hydromechanical properties by Brillouin microscopy. Nat Methods. 2015;12(12):1132-4.

38.    Larson RG. The Structure and Rheology of Complex Fluids. New York: Oxford University Press; 1998.

39.    Wirtz D. Particle-tracking microrheology of living cells: principles and applications. Annu Rev Biophys. 2009;38:301-26.

40.    Li QS, Lee GY, Ong CN, Lim CT. AFM indentation study of breast cancer cells. Biochem Biophys Res Commun. 2008;374(4):609-13.

41.    Massey A, Stewart J, Smith C, Parvini C, McCormick M, Do K, et al. Mechanical properties of human tumour tissues and their implications for cancer development. Nat Rev Phys. 2024;6(4):269-82.

42.    Bastatas L, Martinez-Marin D, Matthews J, Hashem J, Lee YJ, Sennoune S, et al. AFM nano-mechanics and calcium dynamics of prostate cancer cells with distinct metastatic potential. Biochim Biophys Acta. 2012;1820(7):1111-20.

43.    Bausch AR, Ziemann F, Boulbitch AA, Jacobson K, Sackmann E. Local measurements of viscoelastic parameters of adherent cell surfaces by magnetic bead microrheometry. Biophys J. 1998;75(4):2038-49.

44.    Crick FHC. The Physical Properties of Cytoplasm - A Study by Means of the Magnetic Particle Method .1. Experimental. Experimental Cell Research. 1950;1(1):37-80.

45.    Wang Y, Xu C, Jiang N, Zheng L, Zeng J, Qiu C, et al. Quantitative analysis of the cell-surface roughness and viscoelasticity for breast cancer cells discrimination using atomic force microscopy. Scanning. 2016;38(6):558-63.

46.     Rother J, Noding H, Mey I, Janshoff A. Atomic force microscopy-based microrheology reveals significant differences in the viscoelastic response between malign and benign cell lines. Open Biol. 2014;4(5):140046.

47.     Corbin EA, Kong F, Lim CT, King WP, Bashir R. Biophysical properties of human breast cancer cells measured using silicon MEMS resonators and atomic force microscopy. Lab Chip. 2015;15(3):839-47.

48.     Lekka M, Gil D, Pogoda K, Dulinska-Litewka J, Jach R, Gostek J, et al. Cancer cell detection in tissue sections using AFM. Arch Biochem Biophys. 2012;518(2):151-6.

49.     Cross SE, Jin YS, Rao J, Gimzewski JK. Nanomechanical analysis of cells from cancer patients. Nat Nanotechnol. 2007;2(12):780-3.

50.     Nematbakhsh YTP, K.; Teck Lim, C. . Correlating the viscoelasticity of breast cancer cells with their malignancy. Converg Sci Phys Oncol. 2017;3:034003.

51.     Calzado-Martin A, Encinar M, Tamayo J, Calleja M, San Paulo A. Effect of Actin Organization on the Stiffness of Living Breast Cancer Cells Revealed by Peak-Force Modulation Atomic Force Microscopy. ACS Nano. 2016;10(3):3365-74.

52.     Fischer T, Hayn A, Mierke CT. Effect of Nuclear Stiffness on Cell Mechanics and Migration of Human Breast Cancer Cells. Front Cell Dev Biol. 2020;8:393.

53.     Nikkhah M, Strobl JS, Schmelz EM, Agah M. Evaluation of the influence of growth medium composition on cell elasticity. J Biomech. 2011;44(4):762-6.

54.     Fischer TWNH, A.; Mierke, C.T. Matrix and cellular mechanical properties are the driving factors for facilitating human cancer cell motility into 3D engineered matrices. Converg Sci Phys Oncol 2017;3(4).

55.     Omidvar R, Tafazzoli-Shadpour M, Shokrgozar MA, Rostami M. Atomic force microscope-based single cell force spectroscopy of breast cancer cell lines: an approach for evaluating cellular invasion. J Biomech. 2014;47(13):3373-9.

56.     Cross SE, Jin YS, Lu QY, Rao J, Gimzewski JK. Green tea extract selectively targets nanomechanics of live metastatic cancer cells. Nanotechnology. 2011;22(21):215101.

57.     Nguyen AV, Nyberg KD, Scott MB, Welsh AM, Nguyen AH, Wu N, et al. Stiffness of pancreatic cancer cells is associated with increased invasive potential. Integr Biol (Camb). 2016;8(12):1232-45.

58.     Lekka M, Laidler P, Gil D, Lekki J, Stachura Z, Hrynkiewicz AZ. Elasticity of normal and cancerous human bladder cells studied by scanning force microscopy. Eur Biophys J. 1999;28(4):312-6.

59.     Lekka. M.; Lekki JM, M.; Golonka, P.; Stachura, Z.; Cleff, B.; Hrynkiewicz, A.Z. Local elastic properties of cells studied by SFM. Applied Surface Science. 1999;141(3-4):345-9.

60.     Gnachandran K, Kedracka-Krok S, Pabijan J, Lekka M. Discriminating bladder cancer cells through rheological mechanomarkers at cell and spheroid levels. J Biomech. 2022;144:111346.

61.     Ramos JR, Pabijan J, Garcia R, Lekka M. The softening of human bladder cancer cells happens at an early stage of the malignancy process. Beilstein J Nanotechnol. 2014;5:447-57.

62.     Canetta E, Riches A, Borger E, Herrington S, Dholakia K, Adya AK. Discrimination of bladder cancer cells from normal urothelial cells with high specificity and sensitivity: combined application of atomic force microscopy and modulated Raman spectroscopy. Acta Biomater. 2014;10(5):2043-55.

63.     Holuigue H, Lorenc E, Chighizola M, Schulte C, Varinelli L, Deraco M, et al. Force Sensing on Cells and Tissues by Atomic Force Microscopy. Sensors (Basel). 2022;22(6).

64.     Lekka M, Laidler P, Ignacak J, Labedz M, Lekki J, Struszczyk H, et al. The effect of chitosan on stiffness and glycolytic activity of human bladder cells. Biochim Biophys Acta. 2001;1540(2):127-36.

65.     Zeng J, Zhang Y, Xu R, Chen H, Tang X, Zhang S, et al. Nanomechanical-based classification of prostate tumor using atomic force microscopy. Prostate. 2023;83(16):1591-601.

66.     Tang T, Liu X, Yuan Y, Zhang T, Kiya R, Yang Y, et al. Assessment of the electrical penetration of cell membranes using four-frequency impedance cytometry. Microsyst Nanoeng. 2022;8:68.

67.     Xu W, Mezencev R, Kim B, Wang L, McDonald J, Sulchek T. Cell stiffness is a biomarker of the metastatic potential of ovarian cancer cells. PLoS One. 2012;7(10):e46609.

68.     Chen M, Zeng J, Ruan W, Zhang Z, Wang Y, Xie S, et al. Examination of the relationship between viscoelastic properties and the invasion of ovarian cancer cells by atomic force microscopy. Beilstein J Nanotechnol. 2020;11:568-82.

69.     Ketene AN, Schmelz EM, Roberts PC, Agah M. The effects of cancer progression on the viscoelasticity of ovarian cell cytoskeleton structures. Nanomedicine. 2012;8(1):93-102.

70.     Rianna C, Radmacher M. Influence of microenvironment topography and stiffness on the mechanics and motility of normal and cancer renal cells. Nanoscale. 2017;9(31):11222-30.

71.     Rebelo LM, de Sousa JS, Mendes Filho J, Radmacher M. Comparison of the viscoelastic properties of cells from different kidney cancer phenotypes measured with atomic force microscopy. Nanotechnology. 2013;24(5):055102.

72.     Zhang G, Long M, Wu ZZ, Yu WQ. Mechanical properties of hepatocellular carcinoma cells. World J Gastroenterol. 2002;8(2):243-6.

73.     Poon C. Measuring the density and viscosity of culture media for optimized computational fluid dynamics analysis of in vitro devices. J Mech Behav Biomed Mater. 2022;126:105024.

74.     Lekka M, Laidler P. Applicability of AFM in cancer detection. Nat Nanotechnol. 2009;4(2):72; author reply -3.

75.     Tang X, Zhang Y, Mao J, Wang Y, Zhang Z, Wang Z, et al. Effects of substrate stiffness on the viscoelasticity and migration of prostate cancer cells examined by atomic force microscopy. Beilstein J Nanotechnol. 2022;13:560-9.

76.     Mandal K, Asnacios A, Goud B, Manneville JB. Mapping intracellular mechanics on micropatterned substrates. Proc Natl Acad Sci U S A. 2016;113(46):E7159-E68.

77.     Dessard M, Manneville JB, Berret JF. Cytoplasmic viscosity is a potential biomarker for metastatic breast cancer cells. Nanoscale Adv. 2024;6(6):1727-38.

78.     Gal N, Weihs D. Intracellular mechanics and activity of breast cancer cells correlate with metastatic potential. Cell Biochem Biophys. 2012;63(3):199-209.

79.     Guo M, Ehrlicher AJ, Jensen MH, Renz M, Moore JR, Goldman RD, et al. Probing the stochastic, motor-driven properties of the cytoplasm using force spectrum microscopy. Cell. 2014;158(4):822-32.

80.     Li Y, Schnekenburger J, Duits MH. Intracellular particle tracking as a tool for tumor cell characterization. J Biomed Opt. 2009;14(6):064005.

81.     Smelser AM, Macosko JC, O'Dell AP, Smyre S, Bonin K, Holzwarth G. Mechanical properties of normal versus cancerous breast cells. Biomech Model Mechanobiol. 2015;14(6):1335-47.

82.     Alibert C, Pereira D, Lardier N, Etienne-Manneville S, Goud B, Asnacios A, et al. Multiscale rheology of glioma cells. Biomaterials. 2021;275:120903.

83.     Binnig G, Quate CF, Gerber C. Atomic force microscope. Phys Rev Lett. 1986;56(9):930-3.

84.     Radmacher M. Studying the mechanics of cellular processes by atomic force microscopy. Methods Cell Biol. 2007;83:347-72.

85.     Hertz H. Ueber die Berührung fester elastischer Körper. . Angewandte Mathematik (Crelle's Journal). 2009;1882(92).

86.     Sneddon IN. The relation between load and penetration in the axisymmetric boussinesq problem for a punch of arbitrary profile. Science. 1965;3(1):47–57.

87.     Rosenbluth MJ, Lam WA, Fletcher DA. Force microscopy of nonadherent cells: a comparison of leukemia cell deformability. Biophys J. 2006;90(8):2994-3003.

88.     Darling EM, Zauscher S, Block JA, Guilak F. A thin-layer model for viscoelastic, stress-relaxation testing of cells using atomic force microscopy: do cell properties reflect metastatic potential? Biophys J. 2007;92(5):1784-91.

89.     Hosseini K, Taubenberger A, Werner C, Fischer-Friedrich E. EMT-Induced Cell-Mechanical Changes Enhance Mitotic Rounding Strength. Adv Sci (Weinh). 2020;7(19):2001276.

90.     Alcaraz J, Buscemi L, Grabulosa M, Trepat X, Fabry B, Farre R, et al. Microrheology of human lung epithelial cells measured by atomic force microscopy. Biophys J. 2003;84(3):2071-9.

91.     Abuhattum S, Mokbel D, Muller P, Soteriou D, Guck J, Aland S. An explicit model to extract viscoelastic properties of cells from AFM force-indentation curves. iScience. 2022;25(4):104016.

92.     Zbiral B, Weber A, Vivanco MD, Toca-Herrera JL. Characterization of Breast Cancer Aggressiveness by Cell Mechanics. Int J Mol Sci. 2023;24(15).

93.     Staunton JR, Doss BL, Lindsay S, Ros R. Correlating confocal microscopy and atomic force indentation reveals metastatic cancer cells stiffen during invasion into collagen I matrices. Sci Rep. 2016;6:19686.

94.     Khadem H, Mangini M, Farazpour S, De Luca AC. Correlative Raman Imaging: Development and Cancer Applications. Biosensors (Basel). 2024;14(7).

95.     Hochmuth RM. Micropipette aspiration of living cells. J Biomech. 2000;33(1):15-22.

96.     Mietke A, Otto O, Girardo S, Rosendahl P, Taubenberger A, Golfier S, et al. Extracting Cell Stiffness from Real-Time Deformability Cytometry: Theory and Experiment. Biophys J. 2015;109(10):2023-36.

97.     Otto O, Rosendahl P, Mietke A, Golfier S, Herold C, Klaue D, et al. Real-time deformability cytometry: on-the-fly cell mechanical phenotyping. Nat Methods. 2015;12(3):199-202.

98.     Rosendahl P, Plak K, Jacobi A, Kraeter M, Toepfner N, Otto O, et al. Real-time fluorescence and deformability cytometry. Nat Methods. 2018;15(5):355-8.

99.     Gossett DR, Tse HT, Lee SA, Ying Y, Lindgren AG, Yang OO, et al. Hydrodynamic stretching of single cells for large population mechanical phenotyping. Proc Natl Acad Sci U S A. 2012;109(20):7630-5.

100.    Lange JR, Steinwachs J, Kolb T, Lautscham LA, Harder I, Whyte G, et al. Microconstriction arrays for high-throughput quantitative measurements of cell mechanical properties. Biophys J. 2015;109(1):26-34.

101.    Gerum R, Mirzahossein E, Eroles M, Elsterer J, Mainka A, Bauer A, et al. Viscoelastic properties of suspended cells measured with shear flow deformation cytometry. Elife. 2022;11.

102.    Urbanska M, Munoz HE, Shaw Bagnall J, Otto O, Manalis SR, Di Carlo D, et al. A comparison of microfluidic methods for high-throughput cell deformability measurements. Nat Methods. 2020;17(6):587-93.

103.    Matthews HK, Ganguli S, Plak K, Taubenberger AV, Win Z, Williamson M, et al. Oncogenic Signaling Alters Cell Shape and Mechanics to Facilitate Cell Division under Confinement. Dev Cell. 2020;52(5):563-73 e3.

104.    Tavares S, Vieira AF, Taubenberger AV, Araujo M, Martins NP, Bras-Pereira C, et al. Actin stress fiber organization promotes cell stiffening and proliferation of pre-invasive breast cancer cells. Nat Commun. 2017;8:15237.

105.    Soteriou D, Kubankova M, Schweitzer C, Lopez-Posadas R, Pradhan R, Thoma OM, et al. Rapid single-cell physical phenotyping of mechanically dissociated tissue biopsies. Nat Biomed Eng. 2023;7(11):1392-403.

106.    Guck J, Schinkinger S, Lincoln B, Wottawah F, Ebert S, Romeyke M, et al. Optical deformability as an inherent cell marker for testing malignant transformation and metastatic competence. Biophys J. 2005;88(5):3689-98.

107.    Guck J, Ananthakrishnan R, Mahmood H, Moon TJ, Cunningham CC, Kas J. The optical stretcher: a novel laser tool to micromanipulate cells. Biophys J. 2001;81(2):767-84.

108.    Lincoln B, Wottawah F, Schinkinger S, Ebert S, Guck J. High-throughput rheological measurements with an optical stretcher. Methods Cell Biol. 2007;83:397-423.

109.    Berret JF, Farkadi A, Boissier M, Pelous J. Brillouin-scattering study of the orientational glass transition in (KCl)1-x(KCN)x mixed crystals. Phys Rev B Condens Matter. 1989;39(18):13451-6.

110.    Handler C, Testi C, Scarcelli G. Advantages of integrating Brillouin microscopy in multimodal mechanical mapping of cells and tissues. Curr Opin Cell Biol. 2024;88:102341.

111.    Prevedel R, Diz-Munoz A, Ruocco G, Antonacci G. Brillouin microscopy: an emerging tool for mechanobiology. Nat Methods. 2019;16(10):969-77.

112.    Scarcelli G, Yun SH. In vivo Brillouin optical microscopy of the human eye. Opt Express. 2012;20(8):9197-202.

113.    Conrad C, Gray KM, Stroka KM, Rizvi I, Scarcelli G. Mechanical Characterization of 3D Ovarian Cancer Nodules Using Brillouin Confocal Microscopy. Cell Mol Bioeng. 2019;12(3):215-26.

114.    Mahajan V, Beck T, Gregorczyk P, Ruland A, Alberti S, Guck J, et al. Mapping Tumor Spheroid Mechanics in Dependence of 3D Microenvironment Stiffness and Degradability by Brillouin Microscopy. Cancers (Basel). 2021;13(21).

115.    Schlussler R, Kim K, Notzel M, Taubenberger A, Abuhattum S, Beck T, et al. Correlative all-optical quantification of mass density and mechanics of subcellular compartments with fluorescence specificity. Elife. 2022;11.

116.    Scarcelli G, Yun SH. Confocal Brillouin microscopy for three-dimensional mechanical imaging. Nat Photonics. 2007;2:39-43.

117.    Qian H, Sheetz MP, Elson EL. Single particle tracking. Analysis of diffusion and flow in two-dimensional systems. Biophys J. 1991;60(4):910-21.

118.    Panorchan P, Lee JS, Daniels BR, Kole TP, Tseng Y, Wirtz D. Probing cellular mechanical responses to stimuli using ballistic intracellular nanorheology. Methods Cell Biol. 2007;83:115-40.

119.    Squires TM, Mason TG. Fluid Mechanics of Microrheology. Annual Review of Fluid Mechanics. 2010;42(Volume 42, 2010):413-38.

120.    Moffitt JR, Chemla YR, Smith SB, Bustamante C. Recent advances in optical tweezers. Annu Rev Biochem. 2008;77:205-28.

121.    Helgesen G, Pieranski P, Skjeltorp AT. Nonlinear phenomena in systems of magnetic holes. Phys Rev Lett. 1990;64(12):1425-8.

122.    Berret JF. Local viscoelasticity of living cells measured by rotational magnetic spectroscopy. Nat Commun. 2016;7:10134.

123.    Bostoen CL, Berret JF. A mathematical finance approach to the stochastic and intermittent viscosity fluctuations in living cells. Soft Matter. 2020;16(25):5959-69.

124.    Thai LP, Mousseau F, Oikonomou E, Radiom M, Berret JF. Effect of Nanoparticles on the Bulk Shear Viscosity of a Lung Surfactant Fluid. ACS Nano. 2020;14(1):466-75.

125.    Radiom M, Henault R, Mani S, Iankovski AG, Norel X, Berret JF. Magnetic wire active microrheology of human respiratory mucus. Soft Matter. 2021;17(32):7585-95.

126. Dogterom M, Koenderink GH. Actin-microtubule crosstalk in cell biology. Nat Rev Mol Cell Biol. 2019;20(1):38-54.

127. Fletcher DA, Mullins RD. Cell mechanics and the cytoskeleton. Nature. 2010;463(7280):485-92.

128. Kalukula Y, Stephens AD, Lammerding J, Gabriele S. Mechanics and functional consequences of nuclear deformations. Nat Rev Mol Cell Biol. 2022;23(9):583-602.

129. Phuyal S, Romani P, Dupont S, Farhan H. Mechanobiology of organelles: illuminating their roles in mechanosensing and mechanotransduction. Trends Cell Biol. 2023;33(12):1049-61.

130. Vignjevic D, Montagnac G. Reorganisation of the dendritic actin network during cancer cell migration and invasion. Semin Cancer Biol. 2008;18(1):12-22.

131. Dauphin M, Barbe C, Lemaire S, Nawrocki-Raby B, Lagonotte E, Delepine G, et al. Vimentin expression predicts the occurrence of metastases in non small cell lung carcinomas. Lung Cancer. 2013;81(1):117-22.

132. Pachenari M, Seyedpour SM, Janmaleki M, Babazadeh Shayan S, Taranejoo S, Hosseinkhani H. Mechanical properties of cancer cytoskeleton depend on actin filaments to microtubules content: investigating different grades of colon cancer cell lines. J Biomech. 2014;47(2):373-9.

133. Panagiotakopoulou M, Bergert M, Taubenberger A, Guck J, Poulikakos D, Ferrari A. A Nanoprinted Model of Interstitial Cancer Migration Reveals a Link between Cell Deformability and Proliferation. ACS Nano. 2016;10(7):6437-48.

134. Chalut KJ, Paluch EK. The Actin Cortex: A Bridge between Cell Shape and Function. Dev Cell. 2016;38(6):571-3.

135. Taubenberger AV, Girardo S, Traber N, Fischer-Friedrich E, Krater M, Wagner K, et al. 3D Microenvironment Stiffness Regulates Tumor Spheroid Growth and Mechanics via p21 and ROCK. Adv Biosyst. 2019;3(9):e1900128.

136. Chan CJ, Ekpenyong AE, Golfier S, Li W, Chalut KJ, Otto O, et al. Myosin II Activity Softens Cells in Suspension. Biophys J. 2015;108(8):1856-69.

137. Dimitriadis EK, Horkay F, Maresca J, Kachar B, Chadwick RS. Determination of elastic moduli of thin layers of soft material using the atomic force microscope. Biophys J. 2002;82(5):2798-810.

138. Plodinec M, Loparic M, Suetterlin R, Herrmann H, Aebi U, Schoenenberger CA. The nanomechanical properties of rat fibroblasts are modulated by interfering with the vimentin intermediate filament system. J Struct Biol. 2011;174(3):476-84.

139. Mestres I, Atabay A, Escolano JC, Arndt S, Schmidtke K, Einsiedel M, et al. Manipulation of the nuclear envelope-associated protein SLAP during mammalian brain development affects cortical lamination and exploratory behavior. Biol Open. 2024;13(3).

140. Grady ME, Composto RJ, Eckmann DM. Cell elasticity with altered cytoskeletal architectures across multiple cell types. J Mech Behav Biomed Mater. 2016;61:197-207.

141. Baker EL, Bonnecaze RT, Zaman MH. Extracellular matrix stiffness and architecture govern intracellular rheology in cancer. Biophys J. 2009;97(4):1013-21.

142. Wullkopf L, West AV, Leijnse N, Cox TR, Madsen CD, Oddershede LB, et al. Cancer cells' ability to mechanically adjust to extracellular matrix stiffness correlates with their invasive potential. Mol Biol Cell. 2018;29(20):2378-85.

143. Han YL, Pegoraro AF, Li H, Li K, Yuan Y, Xu G, et al. Cell swelling, softening and invasion in a three-dimensional breast cancer model. Nat Phys. 2020;16(1):101-8.

144. Nyga A, Plak K, Krater M, Urbanska M, Kim K, Guck J, et al. Dynamics of cell rounding during detachment. iScience. 2023;26(5):106696.

145. Evers TMJ, Holt LJ, Alberti S, Mashaghi A. Reciprocal regulation of cellular mechanics and metabolism. Nat Metab. 2021;3(4):456-68.

146. Saraswathibhatla A, Indana D, Chaudhuri O. Cell-extracellular matrix mechanotransduction in 3D. Nat Rev Mol Cell Biol. 2023;24(7):495-516.

147. Sievers J, Mahajan V, Welzel PB, Werner C, Taubenberger A. Precision Hydrogels for the Study of Cancer Cell Mechanobiology. Adv Healthc Mater. 2023;12(14):e2202514.

148. Yu W, Sharma S, Rao E, Rowat AC, Gimzewski JK, Han D, et al. Cancer cell mechanobiology: a new frontier for cancer research. J Natl Cancer Cent. 2022;2(1):10-7.

149. Bera K, Kiepas A, Godet I, Li Y, Mehta P, Ifemembi B, et al. Extracellular fluid viscosity enhances cell migration and cancer dissemination. Nature. 2022;611(7935):365-73.

150. Han JW, Sung PS, Jang JW, Choi JY, Yoon SK. Whole blood viscosity is associated with extrahepatic metastases and survival in patients with hepatocellular carcinoma. PLoS One. 2021;16(12):e0260311.

151. Wells RE, Jr., Merrill EW. Shear rate dependence of the viscosity of whole bllod and plasma. Science. 1961;133(3455):763-4.

152. Miteva DO, Rutkowski JM, Dixon JB, Kilarski W, Shields JD, Swartz MA. Transmural flow modulates cell and fluid transport functions of lymphatic endothelium. Circ Res. 2010;106(5):920-31.

153. Gonzalez-Molina J, Zhang X, Borghesan M, Mendonca da Silva J, Awan M, Fuller B, et al. Extracellular fluid viscosity enhances liver cancer cell mechanosensing and migration. Biomaterials. 2018;177:113-24.

154. Maity D, Bera K, Li Y, Ge Z, Ni Q, Konstantopoulos K, et al. Extracellular Hydraulic Resistance Enhances Cell Migration. Adv Sci (Weinh). 2022;9(29):e2200927.

155. Peng X, Janicijevic Z, Lemm S, Hauser S, Knobel M, Pietzsch J, et al. Impact of Viscosity on Human Hepatoma Spheroids in Soft Core-Shell Microcapsules. Adv Healthc Mater. 2024;13(11):e2302609.

156. Bascetin R, Laurent-Issartel C, Blanc-Fournier C, Vendrely C, Kellouche S, Carreiras F, et al. A biomimetic model of 3D fluid extracellular macromolecular crowding microenvironment fine-tunes ovarian cancer cells dissemination phenotype. Biomaterials. 2021;269:120610.

157. Maity D, Li Y, Chen Y, Sun SX. Response of collagen matrices under pressure and hydraulic resistance in hydrogels. Soft Matter. 2019;15(12):2617-26.

158. von Tempelhoff GF, Heilmann L, Hommel G, Pollow K. Impact of rheological variables in cancer. Semin Thromb Hemost. 2003;29(5):499-513.

159. Calibasi Kocal G, Guven S, Foygel K, Goldman A, Chen P, Sengupta S, et al. Dynamic Microenvironment Induces Phenotypic Plasticity of Esophageal Cancer Cells Under Flow. Sci Rep. 2016;6:38221.

160. Dash SK, Patra B, Sharma V, Das SK, Verma RS. Fluid shear stress in a logarithmic microfluidic device enhances cancer cell stemness marker expression. Lab Chip. 2022;22(11):2200-11.

161. Hyler AR, Baudoin NC, Brown MS, Stremler MA, Cimini D, Davalos RV, et al. Fluid shear stress impacts ovarian cancer cell viability, subcellular organization, and promotes genomic instability. PLoS One. 2018;13(3):e0194170.

162. Kim OH, Choi YW, Park JH, Hong SA, Hong M, Chang IH, et al. Fluid shear stress facilitates prostate cancer metastasis through Piezo1-Src-YAP axis. Life Sci. 2022;308:120936.

163. Yankaskas CL, Bera K, Stoletov K, Serra SA, Carrillo-Garcia J, Tuntithavornwat S, et al. The fluid shear stress sensor TRPM7 regulates tumor cell intravasation. Sci Adv. 2021;7(28).

164. Schutt J, Sandoval Bojorquez DI, Avitabile E, Oliveros Mata ES, Milyukov G, Colditz J, et al. Nanocytometer for smart analysis of peripheral blood and acute myeloid leukemia: a pilot study. Nano Lett. 2020;20(9):6572-81.

165. Wang Y, Li Y, Huang J, Zhang Y, Ma R, Zhang S, et al. Correlation between electrical characteristics and biomarkers in breast cancer cells. Sci Rep. 2021;11(1):14294.

166. Cai L, Qin X, Xu Z, Song Y, Jiang H, Wu Y, et al. Comparison of Cytotoxicity Evaluation of Anticancer Drugs between Real-Time Cell Analysis and CCK-8 Method. ACS Omega. 2019;4(7):12036-42.

167. Du Z, Wan H, Chen Y, Pu Y, Wang X. Bioimpedance spectroscopy can precisely discriminate human breast carcinoma from benign tumors. Medicine (Baltimore). 2017;96(4):e5970.

168. Halter RJ, Schned A, Heaney J, Hartov A, Schutz S, Paulsen KD. Electrical impedance spectroscopy of benign and malignant prostatic tissues. J Urol. 2008;179(4):1580-6.

169. Yun JH, Z.T.; Hong, K.H.; Lee, J.H. Ex vivo identification of thyroid cancer tissue using electrical impedance spectroscopy on a needle. Sensors and Actuators B: Chemical. 2018;261:537-44.

170. Blad B, Baldetorp B. Impedance spectra of tumour tissue in comparison with normal tissue; a possible clinical application for electrical impedance tomography. Physiol Meas. 1996;17 Suppl 4A:A105-15.

171. Cordier C, Prevarskaya N, Lehen'kyi V. TRPM7 Ion Channel: Oncogenic Roles and Therapeutic Potential in Breast Cancer. Cancers (Basel). 2021;13(24).

172. Blackiston DJ, McLaughlin KA, Levin M. Bioelectric controls of cell proliferation: ion channels, membrane voltage and the cell cycle. Cell Cycle. 2009;8(21):3527-36.

173. Szlasa W, Zendran I, Zalesinska A, Tarek M, Kulbacka J. Lipid composition of the cancer cell membrane. J Bioenerg Biomembr. 2020;52(5):321-42.

174.    Yalçın ZDS, S., Töral, T.B.; Gündüz, U.; Külah, H. Exploring the relationship between cytoplasmic ion content variation and multidrug resistance in cancer cells via ion-release based impedance spectroscopy. Sensors and Actuators B: Chemical. 2019:180-7.

175.    Petho Z, Najder K, Carvalho T, McMorrow R, Todesca LM, Rugi M, et al. pH-Channeling in Cancer: How pH-Dependence of Cation Channels Shapes Cancer Pathophysiology. Cancers (Basel). 2020;12(9).

176.    Birkeland ES, Koch LM, Dechant R. Another Consequence of the Warburg Effect? Metabolic Regulation of Na(+)/H(+) Exchangers May Link Aerobic Glycolysis to Cell Growth. Front Oncol. 2020;10:1561.

177.    Pedersen SF, Stock C. Ion channels and transporters in cancer: pathophysiology, regulation, and clinical potential. Cancer Res. 2013;73(6):1658-61.

178.    Lee RM, Choi H, Shin JS, Kim K, Yoo KH. Distinguishing between apoptosis and necrosis using a capacitance sensor. Biosens Bioelectron. 2009;24(8):2586-91.

179.    Zhu Z, Frey O, Haandbaek N, Franke F, Rudolf F, Hierlemann A. Time-lapse electrical impedance spectroscopy for monitoring the cell cycle of single immobilized S. pombe cells. Sci Rep. 2015;5:17180.

180.    Bakhtiari S, Manshadi MKD, Candas M, Beskok A. Changes in Electrical Capacitance of Cell Membrane Reflect Drug Partitioning-Induced Alterations in Lipid Bilayer. Micromachines (Basel). 2023;14(2).

181.    Dias C, Nylandsted J. Plasma membrane integrity in health and disease: significance and therapeutic potential. Cell Discov. 2021;7(1):4.

182.    De Ninno A, Reale R, Giovinazzo A, Bertani FR, Businaro L, Bisegna P, et al. High-throughput label-free characterization of viable, necrotic and apoptotic human lymphoma cells in a coplanar-electrode microfluidic impedance chip. Biosens Bioelectron. 2020;150:111887.

183.    Di Gregorio E, Israel S, Staelens M, Tankel G, Shankar K, Tuszynski JA. The distinguishing electrical properties of cancer cells. Phys Life Rev. 2022;43:139-88.

184.    Ibrahim-Hashim A, Estrella V. Acidosis and cancer: from mechanism to neutralization. Cancer Metastasis Rev. 2019;38(1-2):149-55.

185.    Harguindey S, Stanciu D, Devesa J, Alfarouk K, Cardone RA, Polo Orozco JD, et al. Cellular acidification as a new approach to cancer treatment and to the understanding and therapeutics of neurodegenerative diseases. Semin Cancer Biol. 2017;43:157-79.

186.    Gatenby RA, Gawlinski ET, Gmitro AF, Kaylor B, Gillies RJ. Acid-mediated tumor invasion: a multidisciplinary study. Cancer Res. 2006;66(10):5216-23.

187.    Chiche J, Brahimi-Horn MC, Pouyssegur J. Tumour hypoxia induces a metabolic shift causing acidosis: a common feature in cancer. J Cell Mol Med. 2010;14(4):771-94.

188.    Santelices IB, Friesen DE, Bell C, Hough CM, Xiao J, Kalra A, et al. Response to Alternating Electric Fields of Tubulin Dimers and Microtubule Ensembles in Electrolytic Solutions. Sci Rep. 2017;7(1):9594.

189.    Gagliardi LJ, Shain DH. Is intracellular pH a clock for mitosis? Theor Biol Med Model. 2013;10:8.

190. Ohkubo T, Yamazaki J. T-type voltage-activated calcium channel Cav3.1, but not Cav3.2, is involved in the inhibition of proliferation and apoptosis in MCF-7 human breast cancer cells. Int J Oncol. 2012;41(1):267-75.

191. Pratt SJP, Hernandez-Ochoa E, Martin SS. Calcium signaling: breast cancer's approach to manipulation of cellular circuitry. Biophys Rev. 2020;12(6):1343-59.

192. Orrenius S, Zhivotovsky B, Nicotera P. Regulation of cell death: the calcium-apoptosis link. Nat Rev Mol Cell Biol. 2003;4(7):552-65.

193. Leslie TK, James AD, Zaccagna F, Grist JT, Deen S, Kennerley A, et al. Sodium homeostasis in the tumour microenvironment. Biochim Biophys Acta Rev Cancer. 2019;1872(2):188304.

194. Niemtzow RC. Transmembrane Potentials & Characters Immune & Tumor Cell (1st ed.) Boca Raton: CRC Press; 1985.

195. Diaz-Garcia A, Varela D. Voltage-Gated K(+)/Na(+) Channels and Scorpion Venom Toxins in Cancer. Front Pharmacol. 2020;11:913.

196. Payne SL, Ram P, Srinivasan DH, Le TT, Levin M, Oudin MJ. Potassium channel-driven bioelectric signalling regulates metastasis in triple-negative breast cancer. EBioMedicine. 2022;75:103767.

197. Comes N, Serrano-Albarras A, Capera J, Serrano-Novillo C, Condom E, Ramon YCS, et al. Involvement of potassium channels in the progression of cancer to a more malignant phenotype. Biochim Biophys Acta. 2015;1848(10 Pt B):2477-92.

198. Yang M, Brackenbury WJ. Membrane potential and cancer progression. Front Physiol. 2013;4:185.

199. Payne SL, Levin M, Oudin MJ. Bioelectric Control of Metastasis in Solid Tumors. Bioelectricity. 2019;1(3):114-30.

200. Espineda CE, Chang JH, Twiss J, Rajasekaran SA, Rajasekaran AK. Repression of Na,K-ATPase beta1-subunit by the transcription factor snail in carcinoma. Mol Biol Cell. 2004;15(3):1364-73.

201. Li Z, Ruan J, Zhuang X. Effective capture of circulating tumor cells from an S180-bearing mouse model using electrically charged magnetic nanoparticles. J Nanobiotechnology. 2019;17(1):59.

202. Zhou X, Yang G, Guan F. Biological Functions and Analytical Strategies of Sialic Acids in Tumor. Cells. 2020;9(2):273.

203. Al Ahmad MAN, Y.; Mustafa, F.; Rizvi, T.A. . Electrical Characterization of Normal and Cancer Cells. IEEE Access. 2018;6:25979-86.

204. Malecka-Massalska T, Mlak R, Smolen A, Brzozowska A, Surtel W, Morshed K. Capacitance of Membrane As a Prognostic Indicator of Survival in Head and Neck Cancer. PLoS One. 2016;11(11):e0165809.

205. Trainito CI, Sweeney DC, Cemazar J, Schmelz EM, Francais O, Le Pioufle B, et al. Characterization of sequentially-staged cancer cells using electrorotation. PLoS One. 2019;14(9):e0222289.

206.    Lazanas AC, Prodromidis MI. Electrochemical Impedance Spectroscopy-A Tutorial. ACS Meas Sci Au. 2023;3(3):162-93.

207.    Kromer C, Katz A, Feldmann I, Laux P, Luch A, Tschiche HR. A targeted fluorescent nanosensor for ratiometric pH sensing at the cell surface. Sci Rep. 2024;14(1):12302.

208.    Lu H, Chen A, Zhang X, Wei Z, Cao R, Zhu Y, et al. A pH-responsive T(1)-T(2) dual-modal MRI contrast agent for cancer imaging. Nat Commun. 2022;13(1):7948.

209.    Tan JWY, Folz J, Kopelman R, Wang X. In vivo photoacoustic potassium imaging of the tumor microenvironment. Biomed Opt Express. 2020;11(7):3507-22.

210.    Lee CH, Folz J, Tan JWY, Jo J, Wang X, Kopelman R. Chemical Imaging in Vivo: Photoacoustic-Based 4-Dimensional Chemical Analysis. Anal Chem. 2019;91(4):2561-9.

211.    Anemone A, Consolino L, Conti L, Irrera P, Hsu MY, Villano D, et al. Tumour acidosis evaluated in vivo by MRI-CEST pH imaging reveals breast cancer metastatic potential. Br J Cancer. 2021;124(1):207-16.

212.    James AD, Leslie TK, Kaggie JD, Wiggins L, Patten L, Murphy O'Duinn J, et al. Sodium accumulation in breast cancer predicts malignancy and treatment response. Br J Cancer. 2022;127(2):337-49.

213.    Haeverbeke MVS, M.; De Baets, B. Equivalent Electrical Circuits and Their Use Across Electrochemical Impedance Spectroscopy Application Domains IEEE Access. 2022;10:51363-79.

214.    Abasi S, Aggas JR, Garayar-Leyva GG, Walther BK, Guiseppi-Elie A. Bioelectrical Impedance Spectroscopy for Monitoring Mammalian Cells and Tissues under Different Frequency Domains: A Review. ACS Meas Sci Au. 2022;2(6):495-516.

215.    Kaur B, Kumar S, Kaushik BK. Recent advancements in optical biosensors for cancer detection. Biosens Bioelectron. 2022;197:113805.

216.    Mieog JSD, Achterberg FB, Zlitni A, Hutteman M, Burggraaf J, Swijnenburg RJ, et al. Fundamentals and developments in fluorescence-guided cancer surgery. Nat Rev Clin Oncol. 2022;19(1):9-22.

217.    Liang C, Zhang Q, Chen X, Liu J, Tanaka M, Wang S, et al. Human cancer cells generate spontaneous calcium transients and intercellular waves that modulate tumor growth. Biomaterials. 2022;290:121823.

218.    Gawad S, Schild L, Renaud PH. Micromachined impedance spectroscopy flow cytometer for cell analysis and particle sizing. Lab Chip. 2001;1(1):76-82.

219.    Druzhkova I, Lukina M, Dudenkova V, Ignatova N, Snopova L, Gavrina A, et al. Tracing of intracellular pH in cancer cells in response to Taxol treatment. Cell Cycle. 2021;20(16):1540-51.

220.    Bruno F, Granata V, Cobianchi Bellisari F, Sgalambro F, Tommasino E, Palumbo P, et al. Advanced Magnetic Resonance Imaging (MRI) Techniques: Technical Principles and Applications in Nanomedicine. Cancers (Basel). 2022;14(7).

221. Pecoraro M, Messina E, Bicchetti M, Carnicelli G, Del Monte M, Iorio B, et al. The future direction of imaging in prostate cancer: MRI with or without contrast injection. Andrology. 2021;9(5):1429-43.

222. Robinson AJ, Jain A, Sherman HG, Hague RJM, Rahman R, Sanjuan-Alberte P, et al. Toward Hijacking Bioelectricity in Cancer to Develop New Bioelectronic Medicine. Advanced Therapeutics. 2021;4(3):2000248.

223. Fahmy HM, Hamad AM, Sayed FA-z, Abdelaziz YS, Abu Serea ES, Mustafa ABE-D, et al. Dielectric spectroscopy signature for cancer diagnosis: A review. Microwave and Optical Technology Letters. 2020;62(12):3739-53.

224. Loiselle FB, Casey JR. Measurement of Intracellular pH. Methods Mol Biol. 2010;637:311-31.

225. Miller AJ, Smith S. Measuring intracellular ion concentrations with multi-barrelled microelectrodes. Methods Mol Biol. 2012;913:67-77.

226. Yang L, Liu X, Yin B, Deng X, Lin X, Song J, et al. High-Throughput and Real-Time Monitoring of Single-Cell Extracellular pH Based on Polyaniline Microarrays. Anal Chem. 2021;93(41):13852-60.

227. Park HW, Song MS, Sim HJ, Ryu PD, Lee SY. The role of the voltage-gated potassium channel, Kv2.1 in prostate cancer cell migration. BMB Rep. 2021;54(2):130-5.

228. Sizemore G, McLaughlin S, Newman M, Brundage K, Ammer A, Martin K, et al. Opening large-conductance potassium channels selectively induced cell death of triple-negative breast cancer. BMC Cancer. 2020;20(1):595.

229. Buscaglia LAO, O.N.; Carmo J.P. . Roadmap for Electrical Impedance Spectroscopy for Sensing: A Tutorial. IEEE Sensors 2021;21(20):22246-57.

230. Sarac E, Meiwes A, Eigentler T, Forchhammer S, Kofler L, Hafner HM, et al. Diagnostic Accuracy of Electrical Impedance Spectroscopy in Non-melanoma Skin Cancer. Acta Derm Venereol. 2020;100(18):adv00328.

231. Kang G, Kim YJ, Moon HS, Lee JW, Yoo TK, Park K, et al. Discrimination between the human prostate normal cell and cancer cell by using a novel electrical impedance spectroscopy controlling the cross-sectional area of a microfluidic channel. Biomicrofluidics. 2013;7(4):44126.

232. Gharooni M, Alikhani A, Moghtaderi H, Abiri H, Mashaghi A, Abbasvandi F, et al. Bioelectronics of The Cellular Cytoskeleton: Monitoring Cytoskeletal Conductance Variation for Sensing Drug Resistance. ACS Sens. 2019;4(2):353-62.

233. Chiu SP, Batsaikhan B, Huang HM, Wang JY. Application of Electric Cell-Substrate Impedance Sensing to Investigate the Cytotoxic Effects of Andrographolide on U-87 MG Glioblastoma Cell Migration and Apoptosis. Sensors (Basel). 2019;19(10).

234. Nguyen TA, Yin TI, Reyes D, Urban GA. Microfluidic chip with integrated electrical cell-impedance sensing for monitoring single cancer cell migration in three-dimensional matrixes. Anal Chem. 2013;85(22):11068-76.

235. Liang Y, Ji L, Tu T, Zhang S, Liang B, Ye X. In situ continuously monitoring of cancer cell invasion process based on impedance sensing. Anal Biochem. 2021;622:114155.

236. De Leon SE, Pupovac A, McArthur SL. Three-Dimensional (3D) cell culture monitoring: Opportunities and challenges for impedance spectroscopy. Biotechnol Bioeng. 2020;117(4):1230-40.

237. Hong SL, K.; Ha, U.; Kim, H.; Lee, Y.; Kim, Y.; Yoo, H.J. A 4.9 mΩ-Sensitivity Mobile Electrical Impedance Tomography IC for Early Breast-Cancer Detection System IEEE Journal of Solid-State Circuits. 2015;50(1):245-57.

238. Honrado C, Bisegna P, Swami NS, Caselli F. Single-cell microfluidic impedance cytometry: from raw signals to cell phenotypes using data analytics. Lab Chip. 2021;21(1):22-54.

239. Fan W, Xiong Q, Ge Y, Liu T, Zeng S, Zhao J. Identifying the grade of bladder cancer cells using microfluidic chips based on impedance. Analyst. 2022;147(8):1722-9.

240. Cheung K, Gawad S, Renaud P. Impedance spectroscopy flow cytometry: on-chip label-free cell differentiation. Cytometry A. 2005;65(2):124-32.

241. Tang T, Julian T, Ma D, Yang Y, Li M, Hosokawa Y, et al. A review on intelligent impedance cytometry systems: Development, applications and advances. Anal Chim Acta. 2023;1269:341424.

242. Smith JP, Huang C, Kirby BJ. Enhancing sensitivity and specificity in rare cell capture microdevices with dielectrophoresis. Biomicrofluidics. 2015;9(1):014116.

243. Turcan I, Caras I, Schreiner TG, Tucureanu C, Salageanu A, Vasile V, et al. Dielectrophoretic and Electrical Impedance Differentiation of Cancerous Cells Based on Biophysical Phenotype. Biosensors (Basel). 2021;11(10).

244. Varmazyari V, Habibiyan H, Ghafoorifard H, Ebrahimi M, Ghafouri-Fard S. A dielectrophoresis-based microfluidic system having double-sided optimized 3D electrodes for label-free cancer cell separation with preserving cell viability. Sci Rep. 2022;12(1):12100.

245. Bianchi L, Cavarzan F, Ciampitti L, Cremonesi M, Grilli F, Saccomandi P. Thermophysical and mechanical properties of biological tissues as a function of temperature: a systematic literature review. Int J Hyperthermia. 2022;39(1):297-340.

246. Peitzsch C, Gorodetska I, Klusa D, Shi Q, Alves TC, Pantel K, et al. Metabolic regulation of prostate cancer heterogeneity and plasticity. Semin Cancer Biol. 2022;82:94-119.

247. Martinez-Reyes I, Chandel NS. Cancer metabolism: looking forward. Nat Rev Cancer. 2021;21(10):669-80.

248. Gandhi S, Oshi M, Murthy V, Repasky EA, Takabe K. Enhanced Thermogenesis in Triple-Negative Breast Cancer Is Associated with Pro-Tumor Immune Microenvironment. Cancers (Basel). 2021;13(11).

249. Ghosh K, Brandt KR, Reynolds C, Scott CG, Pankratz VS, Riehle DL, et al. Tissue composition of mammographically dense and non-dense breast tissue. Breast Cancer Res Treat. 2012;131(1):267-75.

250.    Valvano JW, Cochran, J.R. & Diller, K.R. Thermal conductivity and diffusivity of biomaterials measured with self-heated thermistors. . Int J Thermophys. 1985;6:301–11.

251.    Vaupel P, Piazena H. Strong correlation between specific heat capacity and water content in human tissues suggests preferred heat deposition in malignant tumors upon electromagnetic irradiation. Int J Hyperthermia. 2022;39(1):987-97.

252.    Braissant O, Keiser J, Meister I, Bachmann A, Wirz D, Gopfert B, et al. Isothermal microcalorimetry accurately detects bacteria, tumorous microtissues, and parasitic worms in a label-free well-plate assay. Biotechnol J. 2015;10(3):460-8.

253.    Pini N, Huo Z, Holland-Cunz S, Gros SJ. Increased Proliferation of Neuroblastoma Cells under Fructose Metabolism Can Be Measured by Isothermal Microcalorimetry. Children (Basel). 2021;8(9).

254.    Fekecs T, Zapf, I., Ferencz, A. et al. Differential scanning calorimetry (DSC) analysis of human plasma in melanoma patients with or without regional lymph node metastases. . J Therm Anal Calorim. 2012;108:149–52.

255.    Garbett NC, Miller JJ, Jenson AB, Miller DM, Chaires JB. Interrogation of the plasma proteome with differential scanning calorimetry. Clin Chem. 2007;53(11):2012-4.

256.    Garbett NC, Mekmaysy CS, Helm CW, Jenson AB, Chaires JB. Differential scanning calorimetry of blood plasma for clinical diagnosis and monitoring. Exp Mol Pathol. 2009;86(3):186-91.

257.    Schneider G, Kaliappan A, Nguyen TQ, Buscaglia R, Brock GN, Hall MB, et al. The Utility of Differential Scanning Calorimetry Curves of Blood Plasma for Diagnosis, Subtype Differentiation and Predicted Survival in Lung Cancer. Cancers (Basel). 2021;13(21).

258.    Wagner P, Bakhshi Sichani S, Khorshid M, Lieberzeit P, Losada-Perez P, Yongabi D. Bioanalytical sensors using the heat-transfer method HTM and related techniques. Tech Mess. 2023;90(12):761-85.

259.    Eersels K, van Grinsven B, Ethirajan A, Timmermans S, Jimenez Monroy KL, Bogie JF, et al. Selective identification of macrophages and cancer cells based on thermal transport through surface-imprinted polymer layers. ACS Appl Mater Interfaces. 2013;5(15):7258-67.

260.    Bers K, Eersels K, van Grinsven B, Daemen M, Bogie JF, Hendriks JJ, et al. Heat-transfer resistance measurement method (HTM)-based cell detection at trace levels using a progressive enrichment approach with highly selective cell-binding surface imprints. Langmuir. 2014;30(12):3631-9.

261.    Eersels K, van Grinsven B, Khorshid M, Somers V, Puttmann C, Stein C, et al. Heat-transfer-method-based cell culture quality assay through cell detection by surface imprinted polymers. Langmuir. 2015;31(6):2043-50.

262.    Oudebrouckx G, Goossens J, Bormans S, Vandenryt T, Wagner P, Thoelen R. Integrating Thermal Sensors in a Microplate Format: Simultaneous Real-Time Quantification of Cell Number and Metabolic Activity. ACS Appl Mater Interfaces. 2022;14(2):2440-51.

263.    Jose M, Oudebrouckx G, Bormans S, Veske P, Thoelen R, Deferme W. Monitoring Body Fluids in Textiles: Combining Impedance and Thermal Principles in a Printed, Wearable, and Washable Sensor. ACS Sens. 2021;6(3):896-907.

264.    Goossens J, Bormans S, Oudebrouckx G, Vandenryt T, Habti SE, Mahdavinasab M, et al. Liquid Identification in a Microplate Format Based on Thermal and Electrical Sensor Data Fusion. IEEE Sensors Journal. 2022;22:19809-17.

265.    Oudebrouckx GN, D.; Vandenryt, T.; Bormans, S.;  Möbius, H.; Thoelen, R. Single element thermal sensor for measuring thermal conductivity and flow rate inside a microchannel. Sensors and Actuators A: Physical. 2021;331:112906.

266.    Cavalli FMG, Remke M, Rampasek L, Peacock J, Shih DJH, Luu B, et al. Intertumoral Heterogeneity within Medulloblastoma Subgroups. Cancer Cell. 2017;31(6):737-54 e6.

267.    Hoadley KA, Yau C, Hinoue T, Wolf DM, Lazar AJ, Drill E, et al. Cell-of-Origin Patterns Dominate the Molecular Classification of 10,000 Tumors from 33 Types of Cancer. Cell. 2018;173(2):291-304 e6.

268.    Goyette MA, Lipsyc-Sharf M, Polyak K. Clinical and translational relevance of intratumor heterogeneity. Trends Cancer. 2023;9(9):726-37.

269.    Marusyk A, Almendro V, Polyak K. Intra-tumour heterogeneity: a looking glass for cancer? Nat Rev Cancer. 2012;12(5):323-34.

270.    Negrini S, Gorgoulis VG, Halazonetis TD. Genomic instability--an evolving hallmark of cancer. Nat Rev Mol Cell Biol. 2010;11(3):220-8.

271.    Kreso A, Dick JE. Evolution of the cancer stem cell model. Cell Stem Cell. 2014;14(3):275-91.

272.    Peitzsch C, Tyutyunnykova A, Pantel K, Dubrovska A. Cancer stem cells: The root of tumor recurrence and metastases. Semin Cancer Biol. 2017;44:10-24.

273.    Skvortsov S, Skvortsova, II, Tang DG, Dubrovska A. Concise Review: Prostate Cancer Stem Cells: Current Understanding. Stem Cells. 2018;36(10):1457-74.

274.    Walcher L, Kistenmacher AK, Suo H, Kitte R, Dluczek S, Strauss A, et al. Cancer Stem Cells-Origins and Biomarkers: Perspectives for Targeted Personalized Therapies. Front Immunol. 2020;11:1280.

275.    Peitzsch C, Nathansen J, Schniewind SI, Schwarz F, Dubrovska A. Cancer Stem Cells in Head and Neck Squamous Cell Carcinoma: Identification, Characterization and Clinical Implications. Cancers (Basel). 2019;11(5).

276.    Philchenkov A, Dubrovska A. Cancer Stem Cells as a Therapeutic Target: Current Clinical Development and Future Prospective. Stem Cells. 2024;42(3):173-99.

277.    Koseer AS, Di Gaetano S, Arndt C, Bachmann M, Dubrovska A. Immunotargeting of Cancer Stem Cells. Cancers (Basel). 2023;15(5).

278.    Cable J, Pei D, Reid LM, Wang XW, Bhatia S, Karras P, et al. Cancer stem cells: advances in biology and clinical translation-a Keystone Symposia report. Ann N Y Acad Sci. 2021;1506(1):142-63.

279.    Krause M, Dubrovska A, Linge A, Baumann M. Cancer stem cells: Radioresistance, prediction of radiotherapy outcome and specific targets for combined treatments. Adv Drug Deliv Rev. 2017;109:63-73.

280.    Li Q, Rycaj K, Chen X, Tang DG. Cancer stem cells and cell size: A causal link? Semin Cancer Biol. 2015;35:191-9.

281.    Liu X, Ye Y, Zhu L, Xiao X, Zhou B, Gu Y, et al. Niche stiffness sustains cancer stemness via TAZ and NANOG phase separation. Nat Commun. 2023;14(1):238.

282.    Pankova D, Jiang Y, Chatzifrangkeskou M, Vendrell I, Buzzelli J, Ryan A, et al. RASSF1A controls tissue stiffness and cancer stem-like cells in lung adenocarcinoma. EMBO J. 2019;38(13):e100532.

283.    Lei KF, Ho YC, Huang CH, Huang CH, Pai PC. Characterization of stem cell-like property in cancer cells based on single-cell impedance measurement in a microfluidic platform. Talanta. 2021;229:122259.

284.    McGranahan N, Swanton C. Clonal Heterogeneity and Tumor Evolution: Past, Present, and the Future. Cell. 2017;168(4):613-28.

285.    Pogrebniak KL, Curtis C. Harnessing Tumor Evolution to Circumvent Resistance. Trends Genet. 2018;34(8):639-51.

286.    Chen HC, Ma Y, Cheng J, Chen YC. Advances in Single-Cell Techniques for Linking Phenotypes to Genotypes. Cancer Heterog Plast. 2024;1(1).

287.    Ge R, Wang Z, Cheng L. Tumor microenvironment heterogeneity an important mediator of prostate cancer progression and therapeutic resistance. NPJ Precis Oncol. 2022;6(1):31.

288.    Kalli M, Stylianopoulos T. Defining the Role of Solid Stress and Matrix Stiffness in Cancer Cell Proliferation and Metastasis. Front Oncol. 2018;8:55.

289.    Pfeifer CR, Alvey CM, Irianto J, Discher DE. Genome variation across cancers scales with tissue stiffness - an invasion-mutation mechanism and implications for immune cell infiltration. Curr Opin Syst Biol. 2017;2:103-14.

290.    Denais CM, Gilbert RM, Isermann P, McGregor AL, te Lindert M, Weigelin B, et al. Nuclear envelope rupture and repair during cancer cell migration. Science. 2016;352(6283):353-8.

291.    Chen X, Chen D, Ban E, Toussaint KC, Janmey PA, Wells RG, et al. Glycosaminoglycans modulate long-range mechanical communication between cells in collagen networks. Proc Natl Acad Sci U S A. 2022;119(15):e2116718119.

292.    Jalilian I, Heu C, Cheng H, Freittag H, Desouza M, Stehn JR, et al. Cell elasticity is regulated by the tropomyosin isoform composition of the actin cytoskeleton. PLoS One. 2015;10(5):e0126214.

293.    Okura K, Matsumoto T, Narita A, Tatsumi H. Mechanical Stress Decreases the Amplitude of Twisting and Bending Fluctuations of Actin Filaments. J Mol Biol. 2023;435(22):168295.

294.    Scott AK, Rafuse M, Neu CP. Mechanically induced alterations in chromatin architecture guide the balance between cell plasticity and mechanical memory. Front Cell Dev Biol. 2023;11:1084759.

295.    Karska J, Kowalski S, Saczko J, Moisescu MG, Kulbacka J. Mechanosensitive Ion Channels and Their Role in Cancer Cells. Membranes (Basel). 2023;13(2).

296.     Zhang Z, Sha B, Zhao L, Zhang H, Feng J, Zhang C, et al. Programmable integrin and N-cadherin adhesive interactions modulate mechanosensing of mesenchymal stem cells by cofilin phosphorylation. Nat Commun. 2022;13(1):6854.

297.     Lin HH, Ng KF, Chen TC, Tseng WY. Ligands and Beyond: Mechanosensitive Adhesion GPCRs. Pharmaceuticals (Basel). 2022;15(2).

298.     Yap AS, Duszyc K, Viasnoff V. Mechanosensing and Mechanotransduction at Cell-Cell Junctions. Cold Spring Harb Perspect Biol. 2018;10(8).

299.     Moose DL, Krog BL, Kim TH, Zhao L, Williams-Perez S, Burke G, et al. Cancer Cells Resist Mechanical Destruction in Circulation via RhoA/Actomyosin-Dependent Mechano-Adaptation. Cell Rep. 2020;30(11):3864-74 e6.

300.     Gargalionis AN, Papavassiliou KA, Basdra EK, Papavassiliou AG. mTOR Signaling Components in Tumor Mechanobiology. Int J Mol Sci. 2022;23(3).

301.     Di-Luoffo M, Ben-Meriem Z, Lefebvre P, Delarue M, Guillermet-Guibert J. PI3K functions as a hub in mechanotransduction. Trends Biochem Sci. 2021;46(11):878-88.

302.     Xiong J, Xiao R, Zhao J, Zhao Q, Luo M, Li F, et al. Matrix stiffness affects tumor-associated macrophage functional polarization and its potential in tumor therapy. J Transl Med. 2024;22(1):85.

303.     Reid SE, Kay EJ, Neilson LJ, Henze AT, Serneels J, McGhee EJ, et al. Tumor matrix stiffness promotes metastatic cancer cell interaction with the endothelium. EMBO J. 2017;36(16):2373-89.

304.     Mierke CT. Extracellular Matrix Cues Regulate Mechanosensing and Mechanotransduction of Cancer Cells. Cells. 2024;13(1).

305.     Wei Y, Geng S, Si Y, Yang Y, Chen Q, Huang S, et al. The Interaction between Collagen 1 and High Mannose Type CD133 Up-Regulates Glutamine Transporter SLC1A5 to Promote the Tumorigenesis of Glioblastoma Stem Cells. Adv Sci (Weinh). 2024;11(3):e2306715.

306.     Ji H, Kong L, Wang Y, Hou Z, Kong W, Qi J, et al. CD44 expression is correlated with osteosarcoma cell progression and immune infiltration and affects the Wnt/beta-catenin signaling pathway. J Bone Oncol. 2023;41:100487.

307.     Chen C, Zhao S, Karnad A, Freeman JW. The biology and role of CD44 in cancer progression: therapeutic implications. J Hematol Oncol. 2018;11(1):64.

308.     Xiong J, Yan L, Zou C, Wang K, Chen M, Xu B, et al. Integrins regulate stemness in solid tumor: an emerging therapeutic target. J Hematol Oncol. 2021;14(1):177.

309.     Ma D, Luo Q, Song G. Matrix stiffening facilitates stemness of liver cancer stem cells by YAP activation and BMF inhibition. Biomater Adv. 2024;163:213936.

310.     Safaei S, Sajed R, Shariftabrizi A, Dorafshan S, Saeednejad Zanjani L, Dehghan Manshadi M, et al. Tumor matrix stiffness provides fertile soil for cancer stem cells. Cancer Cell Int. 2023;23(1):143.

311.     You Y, Zheng Q, Dong Y, Xie X, Wang Y, Wu S, et al. Matrix stiffness-mediated effects on stemness characteristics occurring in HCC cells. Oncotarget. 2016;7(22):32221-31.

312.    Lee J, Cabrera AJH, Nguyen CMT, Kwon YV. Dissemination of Ras(V12)-transformed cells requires the mechanosensitive channel Piezo. Nat Commun. 2020;11(1):3568.

313.    Wang X, Cheng G, Miao Y, Qiu F, Bai L, Gao Z, et al. Piezo type mechanosensitive ion channel component 1 facilitates gastric cancer omentum metastasis. J Cell Mol Med. 2021;25(4):2238-53.

314.    Micek HM, Yang N, Dutta M, Rosenstock L, Ma Y, Hielsberg C, et al. The role of Piezo1 mechanotransduction in high-grade serous ovarian cancer: Insights from an in vitro model of collective detachment. Sci Adv. 2024;10(17):eadl4463.

315.    Zhu B, Qian W, Han C, Bai T, Hou X. Piezo 1 activation facilitates cholangiocarcinoma metastasis via Hippo/YAP signaling axis. Mol Ther Nucleic Acids. 2021;24:241-52.

316.    Katsuta E, Takabe K, Vujcic M, Gottlieb PA, Dai T, Mercado-Perez A, et al. Mechano-Sensing Channel PIEZO2 Enhances Invasive Phenotype in Triple-Negative Breast Cancer. Int J Mol Sci. 2022;23(17).

317.    Chen X, Wanggou S, Bodalia A, Zhu M, Dong W, Fan JJ, et al. A Feedforward Mechanism Mediated by Mechanosensitive Ion Channel PIEZO1 and Tissue Mechanics Promotes Glioma Aggression. Neuron. 2018;100(4):799-815 e7.

318.    Zhang YC, Yang M, Lu CD, Li QY, Shi JN, Shi J. PIEZO2 expression is an independent biomarker prognostic for gastric cancer and represents a potential therapeutic target. Sci Rep. 2024;14(1):1206.

319.    Li X, Hu J, Zhao X, Li J, Chen Y. Piezo channels in the urinary system. Exp Mol Med. 2022;54(6):697-710.

320.    Pardo-Pastor C, Rubio-Moscardo F, Vogel-Gonzalez M, Serra SA, Afthinos A, Mrkonjic S, et al. Piezo2 channel regulates RhoA and actin cytoskeleton to promote cell mechanobiological responses. Proc Natl Acad Sci U S A. 2018;115(8):1925-30.

321.    Wu J, Lewis AH, Grandl J. Touch, Tension, and Transduction - The Function and Regulation of Piezo Ion Channels. Trends Biochem Sci. 2017;42(1):57-71.

322.    Yao M, Tijore A, Cheng D, Li JV, Hariharan A, Martinac B, et al. Force- and cell state-dependent recruitment of Piezo1 drives focal adhesion dynamics and calcium entry. Sci Adv. 2022;8(45):eabo1461.

323.    Liu H, Hu J, Zheng Q, Feng X, Zhan F, Wang X, et al. Piezo1 Channels as Force Sensors in Mechanical Force-Related Chronic Inflammation. Front Immunol. 2022;13:816149.

324.    Li R, Wang D, Li H, Lei X, Liao W, Liu XY. Identification of Piezo1 as a potential target for therapy of colon cancer stem-like cells. Discov Oncol. 2023;14(1):95.

325.    Guo J, Shan C, Xu J, Li M, Zhao J, Cheng W. New Insights into TRP Ion Channels in Stem Cells. Int J Mol Sci. 2022;23(14).

326.    Lv J, Liu Y, Cheng F, Li J, Zhou Y, Zhang T, et al. Cell softness regulates tumorigenicity and stemness of cancer cells. EMBO J. 2021;40(2):e106123.

327.    Peitzsch C, Perrin R, Hill RP, Dubrovska A, Kurth I. Hypoxia as a biomarker for radioresistant cancer stem cells. Int J Radiat Biol. 2014;90(8):636-52.

328. Donato C, Kunz L, Castro-Giner F, Paasinen-Sohns A, Strittmatter K, Szczerba BM, et al. Hypoxia Triggers the Intravasation of Clustered Circulating Tumor Cells. Cell Rep. 2020;32(10):108105.

329. Lukovic J, Han K, Pintilie M, Chaudary N, Hill RP, Fyles A, et al. Intratumoral heterogeneity and hypoxia gene expression signatures: Is a single biopsy adequate? Clin Transl Radiat Oncol. 2019;19:110-5.

330. Pang MF, Siedlik MJ, Han S, Stallings-Mann M, Radisky DC, Nelson CM. Tissue Stiffness and Hypoxia Modulate the Integrin-Linked Kinase ILK to Control Breast Cancer Stem-like Cells. Cancer Res. 2016;76(18):5277-87.

331. Gorodetska I, Offermann A, Puschel J, Lukiyanchuk V, Gaete D, Kurzyukova A, et al. ALDH1A1 drives prostate cancer metastases and radioresistance by interplay with AR- and RAR-dependent transcription. Theranostics. 2024;14(2):714-37.

332. Aiello NM, Kang Y. Context-dependent EMT programs in cancer metastasis. J Exp Med. 2019;216(5):1016-26.

333. Ribatti D, Tamma R, Annese T. Epithelial-Mesenchymal Transition in Cancer: A Historical Overview. Transl Oncol. 2020;13(6):100773.

334. Mani SA, Guo W, Liao MJ, Eaton EN, Ayyanan A, Zhou AY, et al. The epithelial-mesenchymal transition generates cells with properties of stem cells. Cell. 2008;133(4):704-15.

335. Genna A, Vanwynsberghe AM, Villard AV, Pottier C, Ancel J, Polette M, et al. EMT-Associated Heterogeneity in Circulating Tumor Cells: Sticky Friends on the Road to Metastasis. Cancers (Basel). 2020;12(6).

336. Brown MS, Abdollahi B, Wilkins OM, Lu H, Chakraborty P, Ognjenovic NB, et al. Phenotypic heterogeneity driven by plasticity of the intermediate EMT state governs disease progression and metastasis in breast cancer. Sci Adv. 2022;8(31):eabj8002.

337. Wei SC, Fattet L, Tsai JH, Guo Y, Pai VH, Majeski HE, et al. Matrix stiffness drives epithelial-mesenchymal transition and tumour metastasis through a TWIST1-G3BP2 mechanotransduction pathway. Nat Cell Biol. 2015;17(5):678-88.

338. Fattet L, Jung HY, Matsumoto MW, Aubol BE, Kumar A, Adams JA, et al. Matrix Rigidity Controls Epithelial-Mesenchymal Plasticity and Tumor Metastasis via a Mechanoresponsive EPHA2/LYN Complex. Dev Cell. 2020;54(3):302-16 e7.

339. Lopez-Cavestany M, Hahn SB, Hope JM, Reckhorn NT, Greenlee JD, Schwager SC, et al. Matrix stiffness induces epithelial-to-mesenchymal transition via Piezo1-regulated calcium flux in prostate cancer cells. iScience. 2023;26(4):106275.

340. Rice AJ, Cortes E, Lachowski D, Cheung BCH, Karim SA, Morton JP, et al. Matrix stiffness induces epithelial-mesenchymal transition and promotes chemoresistance in pancreatic cancer cells. Oncogenesis. 2017;6(7):e352.

341. Chen YQ, Lan HY, Wu YC, Yang WH, Chiou A, Yang MH. Epithelial-mesenchymal transition softens head and neck cancer cells to facilitate migration in 3D environments. J Cell Mol Med. 2018;22(8):3837-46.

342.	McGrail DJ, Mezencev R, Kieu QM, McDonald JF, Dawson MR. SNAIL-induced epithelial-to-mesenchymal transition produces concerted biophysical changes from altered cytoskeletal gene expression. FASEB J. 2015;29(4):1280-9.

343.	Weissenbruch K, Mayor R. Actomyosin forces in cell migration: Moving beyond cell body retraction. Bioessays. 2024;46(10):e2400055.

344.	Gauthier BR, Lorenzo PI, Comaills V. Physical Forces and Transient Nuclear Envelope Rupture during Metastasis: The Key for Success? Cancers (Basel). 2021;14(1).

345.	Li X, Shi J, Gao Z, Xu J, Wang S, Li X, et al. Biophysical studies of cancer cells' traverse-vessel behaviors under different pressures revealed cells' motion state transition. Sci Rep. 2022;12(1):7392.

346.	Roberts AB, Zhang J, Raj Singh V, Nikolic M, Moeendarbary E, Kamm RD, et al. Tumor cell nuclei soften during transendothelial migration. J Biomech. 2021;121:110400.

347.	Harada T, Swift J, Irianto J, Shin JW, Spinler KR, Athirasala A, et al. Nuclear lamin stiffness is a barrier to 3D migration, but softness can limit survival. J Cell Biol. 2014;204(5):669-82.

348.	Nasrollahi S, Walter C, Loza AJ, Schimizzi GV, Longmore GD, Pathak A. Past matrix stiffness primes epithelial cells and regulates their future collective migration through a mechanical memory. Biomaterials. 2017;146:146-55.

349.	Cambria E, Coughlin MF, Floryan MA, Offeddu GS, Shelton SE, Kamm RD. Linking cell mechanical memory and cancer metastasis. Nat Rev Cancer. 2024;24(3):216-28.

350.	Smit DJ, Pantel K. Circulating tumor cells as liquid biopsy markers in cancer patients. Mol Aspects Med. 2024;96:101258.

351.	Franken A, Kraemer A, Sicking A, Watolla M, Rivandi M, Yang L, et al. Comparative analysis of EpCAM high-expressing and low-expressing circulating tumour cells with regard to their clonal relationship and clinical value. Br J Cancer. 2023;128(9):1742-52.

352.	Mani V, Lyu Z, Kumar V, Ercal B, Chen H, Malhotra SV, et al. Epithelial-to-Mesenchymal Transition (EMT) and Drug Response in Dynamic Bioengineered Lung Cancer Microenvironment. Adv Biosyst. 2019;3(1):e1800223.

353.	Choi HY, Yang GM, Dayem AA, Saha SK, Kim K, Yoo Y, et al. Hydrodynamic shear stress promotes epithelial-mesenchymal transition by downregulating ERK and GSK3beta activities. Breast Cancer Res. 2019;21(1):6.

354.	Lee HJ, Ewere A, Diaz MF, Wenzel PL. TAZ responds to fluid shear stress to regulate the cell cycle. Cell Cycle. 2018;17(2):147-53.

355.	Lee HJ, Diaz MF, Price KM, Ozuna JA, Zhang S, Sevick-Muraca EM, et al. Fluid shear stress activates YAP1 to promote cancer cell motility. Nat Commun. 2017;8:14122.

356.	Yu H, He J, Su G, Wang Y, Fang F, Yang W, et al. Fluid shear stress activates YAP to promote epithelial-mesenchymal transition in hepatocellular carcinoma. Mol Oncol. 2021;15(11):3164-83.

357.	Alvarado-Estrada K, Marenco-Hillembrand L, Maharjan S, Mainardi VL, Zhang YS, Zarco N, et al. Circulatory shear stress induces molecular changes and side population enrichment in primary tumor-derived lung cancer cells with higher metastatic potential. Sci Rep. 2021;11(1):2800.

358.    Chambers AF, Groom AC, MacDonald IC. Dissemination and growth of cancer cells in metastatic sites. Nat Rev Cancer. 2002;2(8):563-72.

359.    Jiang K, Lim SB, Xiao J, Jokhun DS, Shang M, Song X, et al. Deleterious Mechanical Deformation Selects Mechanoresilient Cancer Cells with Enhanced Proliferation and Chemoresistance. Adv Sci (Weinh). 2023;10(22):e2201663.

360.    Gensbittel V, Krater M, Harlepp S, Busnelli I, Guck J, Goetz JG. Mechanical Adaptability of Tumor Cells in Metastasis. Dev Cell. 2021;56(2):164-79.

361.    Matos MA, Cicerone MT. Alternating current electric field effects on neural stem cell viability and differentiation. Biotechnol Prog. 2010;26(3):664-70.

362.    Liu M, Xie D, Zeng H, Zhai N, Liu L, Yan H. Direct-current electric field stimulation promotes proliferation and maintains stemness of mesenchymal stem cells. Biotechniques. 2023;74(6):293-301.

363.    Tajada S, Villalobos C. Calcium Permeable Channels in Cancer Hallmarks. Front Pharmacol. 2020;11:968.

364.    Sheth M, Esfandiari L. Bioelectric Dysregulation in Cancer Initiation, Promotion, and Progression. Front Oncol. 2022;12:846917.

365.    Fernandez Cabada T, Ruben M, El Merhie A, Proietti Zaccaria R, Alabastri A, Petrini EM, et al. Electrostatic polarization fields trigger glioblastoma stem cell differentiation. Nanoscale Horiz. 2022;8(1):95-107.

366.    Zhao H, Zhang W, Tang X, Galan EA, Zhu Y, Sang G, et al. Electrostatic potential difference between tumor and paratumor regulates cancer stem cell behavior and prognose tumor spread. Bioeng Transl Med. 2023;8(2):e10399.

367.    Garg AA, Jones TH, Moss SM, Mishra S, Kaul K, Ahirwar DK, et al. Electromagnetic fields alter the motility of metastatic breast cancer cells. Commun Biol. 2019;2:303.

368.    Dubrovska A, Krause M, Baumann M. 438Biological effect of radiotherapy on cancer cells. In: Pezzella F, Tavassoli M, Kerr DJ, Pezzella F, Tavassoli M, Kerr DJ, editors. Oxford Textbook of Cancer Biology: Oxford University Press; 2019. p. 0.

369.    Atun R, Jaffray DA, Barton MB, Bray F, Baumann M, Vikram B, et al. Expanding global access to radiotherapy. Lancet Oncol. 2015;16(10):1153-86.

370.    Baumann M, Krause M, Overgaard J, Debus J, Bentzen SM, Daartz J, et al. Radiation oncology in the era of precision medicine. Nat Rev Cancer. 2016;16(4):234-49.

371.    McCullough EC, Payne JT. X-ray-transmission computed tomography. Med Phys. 1977;4(2):85-98.

372.    Garcia-Figueiras R, Baleato-Gonzalez S, Padhani AR, Luna-Alcala A, Vallejo-Casas JA, Sala E, et al. How clinical imaging can assess cancer biology. Insights Imaging. 2019;10(1):28.

373.    Garra BS. Imaging and estimation of tissue elasticity by ultrasound. Ultrasound Q. 2007;23(4):255-68.

374. Duhon BH, Thompson K, Fisher M, Kaul VF, Nguyen HT, Harris MS, et al. Tumor biomechanical stiffness by magnetic resonance elastography predicts surgical outcomes and identifies biomarkers in vestibular schwannoma and meningioma. Sci Rep. 2024;14(1):14561.

375. Kwon MR, Youn I, Ko ES, Choi SH. Correlation of shear-wave elastography stiffness and apparent diffusion coefficient values with tumor characteristics in breast cancer. Sci Rep. 2024;14(1):7180.

376. Hennedige TP, Hallinan JT, Leung FP, Teo LL, Iyer S, Wang G, et al. Comparison of magnetic resonance elastography and diffusion-weighted imaging for differentiating benign and malignant liver lesions. Eur Radiol. 2016;26(2):398-406.

377. Reiss-Zimmermann M, Streitberger KJ, Sack I, Braun J, Arlt F, Fritzsch D, et al. High Resolution Imaging of Viscoelastic Properties of Intracranial Tumours by Multi-Frequency Magnetic Resonance Elastography. Clin Neuroradiol. 2015;25(4):371-8.

378. Patel BK, Samreen N, Zhou Y, Chen J, Brandt K, Ehman R, et al. MR Elastography of the Breast: Evolution of Technique, Case Examples, and Future Directions. Clin Breast Cancer. 2021;21(1):e102-e11.

379. Deng Y, Yi Z, Zhang T, Hu B, Zhang L, Rajlawot K, et al. Magnetic resonance elastography of the prostate in patients with lower urinary tract symptoms: feasibility of the modified driver at high multi-frequencies. Abdom Radiol (NY). 2022;47(1):399-408.

380. Iglesias-Garcia J, Larino-Noia J, de la Iglesia-Garcia D, Dominguez-Munoz JE. Endoscopic ultrasonography: Enhancing diagnostic accuracy. Best Pract Res Clin Gastroenterol. 2022;60-61:101808.

381. Xu W, Shi J, Zeng X, Li X, Xie WF, Guo J, et al. EUS elastography for the differentiation of benign and malignant lymph nodes: a meta-analysis. Gastrointest Endosc. 2011;74(5):1001-9; quiz 115 e1-4.

382. Lisotti A, Ricci C, Serrani M, Calvanese C, Sferrazza S, Brighi N, et al. Contrast-enhanced endoscopic ultrasound for the differential diagnosis between benign and malignant lymph nodes: a meta-analysis. Endosc Int Open. 2019;7(4):E504-E13.

383. Ignee A, Jenssen C, Arcidiacono PG, Hocke M, Moller K, Saftoiu A, et al. Endoscopic ultrasound elastography of small solid pancreatic lesions: a multicenter study. Endoscopy. 2018;50(11):1071-9.

384. Shen Y, He J, Liu M, Hu J, Wan Y, Zhang T, et al. Diagnostic value of contrast-enhanced ultrasound and shear-wave elastography for small breast nodules. PeerJ. 2024;12:e17677.

385. DeClerck YA. Desmoplasia: a response or a niche? Cancer Discov. 2012;2(9):772-4.

386. Yaffe MJ. Mammographic density. Measurement of mammographic density. Breast Cancer Res. 2008;10(3):209.

387. Thomas D, Radhakrishnan P. Tumor-stromal crosstalk in pancreatic cancer and tissue fibrosis. Mol Cancer. 2019;18(1):14.

388. Erstad DJ, Sojoodi M, Taylor MS, Jordan VC, Farrar CT, Axtell AL, et al. Fibrotic Response to Neoadjuvant Therapy Predicts Survival in Pancreatic Cancer and Is Measurable with Collagen-Targeted Molecular MRI. Clin Cancer Res. 2020;26(18):5007-18.

389.    Esfahani SA, Ma H, Krishna S, Shuvaev S, Sabbagh M, Deffler C, et al. Collagen type I PET/MRI enables evaluation of treatment response in pancreatic cancer in pre-clinical and first-in-human translational studies. Theranostics. 2024;14(15):5745-61.

390.    Kimura RH, Wang L, Shen B, Huo L, Tummers W, Filipp FV, et al. Evaluation of integrin alphavbeta(6) cystine knot PET tracers to detect cancer and idiopathic pulmonary fibrosis. Nat Commun. 2019;10(1):4673.

391.    Quigley NG, Steiger K, Richter F, Weichert W, Hoberuck S, Kotzerke J, et al. Tracking a TGF-beta activator in vivo: sensitive PET imaging of alphavbeta8-integrin with the Ga-68-labeled cyclic RGD octapeptide trimer Ga-68-Triveoctin. EJNMMI Res. 2020;10(1):133.

392.    Kimura RH, Iagaru A, Guo HH. Mini review of first-in-human integrin αvβ6 PET tracers. Frontiers in Nuclear Medicine. 2023;3.

393.    Liburkin-Dan T, Toledano S, Neufeld G. Lysyl Oxidase Family Enzymes and Their Role in Tumor Progression. Int J Mol Sci. 2022;23(11).

394.    Saatci O, Kaymak A, Raza U, Ersan PG, Akbulut O, Banister CE, et al. Targeting lysyl oxidase (LOX) overcomes chemotherapy resistance in triple negative breast cancer. Nat Commun. 2020;11(1):2416.

395.    Xiao Q, Ge G. Lysyl oxidase, extracellular matrix remodeling and cancer metastasis. Cancer Microenviron. 2012;5(3):261-73.

396.    Benson AB, 3rd, Wainberg ZA, Hecht JR, Vyushkov D, Dong H, Bendell J, et al. A Phase II Randomized, Double-Blind, Placebo-Controlled Study of Simtuzumab or Placebo in Combination with Gemcitabine for the First-Line Treatment of Pancreatic Adenocarcinoma. Oncologist. 2017;22(3):241-e15.

397.    Findlay A, Turner C, Schilter H, Deodhar M, Zhou W, Perryman L, et al. An activity-based bioprobe differentiates a novel small molecule inhibitor from a LOXL2 antibody and provides renewed promise for anti-fibrotic therapeutic strategies. Clin Transl Med. 2021;11(11):e572.

398.    Rowbottom MW, Bain G, Calderon I, Lasof T, Lonergan D, Lai A, et al. Identification of 4-(Aminomethyl)-6-(trifluoromethyl)-2-(phenoxy)pyridine Derivatives as Potent, Selective, and Orally Efficacious Inhibitors of the Copper-Dependent Amine Oxidase, Lysyl Oxidase-Like 2 (LOXL2). J Med Chem. 2017;60(10):4403-23.

399.    Chitty JL, Yam M, Perryman L, Parker AL, Skhinas JN, Setargew YFI, et al. A first-in-class pan-lysyl oxidase inhibitor impairs stromal remodeling and enhances gemcitabine response and survival in pancreatic cancer. Nat Cancer. 2023;4(9):1326-44.

400.    How JL, Z.; Lombardi Story, J.; Neuberg, D.S.; Ravid, K.; Jarolimek, W.; Charlton, B.; Hobbs, G.S. Evaluation of a Pan-Lysyl Oxidase Inhibitor, Pxs-5505, in Myelofibrosis: A Phase I, Randomized, Placebo Controlled Double Blind Study in Healthy Adults. Blood. 2020;136(Suppl 1):16.

401.    Karki T, Rajakyla EK, Acheva A, Tojkander S. TRPV6 calcium channel directs homeostasis of the mammary epithelial sheets and controls epithelial mesenchymal transition. Sci Rep. 2020;10(1):14683.

402.    Wang Y, Deng X, Zhang R, Lyu H, Xiao S, Guo D, et al. The TRPV6 Calcium Channel and Its Relationship with Cancer. Biology (Basel). 2024;13(3).

403.    Fu S, Hirte H, Welch S, Ilenchuk TT, Lutes T, Rice C, et al. First-in-human phase I study of SOR-C13, a TRPV6 calcium channel inhibitor, in patients with advanced solid tumors. Invest New Drugs. 2017;35(3):324-33.

404.    Rosenbaum T, Benitez-Angeles M, Sanchez-Hernandez R, Morales-Lazaro SL, Hiriart M, Morales-Buenrostro LE, et al. TRPV4: A Physio and Pathophysiologically Significant Ion Channel. Int J Mol Sci. 2020;21(11).

405.    Toft-Bertelsen TL, MacAulay N. TRPing to the Point of Clarity: Understanding the Function of the Complex TRPV4 Ion Channel. Cells. 2021;10(1).

406.    Baratchi S, Knoerzer M, Khoshmanesh K, Mitchell A, McIntyre P. Shear Stress Regulates TRPV4 Channel Clustering and Translocation from Adherens Junctions to the Basal Membrane. Sci Rep. 2017;7(1):15942.

407.    Li X, Cheng Y, Wang Z, Zhou J, Jia Y, He X, et al. Calcium and TRPV4 promote metastasis by regulating cytoskeleton through the RhoA/ROCK1 pathway in endometrial cancer. Cell Death Dis. 2020;11(11):1009.

408.    Liu J, Guo Y, Zhang R, Xu Y, Luo C, Wang R, et al. Inhibition of TRPV4 remodels single cell polarity and suppresses the metastasis of hepatocellular carcinoma. Cell Death Dis. 2023;14(6):379.

409.    Fang Y, Liu G, Xie C, Qian K, Lei X, Liu Q, et al. Pharmacological inhibition of TRPV4 channel suppresses malignant biological behavior of hepatocellular carcinoma via modulation of ERK signaling pathway. Biomed Pharmacother. 2018;101:910-9.

410.    Goyal N, Skrdla P, Schroyer R, Kumar S, Fernando D, Oughton A, et al. Clinical Pharmacokinetics, Safety, and Tolerability of a Novel, First-in-Class TRPV4 Ion Channel Inhibitor, GSK2798745, in Healthy and Heart Failure Subjects. Am J Cardiovasc Drugs. 2019;19(3):335-42.

411.    Yee NS. Roles of TRPM8 Ion Channels in Cancer: Proliferation, Survival, and Invasion. Cancers (Basel). 2015;7(4):2134-46.

412.    Tolcher AP, A.; Papadopoulos, K.; Mays, T.; Stephan, t.; Humble D.J.; Frohlich, M.W.; Sims, R.B. Preliminary results from a Phase 1 study of D-3263 HCl, a TRPM8 calcium channel agonist, in patients with advanced cancer. European Journal of Cancer 2010;8 (7, Suppl):119.

413.    Ross TD, Coon BG, Yun S, Baeyens N, Tanaka K, Ouyang M, et al. Integrins in mechanotransduction. Curr Opin Cell Biol. 2013;25(5):613-8.

414.    Jin H, Varner J. Integrins: roles in cancer development and as treatment targets. Br J Cancer. 2004;90(3):561-5.

415.    Eke I, Cordes N. Focal adhesion signaling and therapy resistance in cancer. Semin Cancer Biol. 2015;31:65-75.

416.    Slack RJ, Macdonald SJF, Roper JA, Jenkins RG, Hatley RJD. Emerging therapeutic opportunities for integrin inhibitors. Nat Rev Drug Discov. 2022;21(1):60-78.

417.	Stupp R, Hegi ME, Gorlia T, Erridge SC, Perry J, Hong YK, et al. Cilengitide combined with standard treatment for patients with newly diagnosed glioblastoma with methylated MGMT promoter (CENTRIC EORTC 26071-22072 study): a multicentre, randomised, open-label, phase 3 trial. Lancet Oncol. 2014;15(10):1100-8.

418.	Bell-McGuinn KM, Matthews CM, Ho SN, Barve M, Gilbert L, Penson RT, et al. A phase II, single-arm study of the anti-alpha5beta1 integrin antibody volociximab as monotherapy in patients with platinum-resistant advanced epithelial ovarian or primary peritoneal cancer. Gynecol Oncol. 2011;121(2):273-9.

419.	Elez E, Kocakova I, Hohler T, Martens UM, Bokemeyer C, Van Cutsem E, et al. Abituzumab combined with cetuximab plus irinotecan versus cetuximab plus irinotecan alone for patients with KRAS wild-type metastatic colorectal cancer: the randomised phase I/II POSEIDON trial. Ann Oncol. 2015;26(1):132-40.

420.	O'Day S, Pavlick A, Loquai C, Lawson D, Gutzmer R, Richards J, et al. A randomised, phase II study of intetumumab, an anti-alphav-integrin mAb, alone and with dacarbazine in stage IV melanoma. Br J Cancer. 2011;105(3):346-52.

421.	Raab-Westphal S, Marshall JF, Goodman SL. Integrins as Therapeutic Targets: Successes and Cancers. Cancers (Basel). 2017;9(9).

422.	Bergonzini C, Kroese K, Zweemer AJM, Danen EHJ. Targeting Integrins for Cancer Therapy - Disappointments and Opportunities. Front Cell Dev Biol. 2022;10:863850.

423.	Mikic N, Gentilal N, Cao F, Lok E, Wong ET, Ballo M, et al. Tumor-treating fields dosimetry in glioblastoma: Insights into treatment planning, optimization, and dose-response relationships. Neurooncol Adv. 2024;6(1):vdae032.

424.	Moser JC, Salvador E, Deniz K, Swanson K, Tuszynski J, Carlson KW, et al. The Mechanisms of Action of Tumor Treating Fields. Cancer Res. 2022;82(20):3650-8.

425.	Huang M, Li P, Chen F, Cai Z, Yang S, Zheng X, et al. Is extremely low frequency pulsed electromagnetic fields applicable to gliomas? A literature review of the underlying mechanisms and application of extremely low frequency pulsed electromagnetic fields. Cancer Med. 2023;12(3):2187-98.

426.	Kirson ED, Gurvich Z, Schneiderman R, Dekel E, Itzhaki A, Wasserman Y, et al. Disruption of cancer cell replication by alternating electric fields. Cancer Res. 2004;64(9):3288-95.

427.	Jeong H, Sung J, Oh S-i, Jeong S, Koh EK, Hong S, et al. Inhibition of brain tumor cell proliferation by alternating electric fields. Applied Physics Letters. 2014;105(20).

428.	Giladi M, Schneiderman RS, Voloshin T, Porat Y, Munster M, Blat R, et al. Mitotic Spindle Disruption by Alternating Electric Fields Leads to Improper Chromosome Segregation and Mitotic Catastrophe in Cancer Cells. Sci Rep. 2015;5:18046.

429.	Kirson ED, Dbaly V, Tovarys F, Vymazal J, Soustiel JF, Itzhaki A, et al. Alternating electric fields arrest cell proliferation in animal tumor models and human brain tumors. Proc Natl Acad Sci U S A. 2007;104(24):10152-7.

430.	Timmons JJ, Preto J, Tuszynski JA, Wong ET. Tubulin's response to external electric fields by molecular dynamics simulations. PLoS One. 2018;13(9):e0202141.

431.    Gera N, Yang A, Holtzman TS, Lee SX, Wong ET, Swanson KD. Tumor treating fields perturb the localization of septins and cause aberrant mitotic exit. PLoS One. 2015;10(5):e0125269.

432.    Rominiyi O, Vanderlinden A, Clenton SJ, Bridgewater C, Al-Tamimi Y, Collis SJ. Tumour treating fields therapy for glioblastoma: current advances and future directions. Br J Cancer. 2021;124(4):697-709.

433.    Stupp R, Taillibert S, Kanner A, Read W, Steinberg D, Lhermitte B, et al. Effect of Tumor-Treating Fields Plus Maintenance Temozolomide vs Maintenance Temozolomide Alone on Survival in Patients With Glioblastoma: A Randomized Clinical Trial. JAMA. 2017;318(23):2306-16.

434.    Toms SA, Kim CY, Nicholas G, Ram Z. Increased compliance with tumor treating fields therapy is prognostic for improved survival in the treatment of glioblastoma: a subgroup analysis of the EF-14 phase III trial. J Neurooncol. 2019;141(2):467-73.

435.    Ceresoli GL, Aerts JG, Dziadziuszko R, Ramlau R, Cedres S, van Meerbeeck JP, et al. Tumour Treating Fields in combination with pemetrexed and cisplatin or carboplatin as first-line treatment for unresectable malignant pleural mesothelioma (STELLAR): a multicentre, single-arm phase 2 trial. Lancet Oncol. 2019;20(12):1702-9.

436.    Pless M, Droege C, von Moos R, Salzberg M, Betticher D. A phase I/II trial of Tumor Treating Fields (TTFields) therapy in combination with pemetrexed for advanced non-small cell lung cancer. Lung Cancer. 2013;81(3):445-50.

437.    Rivera F, Benavides M, Gallego J, Guillen-Ponce C, Lopez-Martin J, Kung M. Tumor treating fields in combination with gemcitabine or gemcitabine plus nab-paclitaxel in pancreatic cancer: Results of the PANOVA phase 2 study. Pancreatology. 2019;19(1):64-72.

438.    Ringel-Scaia VM, Beitel-White N, Lorenzo MF, Brock RM, Huie KE, Coutermarsh-Ott S, et al. High-frequency irreversible electroporation is an effective tumor ablation strategy that induces immunologic cell death and promotes systemic anti-tumor immunity. EBioMedicine. 2019;44:112-25.

439.    Zhang N, Li Z, Han X, Zhu Z, Li Z, Zhao Y, et al. Irreversible Electroporation: An Emerging Immunomodulatory Therapy on Solid Tumors. Front Immunol. 2021;12:811726.

440.    Zhong S, Yao S, Zhao Q, Wang Z, Liu Z, Li L, et al. Electricity-Assisted Cancer Therapy: From Traditional Clinic Applications to Emerging Methods Integrated with Nanotechnologies. Advanced NanoBiomed Research. 2023;3(3):2200143.

441.    George AK, Miocinovic R, Patel AR, Lomas DJ, Correa AF, Chen DYT, et al. A Description and Safety Overview of Irreversible Electroporation for Prostate Tissue Ablation in Intermediate-Risk Prostate Cancer Patients: Preliminary Results from the PRESERVE Trial. Cancers (Basel). 2024;16(12).

442.    Tasu JP, Herpe G, Damion J, Richer JP, Debeane B, Vionnet M, et al. Irreversible electroporation to bring initially unresectable locally advanced pancreatic adenocarcinoma to surgery: the IRECAP phase II study. Eur Radiol. 2024;34(10):6885-95.

443.    van Veldhuisen E, Vroomen LG, Ruarus AH, Derksen TC, Busch OR, de Jong MC, et al. Value of CT-Guided Percutaneous Irreversible Electroporation Added to FOLFIRINOX

Chemotherapy in Locally Advanced Pancreatic Cancer: A Post Hoc Comparison. J Vasc Interv Radiol. 2020;31(10):1600-8.

444. Lin M, Zhang X, Liang S, Luo H, Alnaggar M, Liu A, et al. Irreversible electroporation plus allogenic Vgamma9Vdelta2 T cells enhances antitumor effect for locally advanced pancreatic cancer patients. Signal Transduct Target Ther. 2020;5(1):215.

445. Meijerink MR, Ruarus AH, Vroomen L, Puijk RS, Geboers B, Nieuwenhuizen S, et al. Irreversible Electroporation to Treat Unresectable Colorectal Liver Metastases (COLDFIRE-2): A Phase II, Two-Center, Single-Arm Clinical Trial. Radiology. 2021;299(2):470-80.

446. Lepock JR. How do cells respond to their thermal environment? Int J Hyperthermia. 2005;21(8):681-7.

447. Mota-Rojas D, Titto CG, Orihuela A, Martinez-Burnes J, Gomez-Prado J, Torres-Bernal F, et al. Physiological and Behavioral Mechanisms of Thermoregulation in Mammals. Animals (Basel). 2021;11(6).

448. Richter K, Haslbeck M, Buchner J. The heat shock response: life on the verge of death. Mol Cell. 2010;40(2):253-66.

449. Repasky EA, Evans SS, Dewhirst MW. Temperature matters! And why it should matter to tumor immunologists. Cancer Immunol Res. 2013;1(4):210-6.

450. Zhu L, Altman MB, Laszlo A, Straube W, Zoberi I, Hallahan DE, et al. Ultrasound Hyperthermia Technology for Radiosensitization. Ultrasound Med Biol. 2019;45(5):1025-43.

451. Brace C. Thermal tumor ablation in clinical use. IEEE Pulse. 2011;2(5):28-38.

452. Lu J, Chen L, Song Z, Das M, Chen J. Hypothermia Effectively Treats Tumors with Temperature-Sensitive p53 Mutations. Cancer Res. 2021;81(14):3905-15.

453. Kok HP, Cressman ENK, Ceelen W, Brace CL, Ivkov R, Grull H, et al. Heating technology for malignant tumors: a review. Int J Hyperthermia. 2020;37(1):711-41.

454. Lubner MG, Brace CL, Hinshaw JL, Lee FT, Jr. Microwave tumor ablation: mechanism of action, clinical results, and devices. J Vasc Interv Radiol. 2010;21(8 Suppl):S192-203.

455. Dai Q, Cao B, Zhao S, Zhang A. Synergetic Thermal Therapy for Cancer: State-of-the-Art and the Future. Bioengineering (Basel). 2022;9(9).

456. Overgaard J. The current and potential role of hyperthermia in radiotherapy. Int J Radiat Oncol Biol Phys. 1989;16(3):535-49.

457. Hannon G, Tansi FL, Hilger I, Prina-Mello A. The Effects of Localized Heat on the Hallmarks of Cancer. Advanced Therapeutics. 2021;4(7):2000267.

458. Foo CY, Munir N, Kumaria A, Akhtar Q, Bullock CJ, Narayanan A, et al. Medical Device Advances in the Treatment of Glioblastoma. Cancers (Basel). 2022;14(21).

459. Chang D, Lim M, Goos J, Qiao R, Ng YY, Mansfeld FM, et al. Biologically Targeted Magnetic Hyperthermia: Potential and Limitations. Front Pharmacol. 2018;9:831.

460. Chicheł AS, J.; Kubaszewska, M.; Kanikowski M.,. Hyperthermia – description of a method and a review of clinical applications. Reports of Practical Oncology & Radiotherapy. 2007;12(5):267-75.

461. Datta NR, Ordonez SG, Gaipl US, Paulides MM, Crezee H, Gellermann J, et al. Local hyperthermia combined with radiotherapy and-/or chemotherapy: recent advances and promises for the future. Cancer Treat Rev. 2015;41(9):742-53.

462. Jue JS, Coons S, Hautvast G, Thompson SF, Geraats J, Richstone L, et al. Novel Automated Three-Dimensional Surgical Planning Tool and Magnetic Resonance Imaging/Ultrasound Fusion Technology to Perform Nanoparticle Ablation and Cryoablation of the Prostate for Focal Therapy. J Endourol. 2022;36(3):369-72.

463. Bravo M, Fortuni B, Mulvaney P, Hofkens J, Uji IH, Rocha S, et al. Nanoparticle-mediated thermal Cancer therapies: Strategies to improve clinical translatability. J Control Release. 2024;372:751-77.

464. Overgaard J, Gonzalez Gonzalez D, Hulshof MC, Arcangeli G, Dahl O, Mella O, et al. Randomised trial of hyperthermia as adjuvant to radiotherapy for recurrent or metastatic malignant melanoma. European Society for Hyperthermic Oncology. Lancet. 1995;345(8949):540-3.

465. Winter C, Kristiansen G, Kersting S, Roy J, Aust D, Knosel T, et al. Google goes cancer: improving outcome prediction for cancer patients by network-based ranking of marker genes. PLoS Comput Biol. 2012;8(5):e1002511.

466. Kim W, Kim KS, Lee JE, Noh DY, Kim SW, Jung YS, et al. Development of novel breast cancer recurrence prediction model using support vector machine. J Breast Cancer. 2012;15(2):230-8.

467. Bi WL, Hosny A, Schabath MB, Giger ML, Birkbak NJ, Mehrtash A, et al. Artificial intelligence in cancer imaging: Clinical challenges and applications. CA Cancer J Clin. 2019;69(2):127-57.

468. Wang S, Yang DM, Rong R, Zhan X, Fujimoto J, Liu H, et al. Artificial Intelligence in Lung Cancer Pathology Image Analysis. Cancers (Basel). 2019;11(11).

469. Sheth D, Giger ML. Artificial intelligence in the interpretation of breast cancer on MRI. J Magn Reson Imaging. 2020;51(5):1310-24.

470. Chiu HY, Chao HS, Chen YM. Application of Artificial Intelligence in Lung Cancer. Cancers (Basel). 2022;14(6).

471. Wen T, Tong B, Liu Y, Pan T, Du Y, Chen Y, et al. Review of research on the instance segmentation of cell images. Comput Methods Programs Biomed. 2022;227:107211.

472. Gong Z, Zhang J, Guo W. Tumor purity as a prognosis and immunotherapy relevant feature in gastric cancer. Cancer Med. 2020;9(23):9052-63.

473. Zhu W, Xie L, Han J, Guo X. The Application of Deep Learning in Cancer Prognosis Prediction. Cancers (Basel). 2020;12(3).

474. Kehl KL, Xu W, Gusev A, Bakouny Z, Choueiri TK, Riaz IB, et al. Artificial intelligence-aided clinical annotation of a large multi-cancer genomic dataset. Nat Commun. 2021;12(1):7304.

475.     Sun D, Wang M, Li A. A multimodal deep neural network for human breast cancer prognosis prediction by integrating multi-dimensional data. IEEE/ACM Trans Comput Biol Bioinform. 2018.

476.     Hu S, Li, G.;, Xue, L.; Xu, M.; Xiang, A.; Cao, Z. . Machine Learning for Identification of Cancer Cells Based on Electrical Property Using Microfluidic Impedance Flow Cytometry. J Phys: Conf Ser 2024;2809:012024.

477.     Wei J, Gao W, Yang X, Yu Z, Su F, Han C, et al. Machine learning classification of cellular states based on the impedance features derived from microfluidic single-cell impedance flow cytometry. Biomicrofluidics. 2024;18(1):014103.

478.     Ahuja K, Rather GM, Lin Z, Sui J, Xie P, Le T, et al. Toward point-of-care assessment of patient response: a portable tool for rapidly assessing cancer drug efficacy using multifrequency impedance cytometry and supervised machine learning. Microsyst Nanoeng. 2019;5:34.

479.     Englezos D, Voutouri C, Stylianopoulos T. Machine learning analysis reveals tumor stiffness and hypoperfusion as biomarkers predictive of cancer treatment efficacy. Transl Oncol. 2024;44:101944.

480.     Ning Y, Lin K, Fang J, Ding Y, Zhang Z, Chen X, et al. Gastrointestinal pan-cancer landscape of tumor matrix heterogeneity identifies biologically distinct matrix stiffness subtypes predicting prognosis and chemotherapy efficacy. Comput Struct Biotechnol J. 2023;21:2744-58.

481.     Rozen R, Weihs D. Machine-Learning Provides Patient-Specific Prediction of Metastatic Risk Based on Innovative, Mechanobiology Assay. Ann Biomed Eng. 2021;49(7):1774-83.

482.     Jaeschke A, Jacobi A, Lawrence MG, Risbridger GP, Frydenberg M, Williams ED, et al. Cancer-associated fibroblasts of the prostate promote a compliant and more invasive phenotype in benign prostate epithelial cells. Mater Today Bio. 2020;8:100073.

483.     Stylianou A, Gkretsi V, Stylianopoulos T. Transforming growth factor-beta modulates pancreatic cancer associated fibroblasts cell shape, stiffness and invasion. Biochim Biophys Acta Gen Subj. 2018;1862(7):1537-46.

484.     Huang YL, Dickerson LK, Kenerson H, Jiang X, Pillarisetty V, Tian Q, et al. Organotypic Models for Functional Drug Testing of Human Cancers. BME Front. 2023;4:0022.

485.     Schulz A, Meyer F, Dubrovska A, Borgmann K. Cancer Stem Cells and Radioresistance: DNA Repair and Beyond. Cancers (Basel). 2019;11(6).

486.     Tang DG. Understanding cancer stem cell heterogeneity and plasticity. Cell Res. 2012;22(3):457-72.

487.     Sun Y, Li H, Chen Q, Luo Q, Song G. The distribution of liver cancer stem cells correlates with the mechanical heterogeneity of liver cancer tissue. Histochem Cell Biol. 2021;156(1):47-58.

Supplementary information

**Supplementary Table 1**: Most important parameters that can be extracted from impedance and how to derive them

| Parameter | Equation | Main Contributor(s) |
|---|---|---|
| Impedance ($Z$) | $Z(\omega) = Z_{Re}(\omega) + jZ_{im}(\omega)$<br><br>$j = imaginary\ unit$<br>$\omega = 2\pi f\ (angular\ frequeny\ )$<br>$f = frequency\ [Hz]$<br><br>$Z = |Z|\tan\theta;\ Z = |Z|e^{j\theta}$ | Cell membrane, cytoplasm, extracellular fluid/ medium |
| Magnitude ($|Z|$) | $|Z| = \sqrt{R^2 + X^2}$ | Cell membrane, cytoplasm, extracellular fluid/ medium |
| Phase angle ($\theta$) | $\theta = \tan^{-1}\left(\frac{X}{R}\right)$ | Cell membrane, cytoplasm, |
| Resistance ($R$) | $R = |Z|\cos\theta$ | Cytoplasm, extracellular fluid/ medium, ion channels |
| Reactance ($X$) | $X = |Z|\sin\theta$ | Cell membrane |
| Capacitance ($C$) | $C = -\dfrac{1}{\omega X}$ | Cell membrane |
| Permittivity ($\varepsilon$) | $\varepsilon(\omega) = \varepsilon_{Re}(\omega) - j\varepsilon_{Im}(\omega)$ | Cell membrane |
| Conductivity ($\sigma$) | $\sigma = \omega\varepsilon_0\varepsilon_{Im}$<br><br>$\varepsilon_0 = 8.854 \times 10^{-12}\ F/m$<br>$permittivity\ in\ vacuum$ | Cytoplasm, extracellular fluid/ medium |